\documentclass[11pt,a4paper]{article}
\usepackage{graphicx,theorem,mathrsfs,amssymb,color,dsfont}
\usepackage{amsmath}
\usepackage{hyperref}
\usepackage{bm}
\usepackage{bbm}
\usepackage{amssymb}
\usepackage{authblk}
\usepackage{amsfonts}
\usepackage[table]{xcolor}
\usepackage{caption}

\newtheorem{definition}{Definition}[section]
\newtheorem{proposition}{Proposition}[section]
\newtheorem{lemma}{Lemma}[section]
\newtheorem{theorem}{Theorem}[section]

\newtheorem{remark}{Remark}[section]

\definecolor{red}{rgb}{1,0,0}

\begin{document}

%%\title{\bf Characterizing feasible equilibrium solutions for MAL kinetic systems}

\title{\bf Uniqueness of feasible equilibria for mass action law (MAL) kinetic systems}

\author[1]{Antonio A. Alonso\thanks{Author to whom correspondence should be addressed. E-mail: antonio@iim.csic.es}}
\author[2]{G\'abor Szederk\'enyi \thanks{szederkenyi@itk.ppke.hu}}
\affil[1]{Process Engineering Group, IIM-CSIC.
Spanish Council for Scientific Research. Eduardo Cabello 6, 36208 Vigo, Spain}
\affil[2]{Pazmany Peter Catholic University, Faculty of Information Technology and Bionics,
Prater u. 50/a, H-1083 Budapest, Hungary
}

%
%\author{Antonio A. Alonso
%\thanks{Author to whom correspondence should be addressed. E-mail: antonio@iim.csic.es},
%G\'abor Szederk\'enyi}

\maketitle
\date{}

%\begin{center}
%\textit{Process Engineering Group, IIM-CSIC}\\
%\textit{Spanish Council for Scientific Research.}\\
%\textit{Eduardo Cabello 6, 36208 Vigo, Spain}\\
%\end{center}

\begin{abstract}

This paper studies the relations among system parameters, uniqueness, and stability of equilibria, for kinetic systems given in the form of polynomial ODEs. Such models are commonly used to describe the dynamics of nonnegative systems, with a wide range of application fields such as chemistry, systems biology, process modeling or even transportation systems. Using a flux-based description of kinetic models, a canonical representation of the set of all possible feasible equilibria is developed.

The characterization is made in terms of strictly stable compartmental matrices to define the so-called family of solutions. Feasibility is imposed by a set of constraints, which are linear on a log-transformed space of complexes, and relate to the kernel of a matrix, the columns of which span the stoichiometric subspace. One particularly interesting representation of these constraints can be expressed in terms of a class of monotonous decreasing functions. This allows  connections to be established with classical results in CRNT that relate to the existence and uniqueness of equilibria along positive stoichiometric compatibility classes.

In particular, monotonicity can be employed to identify regions in the set of possible reaction rate coefficients leading to complex balancing, and to conclude uniqueness of equilibria for a class of positive deficiency networks. The latter result might support constructing an alternative proof of the well-known deficiency one theorem. The developed notions and results are illustrated through examples.
\end{abstract}

{\bf Keywords:} Chemical reaction networks, kinetic systems, mass action
law, network deficiency, feasible equilibrium, complex balanced equilibrium

\begin{small}
This manuscript was published as: A. A. Alonso and G. Szederk\'enyi. Uniqueness of feasible equilibria for mass action law (MAL) kinetic systems. \textit{Journal of Process Control}, 48: 41--71, 2016. DOI link: \url{http://dx.doi.org/10.1016/j.jprocont.2016.10.002} 
\end{small}

\newpage

\baselineskip=1.25\normalbaselineskip

\section*{Nomenclature}

\smallskip
%\begin{table}[htbp]
{\footnotesize \centering
    \begin{tabular}{lll}
    Notation                &       Description & Defining/introducing eqn. \\
     & & or (sub)section\\ \hline
    ${\mathbb{R}}^{n}$ & $n$-dimensional real space & \\
    ${\mathbb{R}}^{n}_{>0}$ (${\mathbb{R}}^{n}_{<0}$)  & $n$-dimensional positive (resp. negative) orthant & \\
    ${\mathbb{R}}^{n}_{\geq 0}$ (${\mathbb{R}}^{n}_{\leq 0}$)  & $n$-dimensional non-negative (resp. non-positive) orthant & \\
    ${\textbf{x}} > 0$ (${\textbf{x}} <0$)  & each element of the vector ${\textbf{x}}$ is positive (resp. negative) & \\
    ${\textbf{x}} \geq 0$ (${\textbf{x}}  \leq 0$)  & each element of the vector ${\textbf{x}}$ is non-negative (resp. non-positive) & \\
       ${\mathbf{1}}_{n} \in \mathbb{R}^n$ & The $n$-dimensional vector with each element being one &  \\
    $\bm{\varepsilon}_i$ & the $i$th standard basis vector in $\mathbb{R}^n$ & sec. 2, sec.3 \\
    ${\cal{D}} (\textbf{x})$ & diagonal matrix ${\cal{D}} (\textbf{x}) \in{\mathbb{R}}^{n \times n}$ with components of $\textbf{x}$ in the diagonal  & eq. \eqref{eq:Aux_W_diff_Flux},\eqref{eq:hDp}\\
    $m$ & number of species & sec. 2  \\
    $n$ & number of complexes & sec. 2\\
    $\rho$ & number irreversible chemical reaction steps in the network & sec. 2\\
    $R_{ij}$ & rate of the reaction from complex $i$ to complex $j$ & eq. \eqref{eq:reaction_rate}\\
    $k_{ij}$ & rate coefficient of the reaction from complex $i$ to complex $j$ & eq. \eqref{eq:reaction_rate}\\
%%    $\mathcal{L}$ & the set of all linkage classes & sec. 2\\
    $\ell$ & the number of linkage classes & sec. 2 \\
    $\lambda$ & integer for indexing linkage classes & sec. 2 \\
    $\mathcal{L}_{\lambda}$ & the $\lambda$th linkage class & sec. 2\\
    $N_{\lambda}$ & the number of complexes in linkage class no. $\lambda$   & sec. 2 (footnote 1) \\
    $j_{\lambda}$ & index of the reference complex in linkage class no. $\lambda$ & subsec. 2.1\\
    $\textbf{c}$ & vector of concentrations (state variables) & sec. 2\\
    $Y$ & $m\times n$ dimensional molecularity matrix & eq. \eqref{subeq:canonical_form_AK}\\
    $\bm{\psi}:\mathbb{R}^m\rightarrow\mathbb{R}^n$ & monomial function of the kinetic dynamics, $\psi_i(c)=\prod_{j=1}^{m}c_j^{Y_{ji}}$ & eq. \eqref{eq:reaction_rate}  \\
    $\phi_i$ & net reaction rate corresponding to complex $i$ & eq. \eqref{eq:this_is_flux}\\
    $\Delta$ & subspace containing im($A_k$) & eq. \eqref{eq:subspace_DELTA} \\
    $\Xi$ & stoichiometric subspace & eq. \eqref{eq:stoichiometric_subsp}\\
    $s$ & dimension of the stoichiometric subspace & subsec. 2.2\\
    $\delta$ & deficiency of the reaction network &  eq. \eqref{eq:deficiency}\\
%    $\Omega(\textbf{c}_0)$ & stoichiometric compatibility for a reference concentration $\textbf{c}_0$ & eq. \eqref{eq:eqpolyhedron} \\
\end{tabular}}

%

%%----------------------------------------------------------------------------------------------------------------
\section{Introduction}
%%----------------------------------------------------------------------------------------------------------------
%%
Deterministic reaction networks obeying  mass action law (MAL) kinetics form an important subclass of kinetic systems, which in spite of their apparent
simplicity, are able to describe a rich variety of dynamical behavior, that includes multiple equilibria conditions, oscillations or even chaos \cite{Erdi:89, Chellaboina:09}.
Such networks are typically employed to describe the dynamics of open or closed chemical reaction systems, but over the last years they proved useful in modeling other system classes as well.

Reaction networks belong to the class of nonnegative (or positive) systems, the main characteristics of which being that the non-negative orthant is invariant for the dynamics. The application field of nonnegative systems extends far beyond chemistry, and includes dynamical models whose state variables are naturally nonnegative, as it is the case of biological systems in their many scales (from cells to ecological systems), or systems that can be transformed to be nonnegative, such as certain process models (e.g. heat exchangers, distillation columns, convection networks), economic, transportation or stochastic models \cite{Farina2000}. With an appropriate selection of coordinates, even many classical mechanical and electrical models can be described in the nonnegative framework.

The main specialty of reaction networks within nonnegative polynomial models is the lack of so-called cross-effects, what defines an additional constraint between the monomial coefficients and exponents \cite{Erdi:89}. Still, the class of reaction network models is quite wide, and many non-chemical models can be brought into a kinetic form using simple transformations \cite{Chellaboina:09,Samardzija1989}. Widely used examples of kinetic systems are compartmental systems \cite{Haddad2010} and Lotka-Volterra models into which most smooth nonlinear ODEs can be embedded \cite{HerFaiBre:98}.  These facts,  clearly underline the importance of reaction network models and motivate us to attempt to look at general dynamical models through the glasses of kinetic systems.

The study of the relationships between chemical reaction structure and dynamic behaviour is the purpose of Chemical Reaction Network Theory (CRNT), a program formally proposed and developed in \cite{aris:65,aris:69, krambeck:70}. One of the earliest results on the relation between the solutions of nonlinear dynamical systems (including kinetic systems) and their associated directed graphs is published in \cite{Volpert1972}. Important cycle-related conditions on the stability of kinetic systems were given in \cite{Clarke1975}. An extensive stability analysis of reaction networks using algebraic and graph-theoretical tools can be found in \cite{Clarke1980}.  Thermodynamically motivated Lyapunov-function-based stability analysis of kinetic systems, considering certain frequently applied model-simplification steps, is proposed in \cite{Hangos2010}.

The seminal works in \cite{Horn-Jackson:72,Feinberg:72} (collected in their most comprehensive form in \cite{Feinberg:87}) explored the dynamic properties of MAL complex chemical systems, and contributed to equip CRNT with a mathematical formalism that has prevailed to present. It is important to remark here, that several different network structures may correspond to the same kinetic differential equations \cite{Horn-Jackson:72,Craciun2008}. Therefore, important network properties such as deficiency, weak reversibility or complex balancing, may vary among the possible reaction structures belonging to the same ODE model (see, e.g.  \cite{Szederkenyi2011a,Johnston2013}).

One fundamental problem in CRNT
is to decide from the structure or parameters of the network, whether it can exhibit or not multiple equilibria.
An important early result in this field is the rigorous proof of the existence and uniqueness of thermodynamic equilibrium in a mixture of chemically reacting ideal gases \cite{Zeldovich2014}. The motivation in \cite{Shapiro1965} was the computational analysis of large thermodynamical models. The work contains fundamental results about the existence and uniqueness of compositions minimizing the free energy.

In answering the questions about the properties of equilibria, the concept of
network deficiency (a number that relates to reaction network structure and stoichiometry) has become central to characterize the network behavior. Two essential results of CRNT are the well-known deficiency
zero and deficiency one theorems \cite{Feinberg:87, Feinberg:95}
which
(besides other important results) establish conditions for networks to have exactly one equilibrium point in each positive stoichiometric compatibility class \cite{Feinberg:87}.
This suggests network
robustness with respect to parameter variability, and underlines the
importance of the kinetic system class in general nonlinear systems
theory.

CRNT has received renewed interest over the last years, particularly in the area of systems biology, because of its potential to explore and to analyse complex behavior and functionality in biological systems (e.g. \cite{Conradi:07, MinchevaCracium:08, Conradietal:12, Otero:12}). Most efforts were dedicated to investigate the relationships between reaction network structure and dynamic behaviour. In this regard, special mention should be made of the so-called injectivity property, investigated as a condition that relates to the singularity (or not) of the determinant of the Jacobian associated to a given dynamic system \cite{CraciunFeinberg:05}. Algebraic and graph theoretical methods have been devised to check injectivity, and therefore uniqueness of equilibria \cite{CraciunFeinberg:05, CraciunFeinberg:06}. In the same direction, extensions to cope with instabilities have been developed in \cite{MinchevaCracium:08}.  From different perspectives, a number of necessary and sufficient conditions for a given network structure and stoichiometry to accommodate multiple equilibria have been also recently proposed in \cite{Conradietal:12, Otero:12}.

A particularly interesting class of chemical networks are the reversible ones, either in the strict thermodynamic sense, in which every elementary reaction step is reversible, or in a weak reversibility context. Reversibility leads to a particular set of positive equilibria which is known as detailed balance if each reaction step is equilibrated by a reverse one, or complex balanced if the network is weakly reversible.

At this point, it must be remarked that equilibrium should be understood along the sequel in the sense given in dynamic systems, irrespectively of whether it corresponds to thermodynamic equilibrium or to a particular steady-state on a chemical reactor. Note, however, that in agrement with thermodynamics, instabilities in the dynamics of reaction systems (when taking place on a homogeneous medium in isothermal conditions) require the reaction domain to be open to mass exchange with the environment.

Because of microreversibility, most chemical systems, when closed to mass and energy exchanges with the environment, satisfy the principle of detailed balance equilibrium, resulting into stable equilibria \cite{Gorban_Shahzad:11}. As discussed in \cite{Gorban_Yablonsky:11} and  \cite{Gorban_etal:13}, irreversibility can be allowed within a reaction network, as limit cases of reversible steps under a thermodynamic consistency condition (known as the Wegscheider condition) which necessarily assumes microreversibility.

The notion of complex balancing (also known as cyclic balancing or semi-detailed balancing), on the other hand, generalizes the detailed balance condition to any weakly reversible network. The structure of complex balanced systems has been explored in \cite{Craciunetal:09} and shown to be a toric variety with unique and stable equilibrium points (see also \cite{vanderchaftetal:13}). Extensions to cope with more general classes of kinetic systems have been investigated in  \cite{MullerRegensburger:12, Perezmillanetal:12}.

It is important to mention here the recent fundamental results on the proof of the Global Attractor Conjecture which says that any equilibrium point of a complex balanced mass action system is globally stable. A proof for the single linkage class case was given in \cite{anderson:2011}, while a possible general proof based on differential inclusions was described in \cite{craciun:2015}.

CRNT, as it stands nowadays within the field of applied mathematics, offers an extraordinary potential in system's theory for analysis and design of complex dynamic systems of polynomial type, what in turn may cover a wide spectrum of chemical and biological systems.
Unfortunately, many of its results remain at a large extent unexploited, when not unnoticed, in the fields of process systems and engineering.

Among the reasons that hamper application might be certain advanced mathematical tools and the intensive use of graph theory that are often not well-enough known to engineers. Some practical questions that demand attention relate to the link between dynamic behavior of a given mechanism and parameter sets (reaction rate coefficients), or to the design of a chemical/biochemical network with some pre-specified behaviour (e.g bistable, oscillatory, etc).

In this contribution we present some conditions that ensure feasibility of equilibrium solutions for weakly reversible mass action law (MAL) systems. They are linked to the notion of “family of solutions”, a concept originally derived in \cite{Otero:09, Otero:12} to study multiplicity phenomena as a function of network parameters.

In deriving what it will be referred in the sequel as feasibility conditions, we exploit a flux-based form of the model equation. Within such structure, the time evolution of the species concentration vector is expressed as the product of a matrix denoted by $S$, whose columns span the stoichiometric subspace of the reaction system, and a vector function that is related to concentrations through a class of stable Metzler matrices \cite{Arrow:89}.

As we will show, feasibility relates to the orthogonality between a log-transformed vector function of reaction complexes and the kernel of matrix $S$. Based on this observation, feasibility conditions will be expressed in terms of certain functions that can be employed to identify admissible equilibria within the positive orthant of the concentration space. It will be shown that such functions are monotonous in their respective argument and take the zero within their domain, what will allow us to establish links with existence and uniqueness of equilibria along positive stoichiometric compatibility classes for MAL kinetic systems. In this context, connections between monotonicity and two classical results in CRNT theory that relate to complex balanced equilibrium \cite{Horn:72,Horn-Jackson:72}, and to a class of positive deficiency networks \cite{Feinberg:79,Feinberg:87,Feinberg:95}, will be discussed.

Finally, it must be remarked that the potential interest of the notion of complex balancing in the context of process control is to characterize stable operation regimes in open systems, where the principle of detailed balance does not necessarily hold. This may allow, for instance, the selection or manipulation of exchange fluxes so to preserve stability of the resulting (open to the environment) reaction system, via appropriate process optimization and/or feed-back control (see e.g. \cite{Liptaketal16}). Future directions may also involve the detection or design of networks having multiple equilibria.

The paper is organized as follows: Section \ref{sec:formal_description_RN}
introduces a formal description of chemical reaction networks. The graph structure underlying a reaction network, and its algebraic counterpart, will be described in Section
\ref{sec:linkandCompMat}. Section \ref{sec:canonical_representation} presents a  flux-based form canonical representation of the equilibrium set, that includes some feasibility conditions. Relationships between network structure and monotonicity of feasibility conditions will be established in Section \ref{sec:Domain_properties_Feasible}.
Connections between monotonicity of feasibility functions and some classical results on uniqueness and stability of equilibria will be discussed in Section \ref{sec:SNetClUnique}.

%%----------------------------------------------------------------------------------------------------------------

\section{Preliminaries: Reaction Network Structure and Dynamics}\label{sec:formal_description_RN}
%%----------------------------------------------------------------------------------------------------------------
Let $m$ be the number of chemical species which react by $\rho$ irreversible chemical reaction steps,
and $\textbf{c} \in \mathbb{R}^{m}$ the corresponding vector of
species concentrations, defined as mole
number per unit of volume. Each reaction step transforms some set
of chemicals, usually referred to as reactants, into a set of
reaction products. In CRNT, reactants and reaction products receive
the name of \emph{reaction complexes}. Complexes and reaction steps
describe a graph where complexes correspond to nodes and reaction
steps to directed edges.

Formally, a graph involving $n$
complexes $\{\mathcal{C}_1,\ldots,\mathcal{C}_n\}$ linked by
irreversible reaction steps can be constructed by associating to
each complex $i$ a set ${\cal{I}}_i$ with $n$ integer elements, and a vector $\bm{y}_i$. The elements of the set $\mathcal{I}_i$ are the indices of the complexes that are directly  reachable (i.e. by one reaction step)  from ${\mathcal{C}}_i$. From now on, we will refer to each complex ${\mathcal{C}}_i$ by the corresponding index $i$. Vector $\bm{y}_i \in \mathbb{R}^m$ has as entries the
(positive) stoichiometric coefficients of the molecular species that
participate in complex $i$.

The graph structure is then built by linking every complex $i$ to
$j\in{\cal{I}}_i$. This process results in a number $\ell$ of
connected components known in CRNT as {\it linkage classes}. For
each linkage class $\lambda=1, ...,\ell$, we define the set
${\cal{L}}_{\lambda}$ which contains as elements the
indexes of the complexes that belong to that linkage
class\footnote{To be precise, the set ${\cal{L}}_{\lambda}$ is
that containing as elements ${\cal{L}}_{\lambda} = \{i_1, i_2, ...,
i_{N_{\lambda}}\}$, with $N_{\lambda} =
{\cal{N}}({\cal{L}}_{\lambda})$, being $i_j$ the cardinality
associated to complex $\mathcal{C}_{i_{j}}$, and ${\cal{N}}(\cdot)$
the operator which indicates the number of elements in the set.}.
%The complexes that belong to the  network will be in the set
%%
%\[
%{\cal{L}} = {\bigcup}_{\lambda =1}^{\ell} {\cal{L}}_{\lambda}.
%\]
%

Complexes are connected within a linkage class by sequences of
irreversible reaction steps that define \textit{directed paths}. Two
complexes are strongly linked if they can be mutually reached from each other
by directed paths (trivially, every complex is strongly linked to
itself). A maximal set of pairwise strongly linked complexes defines a
\textit{strong terminal linkage class} if no other complex can be
reached from its nodes. In this work we will consider only networks in which every linkage class contains just one strong terminal linkage class.

A linkage class $\cal{L}_{\lambda}$ is
said to be \textit{weakly reversible} if any pair of its complexes is
strongly linked. Weakly reversible networks are those composed by
weakly reversible linkage classes.
%Clearly, a CRN is weakly
%reversible if and only if all components of the reaction graph are
%strongly connected components.
%
A particular type of weakly reversible linkage class is a reversible linkage class if each reaction step is itself reversible, so that for every $i$ and $j\in{\cal{I}}_i$, we have that $i\in{\cal{I}}_j$.
%To each linkage class $\lambda$ we associate an $n$-dimensional
%vector $\bm{\omega}_{\lambda}$ which is defined as
%follows:\footnote{Vector $\bm{\omega}_{\lambda}$ is referred in
%classical CRNT as the \emph{characteristic function} of linkage
%class $\lambda$.}
%%
%\begin{equation}\label{eq:charact_function}
%\bm{\omega}_{\lambda} = \sum_{i \in {\cal{L}}_{\lambda}}
%\bm{\varepsilon}_i,
%\end{equation}
%%
%where $\bm{\varepsilon}_i \in \mathbb{N}^n$ denotes the $i$th standard
%unit vector employed to represent axes on a cartesian coordinate system.
%Vectors $\bm{\omega}_{\lambda}$ ($\lambda =1,...,\ell$) are
%orthogonal to each other since by construction, sets
%${\cal{L}}_{\lambda}$ are disjoint.
The rate $R_{ij}$, at which a
set of reactants in complex $i$ is
transformed into a set of products in complex $j$, will be
assumed to be mass action, so that:
\begin{equation}
\label{eq:reaction_rate}
 R_{ij} (\textbf{c}) = k_{ij} \psi_i (\textbf{c}), ~~ \textnormal{with}~~\psi_i (\textbf{c}) = \prod_{j=1}^m c_j^{y_{ji}} \equiv
{\textbf{c}}^{\bm{y}_{i}},
\end{equation}
where $\bm{y}_{i}$ is the stoichiometric vector corresponding to complex ${\mathcal{C}}_i$. The reaction systems we consider in this work will take place under isothermal conditions, what makes any reaction rate parameter $k_{ij} (> 0)$ constant.
Whenever $\textbf{c}$ is a strictly positive vector, the following alternative representation for $\psi_i (c)$ may be more convenient:
\begin{equation}\label{eq:lnpsi}
            \ln \psi_i (\textbf{c})  = {\bm{y}}_{i}^{\mathrm{T}} \ln
            \textbf{c},
\end{equation}
where the natural logarithm operator $\ln(\cdot)$ acts on any
vector element-wise. Let $\bm{\psi} : \mathbb{R}_{> 0}^{m} \rightarrow \mathbb{R}_{> 0}^{n}$ be the vector
containing as entries the monomials described in
(\ref{eq:reaction_rate}), then the previous expression can be written in matrix form as:
\begin{equation}\label{eq:lnpsi_Vector}
            \ln \bm{\psi} (\textbf{c})  = Y ^{\mathrm{T}} \ln
            \textbf{c},
\end{equation}
where $Y\in\mathbb{R}^{m\times n}$  is the so-called \textit{molecularity matrix} which collects as columns the stoichiometric vectors ${\bm{y}}_{i} \in \mathbb{R}^{m}$ associated to the complexes of the network.

\subsection{The dynamics of reaction networks}\label{secsec:dynamics_RN}

Following the classical work by Feinberg \cite{Feinberg:72}, the time evolution of species concentrations on a well-mixed reaction medium at constant temperature can be described by a set of ordinary differential
equations that we write as:
\begin{equation}\label{subeq:canonical_form_AK}
            \dot{\textbf{c}} = Y \cdot A_k (\bm{\psi} (\textbf{c})) = Y \cdot \sum_{\lambda}
            A_k^{\lambda}(\bm{\psi} (\textbf{c})),
\end{equation}
where $A_k^{\lambda}: \mathbb{R}^{n} \rightarrow \mathbb{R}^{n}$ is a linear operator defined as:
\begin{equation}\label{eq:Ak_by_fluxes_A_Feinb}
A_k^{\lambda}(\bm{\psi}) \equiv \sum_{i \in {\cal{L}}_{\lambda}} \psi_i  \sum_{j \in \mathcal{I}_i} k_{ij} \cdot
(\bm{\varepsilon}_j - \bm{\varepsilon}_i),
\end{equation}
with $\bm{\varepsilon}_i \in \mathbb{R}^n$ denoting the $i$th standard unit vector employed to represent axes on a cartesian coordinate system.
Let us define the net reaction rate flux around
a complex $i$, as the signed sum of in- and out-flowing fluxes, i.e. as a function $\phi_{i}: \mathbb{R}_{\geq 0}^{n} \rightarrow \mathbb{R}$ of the form:
\begin{equation}\label{eq:this_is_flux}
\phi_{i} (\bm{\psi}) = \sum_{\{j | i \in {\cal{I}}_{j}\}} R_{ji} (\bm{\psi}) -
\sum_{j \in {\cal{I}}_{i}} R_{ij} (\bm{\psi}),
\end{equation}
where the first summation at the right hand side extends to all source complexes $j$ in the network from which there exists a reaction step to product complex $i$, and is
represented by $\{j | i \in {\cal{I}}_{j}\}$.

We can express  $A_k^{\lambda} (\bm{\psi})$ in (\ref{eq:Ak_by_fluxes_A_Feinb}) in terms of fluxes (\ref{eq:this_is_flux}), by selecting any reference complex $j_{\lambda} \in {\mathcal{L}}_{\lambda}$, and adding and subtracting $\bm{\varepsilon}_{j_{\lambda}}$ from the right hand side of (\ref{eq:Ak_by_fluxes_A_Feinb}) so that:
\[
A_k^{\lambda}(\bm{\psi}) =  \sum_{i \in {\cal{L}}_{\lambda}} \psi_i \sum_{j \in \mathcal{I}_i} k_{ij} \cdot
(\bm{\varepsilon}_j - \bm{\varepsilon}_{j_{\lambda}}) -
\sum_{i \in {\cal{L}}_{\lambda}} \left( \sum_{j \in \mathcal{I}_i} k_{ij} \psi_i   \right) \cdot
(\bm{\varepsilon}_i - \bm{\varepsilon}_{j_{\lambda}}).
\]
After switching subindexes, re-ordering the summations for the first term at the right hand side and making use of (\ref{eq:this_is_flux}), we get the following equivalent expressions:
\begin{eqnarray}
A_k^{\lambda}(\bm{\psi}) &=& \sum_{i \in {\cal{L}}_{\lambda}} \left( \sum_{\{j | i \in {\cal{I}}_{j}\}} k_{ji} \psi_j   \right) \cdot
(\bm{\varepsilon}_i - \bm{\varepsilon}_{j_{\lambda}})
 - \sum_{i \in {\cal{L}}_{\lambda}} \left( \sum_{j \in \mathcal{I}_i} k_{ij} \psi_i   \right) \cdot
(\bm{\varepsilon}_i - \bm{\varepsilon}_{j_{\lambda}}) \nonumber \\
&=& \sum_{i \in {\cal{L}}_{\lambda}} \phi_{i} (\bm{\psi}) \cdot (\bm{\varepsilon}_i - \bm{\varepsilon}_{j_{\lambda}}). \label{eq:Ak_by_fluxes_A}
\end{eqnarray}
For convenience, the reference complex will be chosen from the corresponding strong terminal linkage class.
Since vectors
$\bm{\varepsilon}_i$ are orthogonal, by using
(\ref{eq:Ak_by_fluxes_A}), we have that $\phi_{i} (\bm{\psi} )=
\bm{\varepsilon}_{i}^{\mathrm{T}} A_k^{\lambda}(\bm{\psi})$ for
every $i \in {\cal{L}}_{\lambda}$. Let $\bm{\omega}_{\lambda} = \sum_{i \in {\cal{L}}_{\lambda}} \bm{\varepsilon}_i$,
 then we also have that
$\bm{\omega}_{\lambda}^{T}A_k^{\lambda}(\bm{\psi}) = 0$ and therefore:
\begin{equation}\label{eq:linear_dependent_fluxes}
\sum_{i\in\mathcal{L}_{\lambda}} \phi_{i} (\bm{\psi}) =
\left(\sum_{i\in\mathcal{L}_{\lambda}}
\bm{\varepsilon}_{i}\right)^{T} A_k^{\lambda}(\bm{\psi}) = 0.
\end{equation}
Note that fluxes in (\ref{eq:this_is_flux})  (as well as  the linear operator in (\ref{eq:Ak_by_fluxes_A_Feinb})) are implicitly
dependent on the reaction rate coefficients associated to the reaction steps in the linkage class.
By inspection of (\ref{eq:Ak_by_fluxes_A}), it can be
concluded that the image of $A_k(\bm{\psi})$
lies on the
subspace $\Delta$ defined as follows:
\begin{equation}\label{eq:subspace_DELTA_TOTAL}
\Delta = \Delta_1 + \dots + \Delta_{\lambda} + \dots + \Delta_{\ell},
\end{equation}
%%
%\begin{equation}\label{eq:subspace_DELTA_TOTAL}
%\Delta = \text{span}\{\Delta_1 \} \cup \cdots \cup \text{span}\{\Delta_{\lambda} \} \cup \cdots \cup \text{span}\{\Delta_\ell \},
%\end{equation}
where
\begin{equation}\label{eq:subspace_DELTA}
\begin{array}{ccc} \Delta_{\lambda}=\textnormal{span}\{\bm{\varepsilon}_i -
\bm{\varepsilon}_{j_{\lambda}}~|~ i \in \mathcal{L}_{\lambda} \}
& \textnormal{for} & \lambda = 1, \cdots, \ell
\end{array}.
\end{equation}
and the sum of vector spaces $V_1$ and $V_2$ is defined as:
\[
V_1 + V_2 = \{v_1 + v_2 ~|~v_1\in V_1,~v_2\in V_2 \}.
\]

Since vectors in $\{\bm{\varepsilon}_i -
\bm{\varepsilon}_{j_{\lambda}}~|~ i\in\mathcal{L}_{\lambda} \}$ are linearly independent,  they form a basis for the
subspace ${\Delta}_{\lambda}$, thus $\textnormal{dim} ({\Delta}_{\lambda}) =
N_{\lambda}-1$. In addition, since the subspaces
${\Delta}_{\lambda}$ are orthogonal:
\[
\textnormal{dim} (\Delta) = \sum_{\lambda} (N_{\lambda} -1) = n -\ell.
\]
This implies that $A_k(\bm{\psi}) = 0$ if and only if $\phi_{i} (\bm{\psi}) = 0$ for all
$i \in {\cup}_{\lambda} {\mathcal{L}}_{\lambda}$. Consequently, if a positive concentration vector $\textbf{c}$ exists compatible with a zero flux condition for every complex in
the network, that vector should be an equilibrium for system (\ref{subeq:canonical_form_AK}). Such equilibrium condition, known as \emph{complex balanced} \cite{Horn:72}, is formally defined as follows:
\begin{definition}\label{def:Complex_Balance_Condition}
 {\bf (Complex Balanced Equilibrium)} Any vector ${\mathbf{c}}^{*} > 0$  such that $\phi_i (\bm{\psi} ({\mathbf{c}}^{*}) ) = 0$ (Eqn (\ref{eq:this_is_flux})) for every $i = 1, \cdots, n$ is called a complex balanced equilibrium solution.
\end{definition}
A subclass of complex balanced equilibrium, particularly meaningful from a thermodynamic point of view as it relates to microreversibility (\cite{VanKampen:81, Gorban_Shahzad:11}), is the detailed balance equilibrium which we define next:
\begin{definition}\label{def:Detailed_Balanced_Condition}
 {\bf (Detailed Balance Equilibrium)} If the network is reversible (i.e. for every $i$ and $j\in{\cal{I}}_i$, we have that $i\in{\cal{I}}_j$) any vector ${\mathbf{c}}^{*} > 0$ such that $R_{ij} ({\mathbf{c}}^{*}) = R_{ji} ({\mathbf{c}}^{*})$ (where $R_{ij} (\textbf{c})$ is of the form (\ref{eq:reaction_rate})) is called as a detailed balance equilibrium solution.
\end{definition}

%%
%%-------------------------------------------------------------------------------------------
%% DESCRIBE INVARIANCE THROUGH THE SIMPLEX
%% %%-------------------------------------------------------------------------------------------

\subsection{The stoichiometric subspace}\label{secsec:reaction_simplex}

Similarly to the subspace $\Delta$, we define the stoichiometric subspace $\Xi$ as:
\[
\Xi = \Xi_1 + \dots + \Xi_{\lambda} + \dots + \Xi_{\ell},
\]
where:
\begin{equation}\label{eq:stoichiometric_subsp}
\begin{array}{ccc} \Xi_{\lambda}=\textnormal{span}\{\bm{y}_i - \bm{y}_{j_{\lambda}}~|~ i\in\mathcal{L}_{\lambda} \}
& \textnormal{for} & \lambda = 1, \cdots, \ell
\end{array}.
\end{equation}
In what follows it will be more convenient to collect the elements
from each of the sets $\{\bm{y}_i - \bm{y}_{j_{\lambda}}~|~
i\in\mathcal{L}_{\lambda} \}$ and their union,
column-wise in matrices $S_{\lambda} \in {\mathbb{R}}^{m \times
(N_{\lambda} - 1)}$ and $S \in {\mathbb{R}}^{m\times (n-\ell)}$,
respectively, so that:
\begin{equation}\label{eq:MatrixSNet}
 S = [\begin{array}{ccccc} S_1 & \cdots & S_{\lambda} & \cdots & S_{\ell} \end{array}].
\end{equation}
Let $s = \textnormal{dim} (\Xi)$, which eventually coincides with the rank of $S$, then it follows from the rank-nullity theorem that the
dimension of the kernel (null space) of $S$ will be:
\begin{equation}\label{eq:deficiency}
\delta = n - \ell - s.
\end{equation}
This number is known in CRNT as the \emph{deficiency} of the
network. In a similar way, we can define the deficiency of each
linkage class as  the dimension of the kernel of
$S_{\lambda}$ so that ${\delta}_{\lambda} = N_{\lambda} -
1 - s_{\lambda}$, where $s_{\lambda} = \textnormal{dim}
({\Xi}_{\lambda})$. Since $s \leq \sum_{\lambda} {s}_{\lambda}$, it is not difficult to conclude that linkage class and network deficiencies relate as:
\begin{equation}\label{eq:Delta_vs_DelatL}
\delta \geq \sum_{\lambda} {\delta}_{\lambda}.
\end{equation}
Let $\{{\textbf{g}}^{r} ~|~ r=1, \ldots, \delta \}$ be a basis for
the kernel of $S$, and express each vector ${\textbf{g}}^{r} \in
{\mathbb{R}}^{n - \ell}$ in
terms of $\ell$ sub-vectors ${\textbf{g}}^{r}_{\lambda} \in
{\mathbb{R}}^{N_{\lambda} -1}$ (one per linkage class), so that:
\begin{equation}\label{eq:BasisKernel}
({\textbf{g}}^{r})^T = [\begin{array}{ccccc}
           ({\textbf{g}}^{r}_{1})^T & \cdots & ({\textbf{g}}^{r}_{\lambda})^T & \cdots &
           ({\textbf{g}}^{r}_{\ell})^T
         \end{array}
 ], ~\textnormal{for}~ r = 1, \cdots, \delta.
\end{equation}
Using the above description, equation $S {\textbf{g}}^{r} = 0$ can be
re-written as:
\begin{equation}\label{eq:formalSg}
\sum_{\lambda}  {S}_{\lambda} {\textbf{g}}^{r}_{\lambda} = 0
~~\textnormal{for} ~~ r = 1,\ldots,\delta.
\end{equation}
We will be particularly interested in solutions of system
(\ref{subeq:canonical_form_AK}) on the convex region resulting from
the intersection of the non-negative (respectively positive) orthant in the concentration space
and a certain linear variety associated to the stoichiometric subspace
$\Xi$, the result known in CRNT as a stoichiometric (respectively, positive stoichiometric) compatibility class. Given a reference
concentration vector $\textbf{c}_0$, the stoichiometric compatibility class can be formally defined as:
\begin{equation}\label{eq:eqpolyhedron}
\Omega(\textbf{c}_0) = \{\textbf{c} \in \mathbb{R}^{m}  ~~ | ~~
\textbf{c}\geq 0, {P}^\mathrm{T} (\textbf{c}-\textbf{c}_0) = 0 \},
\end{equation}
where ${P} \in \mathbb{R}^{m \times (m-s)}$ is a full rank matrix
whose columns span the orthogonal complement $\Xi^{\perp}$. The corresponding positive stoichiometric compatibility class can expressed as ${\Omega}^{+}(\textbf{c}_0) = {\Omega}(\textbf{c}_0) \cap {\mathbb{R}}_{>0}^{n}$.
In passing, let us define the function $\bm{\sigma}: {\mathbb{R}}^m \rightarrow {\mathbb{R}}^{m - s}$ as $\bm{\sigma} (\textbf{c}) = {P}^\mathrm{T} \textbf{c}$. Such function is constant along trajectories (\ref{subeq:canonical_form_AK}), since by combining (\ref{eq:Ak_by_fluxes_A}) and  (\ref{subeq:canonical_form_AK}) we have that:
\begin{equation}\label{eq:dotC_first}
\dot{\textbf{c}} = \sum_{\lambda} \sum_{i\in\mathcal{L}_{\lambda}} \phi_i
(\bm{\psi} (\textbf{c}))(\bm{y}_i - \bm{y}_{j_{\lambda}}),
\end{equation}
and the columns of $P$ are orthogonal to $S$, hence $ \dot{\bm{\sigma}} =  {P}^\mathrm{T} \dot{\textbf{c}} = 0 $. In other words, $\bm{\sigma}$ is an invariant of motion for system (\ref{subeq:canonical_form_AK}). From this observation it is not difficult to conclude that any trajectory that starts in a compatibility class $\Omega(\textbf{c}_0)$ will remain there.

\subsection{Some examples of chemical reaction networks}
\label{secsec:example_1_WR}

{\bf A reversible chemical reaction network}

 Let us consider a reaction
network involving $m=6$ molecular species we label as $\{ M_1, ...,
M_6 \}$, each of them constituted by a combination of three types of
functional groups (or atoms) we denote as $A$, $B$ and $C$.
The (reversible) chemical
reaction steps that take place are:
\begin{equation} \label{eq:RN_Steps_1_Rev}
\begin{array}{l}
  A_2 B  + C  \leftrightarrows  AC  +  AB\\
  AB   +  2C  \leftrightarrows  AC_2 B\\
  AC_2 B  \leftrightarrows  AC  +  CB\\
\end{array}
\end{equation}
Molecular species and functional groups are related as follows: $M_1
\equiv A_2 B$, $M_2 \equiv A C$, $M_3 \equiv AB$, $M_4 \equiv C$,
$M_5 \equiv AC_2 B$, $M_6 \equiv CB$. The network consists of $n=5$ complexes:
\[
\{ {\cal{C}}_1, {\cal{C}}_2, {\cal{C}}_3, {\cal{C}}_4, {\cal{C}}_5
\} \equiv \{ M_1 + M_4, M_2 + M_3, M_3 + 2 M_4, M_5, M_2 + M_6 \}.
\]
Making use of the formal description previously discussed, the sets ${\cal{I}}_i$ that
indicate which complexes are reached from complex $i$ become, for
this example:
\begin{equation}
\begin{array}{lllll}
  {\cal{I}}_1 = \{ 2 \}& {\cal{I}}_2 = \{ 1 \} & {\cal{I}}_3 = \{ 4 \} & {\cal{I}}_4 = \{3, 5 \} & {\cal{I}}_5 = \{ 4 \}.\\
\end{array}
\end{equation}
The corresponding stoichiometric vectors ${\bm{y}}_i$ associated to each
complex are written as columns in the molecularity matrix $Y$:
\begin{equation}
Y = \left[
\begin{array}{ccccc}
  1 & 0 & 0 & 0 & 0\\
  0 & 1 & 0 & 0 & 1\\
  0 & 1 & 1 & 0 & 0\\
  1 & 0 & 2 & 0 & 0\\
  0 & 0 & 0 & 1 & 0\\
  0 & 0 & 0 & 0 & 1\\
\end{array}
\right].
\end{equation}
\begin{figure}[ht]
\begin{center}
$\begin{array}{@{}c@{}}
\includegraphics[scale=0.45]{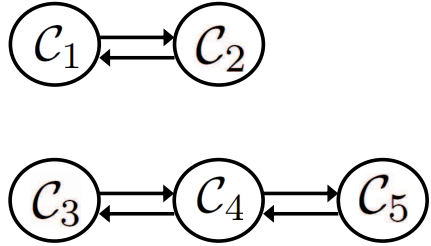}
\end{array}$
\end{center}
\caption{Graph representation for the reaction network described by reversible steps
(\ref{eq:RN_Steps_1_Rev}).} \label{fig:Figure_Example_Sec2}
\end{figure}
The  graph representation is depicted in Figure
\ref{fig:Figure_Example_Sec2}, and comprises two linkage classes
${\cal{L}}_{1} = \{ 1, 2 \}$ and ${\cal{L}}_{2} = \{ 3, 4, 5\}$.
%,
%with the corresponding vectors $\bm{\omega}_{\lambda}$ in (\ref{eq:charact_function}):
%%
%\[
%\bm{\omega}_{1} = \left(
%                                             \begin{array}{ccccc}
%                                               1 & 1 & 0 & 0 & 0\\
%                                             \end{array}
%                                           \right), \qquad
%\bm{\omega}_{2} = \left(
%                                             \begin{array}{ccccc}
%                                               0 & 0 & 1 & 1 & 1\\
%                                             \end{array}
%                                           \right).
%\]
%
%Selecting $j_1 = 1$ and $j_2 = 3$ as reference complexes for linkage
%class $\lambda = 1$ and $\lambda = 2$, respectively,
The net reaction fluxes (\ref{eq:this_is_flux}) around each complex are:
\begin{equation}\label{eq:NetFlux_Example_1}
\begin{array}{l}
  {\phi}_1 (\bm{\psi}) = k_{21} {\psi}_2  - k_{12} {\psi}_1\\
  {\phi}_2 (\bm{\psi}) = k_{12} {\psi}_1  - k_{21} {\psi}_2\\
  {\phi}_3 (\bm{\psi}) = k_{43} {\psi}_4  - k_{34} {\psi}_3\\
  {\phi}_4 (\bm{\psi}) = k_{34} {\psi}_3  + k_{54} {\psi}_5 - (k_{43} + k_{45}) {\psi}_4 \\
  {\phi}_5 (\bm{\psi}) = k_{45} {\psi}_4  - k_{54} {\psi}_5\\
\end{array}
\end{equation}
Note that if $j_1 = 1$ and $j_2 = 3$ are chosen as reference complexes, by relation (\ref{eq:linear_dependent_fluxes}), the fluxes associated to the reference become:
\[
{\phi}_1 (\bm{\psi}) = - {\phi}_2 (\bm{\psi}) \qquad \qquad \textnormal{and} \qquad \qquad {\phi}_3 (\bm{\psi}) = - ({\phi}_4 (\bm{\psi}) + {\phi}_5 (\bm{\psi}) ).
\]
The image of $A_{k}$ coincides with
subspace $\Delta$ (Eqn (\ref{eq:subspace_DELTA_TOTAL})) with $\Delta_{1}$ and $\Delta_{2}$ of the form:
\begin{equation}
\begin{array}{l}
\Delta_{1}=\textnormal{span}\{(\bm{\varepsilon}_2 - \bm{\varepsilon}_{1}) \} \\
\Delta_{2}=\textnormal{span}\{(\bm{\varepsilon}_4 - \bm{\varepsilon}_{3}), (\bm{\varepsilon}_5 - \bm{\varepsilon}_{3})\} \\
\end{array}.
\end{equation}
Matrices $S_{\lambda}$, employed in Section
\ref{secsec:reaction_simplex} to define (column-wise) the
corresponding stoichiometric subspaces
(\ref{eq:stoichiometric_subsp}), are of the form:
\[
S_{1} = \left[\begin{array}{r}
  -1  \\
   1  \\
   1  \\
  -1  \\
   0  \\
   0  \\
\end{array} \right], ~~ S_2 =\left[\begin{array}{rr}
        0 &  0 \\
        0 &  1 \\
       -1 & -1 \\
       -2 & -2 \\
        1 &  0 \\
        0 &  1 \\
     \end{array} \right].
\]
The dimension of the stoichiometric subspace, which coincides with the rank of
matrix $S = \left[\begin{array}{cc}
              S_1 & S_2
            \end{array} \right]$,
is $s = 3$. Hence, network deficiency is $\delta = 5 - 2 - 3 \equiv 0$ what means that no vector ${\textbf{g}}^{r}$ other
than the zero vector exists such that $S {\textbf{g}}^{r} = 0$.

The numbers of atoms (or functional groups) $A$-$C$ remain constant,
provided that reactions take place on a closed domain (i.e. no mass exchanges with the environment occur). Let $([A]_{0}, [ B ]_{0}, [ C ]_{0})$ be total concentrations for $A$, $B$ and $C$ on the closed and homogeneous domain. Mole-number balances result in the following set of linear
relations:
\[
\begin{array}{l}
\left[ A \right]_{0} = 2 [A_2 B] + [A C] + [AB] + [AC_2 B] \\
\left[ B \right]_{0} =  [A_2 B]  + [AB] + [AC_2 B] + [CB] \\
\left[ C \right]_{0} =  [A C]  + [C] + 2 [AC_2 B] + [CB] \\
\end{array}
\]
where brackets indicate chemical species concentrations. Previous
relations can be written in matrix form as follows:
\[
{P}^\mathrm{T} (\textbf{c}-\textbf{c}_0) = 0 ~~\textnormal{with} ~~ {P}
= \left[
\begin{array}{rrrrrr}
             2 & 1 & 1 & 0 & 1 & 0\\
             1 & 0 & 1 & 0 & 1 & 1\\
             0 & 1 & 0 & 1 & 2 & 1\\
           \end{array}\right]^T,
\]
where $\textbf{c}$ is the vector of chemical species concentrations
and ${P}^T {\textbf{c}}_0$ is the (constant) concentration of functional
groups/atoms. It must be noted that the above expression is employed
in (\ref{eq:eqpolyhedron}) to characterize the set of compatibility
classes. As discussed in Section \ref{secsec:reaction_simplex},  because $\textnormal{rank} ({P} )\equiv m -s = 3$ (full rank), the
columns of ${P}$ define a basis for the orthogonal complement of the
stoichiometric subspace.

{\bf An irreversible  network}

\begin{figure}[ht]
\begin{center}
$\begin{array}{@{}c@{}}
\includegraphics[scale=0.45]{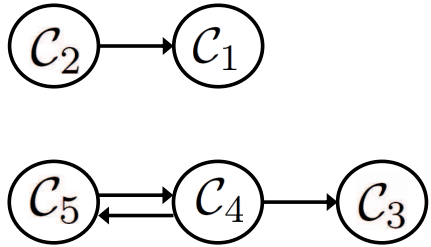}
\end{array}$
\end{center}
\caption{Graph representation for the two linkage class,
irreversible reaction network (\ref{eq:irrreactPlague}).} \label{fig:Figure_Example_Sec2_NWR}
\end{figure}

Let us consider the following set of irreversible reactions:
\begin{equation}\label{eq:irrreactPlague}
\begin{array}{c}
   S  + I  \rightarrow  2 I \\
  I  \leftrightarrows R \rightarrow S
\end{array}
\end{equation}
This network can be interpreted as an extension of the SIR epidemic model \cite{kermacketal:27} which describes the effect of a disease on a large population. Individuals on the population are classified either as those susceptible to the disease ($S$), infected ($I$) or those recovered from the disease ($R$). In this extension, individuals under recovering may evolve either to those susceptible to the disease or directly infected again. In the CRNT formalism, the network comprises three species, with concentrations $[S]$, $[I]$ and $[R]$, and $5$
complexes, numbered as:
\[
\{ {\cal{C}}_1, {\cal{C}}_2, {\cal{C}}_3, {\cal{C}}_4, {\cal{C}}_5
\} \equiv \{ 2I, S + I, S, R, I \}.
\]
Graph structure is depicted in Figure
\ref{fig:Figure_Example_Sec2_NWR}. Molecularity matrix $Y$ for this
network reads:
\begin{equation}
Y = \left[
\begin{array}{ccccc}
  0 & 1 & 1 & 0 & 0\\
  2 & 1 & 0 & 0 & 1\\
  0 & 0 & 0 & 1 & 0\\
\end{array}
\right].
\end{equation}
Choosing as reference complexes $j_1 = 1$ and $j_2 = 3$, the $S_{\lambda}$ matrices become:
\[
S_{1} = \left[\begin{array}{r}
    1  \\
   -1  \\
    0  \\
\end{array} \right], ~~ S_2 =\left[\begin{array}{rr}
        -1 & -1 \\
         0 &  1 \\
         1 &  0 \\
     \end{array} \right].
\]
Because both matrices are full rank, the dimension of ${\Xi}_1$
and ${\Xi}_2$ is $s_1 = 1$ and $s_2 = 2$, respectively. Thus
${\delta}_1 = N_1 - 1 - s_1 = 0$ and ${\delta}_2 = N_2 - 1 - s_2 =
0$. However, matrix $S = \left(\begin{array}{cc}
              S_1 & S_2
            \end{array} \right)$
is rank deficient since vector $S_1$ is parallel to the vector
in the second column of $S_2$. Consequently $s = 2$, and
$\delta = 5 - 2 - 2 = 1$, verifying inequality
(\ref{eq:Delta_vs_DelatL}). A basis for the kernel of $S$ is given by the vector ${\textbf{g}}^{1} = (\begin{array}{ccc} 1 & 0 &
1\end{array} )^T$.

\begin{figure}[ht]
\begin{center}
$\begin{array}{@{}c@{}}
\includegraphics[scale=0.45]{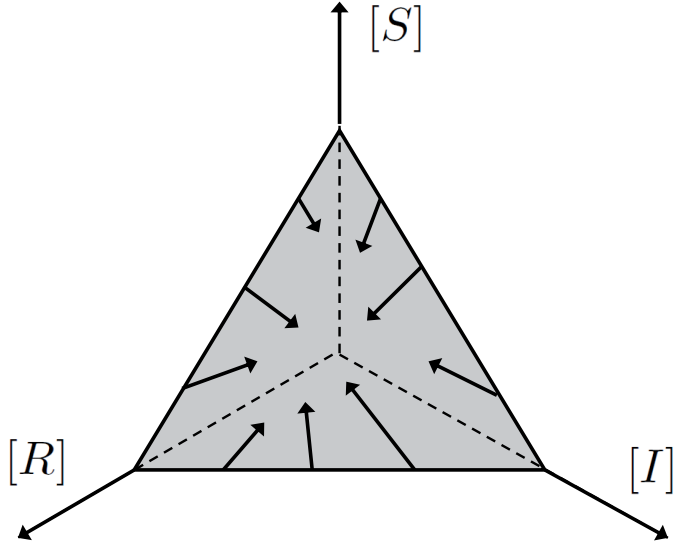}
\end{array}$
\end{center}
\caption{Representation of a compatibility class for the network example (\ref{eq:irrreactPlague}).} \label{fig:Example_NWR_PolyH}
\end{figure}

For this network, the basis that spans the orthogonal complement of
the stoichiometric subspace corresponds to ${P} = (\begin{array}{ccc} 1 & 1 &
1\end{array} )^T$. Each compatibility class is given by $\Omega(\textbf{c}_0)$ (see (\ref{eq:eqpolyhedron})), the region  of
non-negative concentrations ($\textbf{c}\geq 0$) satisfying:
\[
[S] + [I] + [R] = {P}^T \textbf{c}_0,
\]
with $\textbf{c}_0 \geq 0$ being a constant vector. A representation of a  compatibility class for
the reaction network considered is presented in Figure
\ref{fig:Example_NWR_PolyH}, with $\textbf{c} = ([S], [I], [R])^T$.

\section{Linkage classes, Graphs and Compartmental Matrices}
\label{sec:linkandCompMat}

The nature of equilibrium solutions in chemical
reaction networks (e.g. positivity, instability of equilibrium points, or
the possibility of multiple equilibria) is at a large extent
determined by the graph structure of each linkage class and the
properties of some matrices associated to it, that belong to the class of compartmental matrices \cite{Fife1972}.

In this section, we describe such matrices and discuss their properties,  with emphasis on invertibility and non-negativity of their inverses. The main results, summarized in Lemma \ref{prop:E_Metzler}, will be extensively employed in the sequel. For the sake of completeness, we introduce a derivation from scratch, while establishing connections with known facts in the field of positive linear systems and non-negative matrices.

To that purpose, we introduce a graph related to the graph description of a linkage class, and a matrix associated to it. The properties of this matrix will be studied by constructing an auxiliary linear dynamic
system and examining the corresponding equilibrium.

A directed graph $\mathcal{G} = \{ (\mathcal{V} \bigcup v_E),
\mathcal{E}\}$ is constructed by a set of vertices containing a distinguished vertex $v_E$ and a set $\mathcal{E}$ of edges.
The first set of vertices $\mathcal{V}=\{v_1, v_2, \dots, v_n \}$, with
indexes ${\mathcal{L}} = \{ 1, \dots, n \}$ will be referred to as
nodes, while the remaining vertex $v_E$ will represent the
`environment'. To any directed edge $v_i\rightarrow v_j$ for $i,j
\in \mathcal{L}$ and $i\ne j$ in $\mathcal{G}$ (i.e.
$(v_i,v_j)\in\mathcal{E}$) there corresponds a scalar weight
$V_{ij}>0$. In addition, for every $i \in \mathcal{L}$, such that
$(v_i,v_E)\in\mathcal{E}$ or $(v_E,v_i)\in\mathcal{E}$, we associate
scalar weights $b_i > 0 $ and $a_i
> 0$, respectively.  Such weights will be collected as entries in (non-negative) vectors
$\textbf{a}, \textbf{b} \in {\mathbb{R}}^n$, with $a_i = 0$ (respectively, $b_i = 0$) if there is no directed edge $v_E \rightarrow v_i$ (respectively, $v_i \rightarrow v_E$ ). As in the description
of linkage classes (Section \ref{sec:formal_description_RN}) we
say that two nodes are strongly linked if they can be reached from
each other by directed paths. The maximal set of strongly linked nodes (not passing through the environment) in $\mathcal{L}$ from which
there are no outgoing edges to other nodes, defines a strong
terminal set, we will refer to as ${\mathcal{L}}_{\textnormal{p}}$.

\begin{figure*}[ht]
\begin{center}
$\begin{array}{@{}cc@{}}
\mathrm{(a)} & \mathrm{(b)} \\
\includegraphics[scale=0.40]{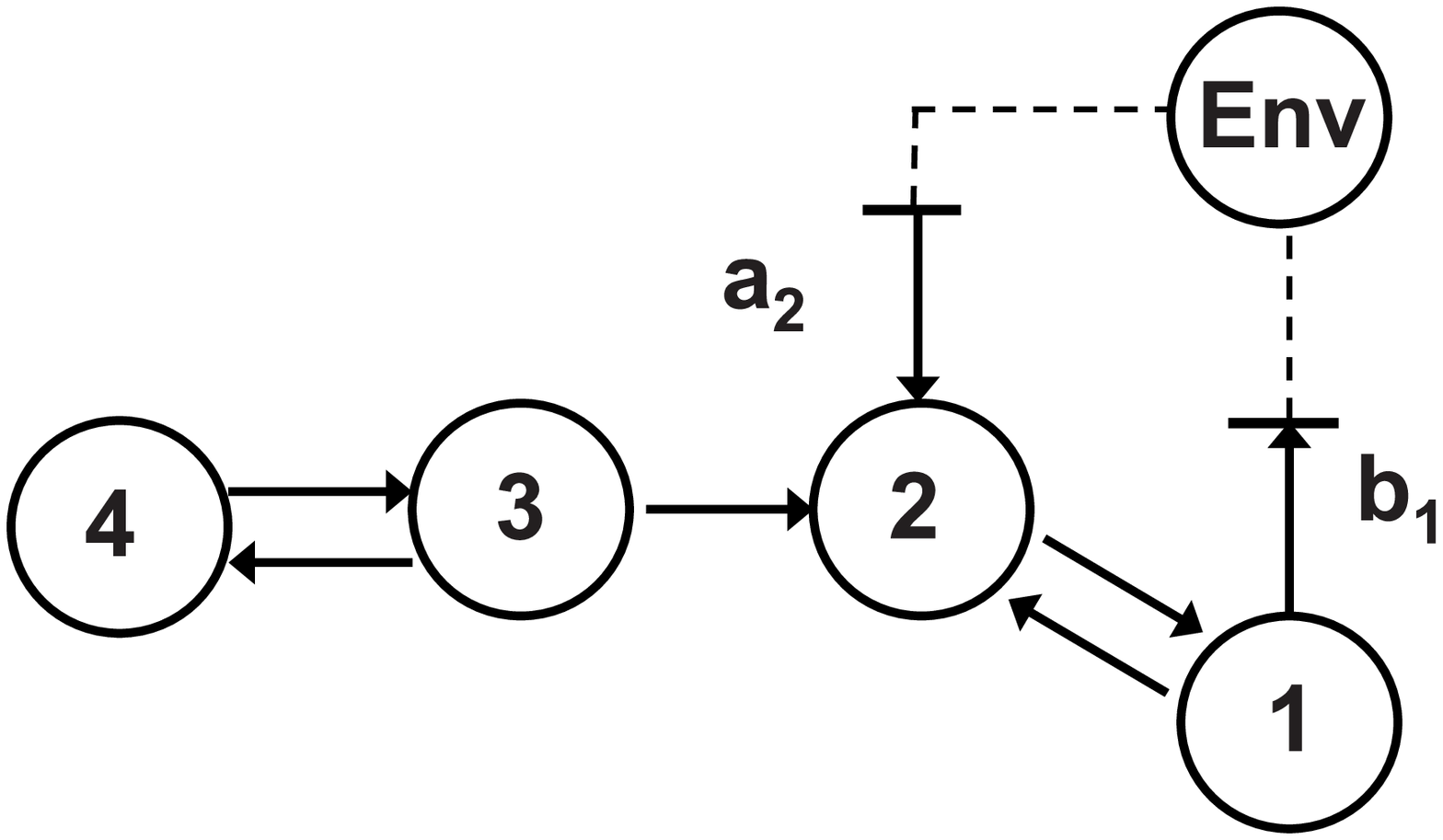} &
\includegraphics[scale=0.40]{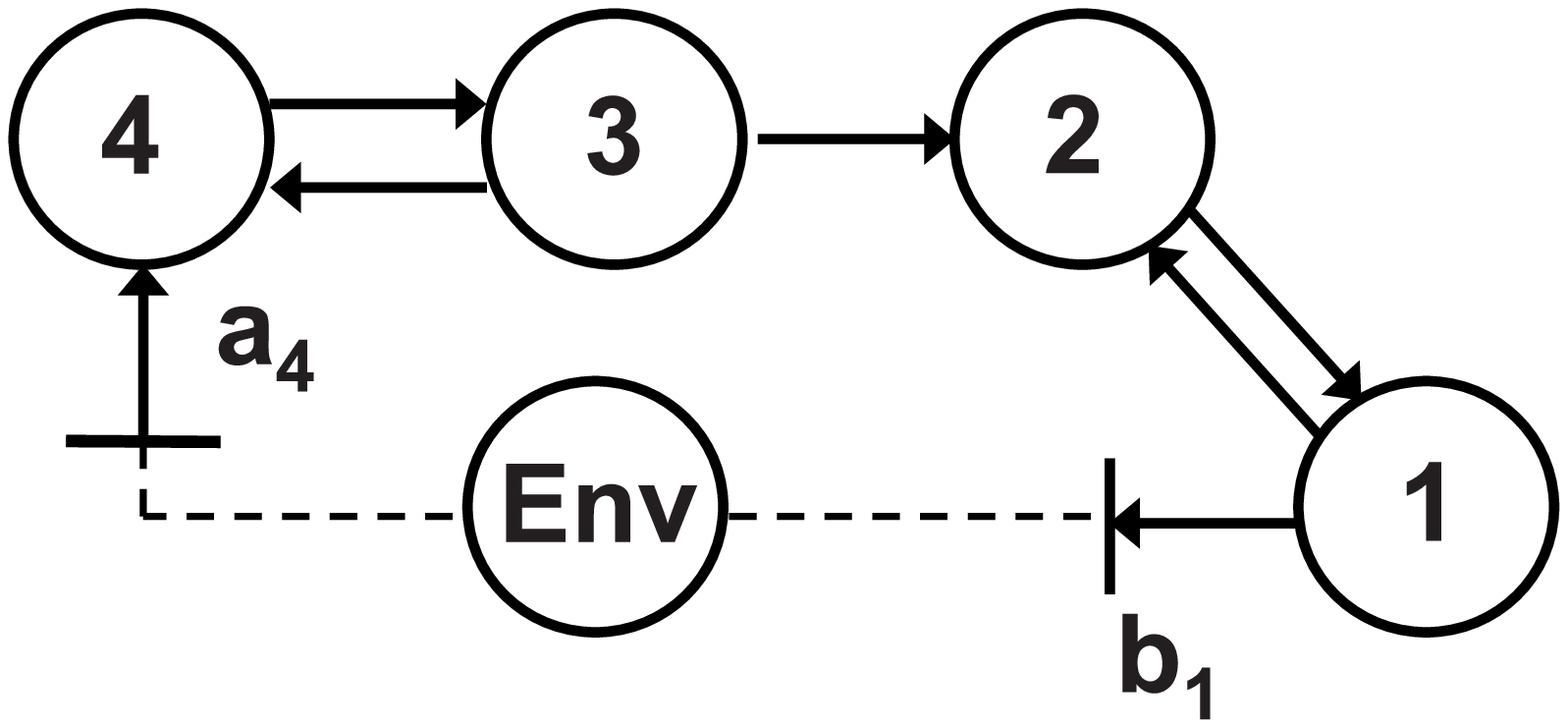}
\end{array}$
\end{center}
\caption{Some typical examples of directed graphs. For both cases, the terminal set is
${\mathcal{L}}_{\textnormal{p}} = \{ 1, 2\}$ whereas the set of non-terminal nodes is ${\mathcal{L}}_{\textnormal{q}} = \{ 3, 4 \}$.
(a) Exchange with
the environment takes place in the strong terminal set through input
$a_2$ (second coordinate of vector $\textbf{a}$) and output $b_1$ (first coordinate of vector $\textbf{b}$). Node \textbf{Env} represents the environment. (b) The input from the environment enters the system through a
non-terminal node (coordinate $a_4$ of vector $\textbf{a}$) while the output to the environment leaves from a terminal node (coordinate $b_1$ of vector $\textbf{b}$).} \label{fig:Figures_graph_ab1214_1}
\end{figure*}

Since in this work we are interested in linkage classes with just one strong terminal linkage class, the graphs we consider will only contain one strong terminal set. The set of non-terminal nodes is defined as ${\mathcal{L}}_{\textnormal{q}}
= {\mathcal{L}} \setminus {\mathcal{L}}_{\textnormal{p}}$.
%Graphs where all
%nodes plus the environment constitute a strong terminal set will be
%referred to as \emph{weakly reversible graphs}.
Some
examples of directed graphs are illustrated in Figure
\ref{fig:Figures_graph_ab1214_1}.
We associate to the graph $\mathcal{G}$ a state vector $\textbf{z}
\in {\mathbb{R}}^{n}$, where each component $z_i$ corresponds to a node. For each node $i=1,\dots,n$, we define an internal
net flux ${\phi}_i: {\mathbb{R}}^{n} \rightarrow {\mathbb{R}}$ and a net exchange with the environment ${\varphi}_i: {\mathbb{R}} \rightarrow {\mathbb{R}}$. Each flux is of the form:
\begin{equation}\label{eq:graph_GenFluX}
{\phi}_i (\textbf{z}) = \sum_{\{j | i \in {\cal{I}}_{j}\}} V_{ji} z_j - z_i \sum_{j
\in {\mathcal{I}}_i} V_{ij},
\end{equation}
where as in Section \ref{sec:formal_description_RN}, the indexes of the nodes that
are directly reached from node $i$ are grouped in a set ${\mathcal{I}}_i$, and $\{j | i \in {\cal{I}}_{j}\}$ refers to all nodes $j$ with edges directed to $i$. In addition, the net exchange with the environment is expressed as ${\varphi}_i (z_i) = a_i - b_i z_i$. From (\ref{eq:graph_GenFluX}), and similarly to expression (\ref{eq:linear_dependent_fluxes}), it is straightforward to see that:
\begin{equation}\label{eq:Sum_FluxMat}
\sum_{i \in {\mathcal{L}}} {\phi}_i (\textbf{z}) = 0.
\end{equation}
We consider that for each node $i$, the state $z_i$ will evolve in time as a function of the
corresponding internal and external net fluxes, so that ${\dot{z}}_i = {\phi}_{i} (\textbf{z}) + {\varphi}_i (z_i)$. Combining this expression with (\ref{eq:Sum_FluxMat}), the dynamics of the system can be described as:
\begin{equation}\label{eq:Aux_W_diff_Flux}
\dot{\textbf{z}} = \sum_{i=1}^{n} {\phi}_{i} (\textbf{z})
(\bm{\varepsilon}_i - \bm{\varepsilon}_1) - B \textbf{z} +
\textbf{a},
\end{equation}
where $B = {\cal{D}} (\textbf{b})$, a diagonal matrix with the components of
${\textbf{b}}$ in the diagonal. Equation
(\ref{eq:Aux_W_diff_Flux}) can be re-written in the alternative form:
\begin{equation}\label{eq:Aux_W_diff}
\dot{\textbf{z}} = W \textbf{z} +  \textbf{a}
\end{equation}
where matrix $W = V^T- B$, and $V \in {\mathbb{R}}^{n \times n}$
is the matrix that contains as off-diagonal components the coefficients in (\ref{eq:graph_GenFluX}), with $V_{ij} = 0$ if there is no directed edge $v_i \rightarrow v_j$. The diagonal elements for $V$ are of the form:
\begin{align*}
V_{ii} = - \sum_{\begin{array}{c} j=1\\ j\ne i\end{array}}^n V_{ij}.
\end{align*}
Both matrices $V^T$ and $W$ are compartmental \cite{Fife1972},
\footnote{ A matrix $A \in {\mathbb{R}}^{n \times n}$ is
compartmental if: (i) $A_{ij} \ge 0$, for $i,j=1,\dots,n$, $i\ne j$,
(ii) $\sum_{i=1}^n A_{ij} \leq 0$, for $j=1,\dots,n$.
%\begin{enumerate}
%\item $A_{ij} \ge 0$, for $i,j=1,\dots,n$, $i\ne j$ \\
%\item $\sum_{i=1}^n A_{ij} \leq 0$, for $j=1,\dots,n$.
%\end{enumerate}
} and belong to the class of Metzler matrices \cite{Arrow:89}.
It is known from e.g. \cite{Berman_Plemmons:94}, that the
eigenvalues of a compartmental matrix are either zero or they have
negative real parts. Such conclusion can be also reached from the structure of compartmental matrices
by applying the Gershgorin disc theorem
\cite{Golub_VanLoan:96}.
\begin{definition}\label{def:C-Metzler}
We say that the matrix $W \in {\mathbb{R}}^{n \times n}$ associated to the system  (\ref{eq:Aux_W_diff}) is
C-Metzler if its entries are of the form:
\begin{equation}\label{eq:entries_E}
\begin{array}{ll}
W_{ij} \geq 0 & \textnormal{for}~~ i \neq j\\
 ~~ & ~~\\
W_{ii} = -(b_i + {\sum}_{j \neq i} W_{ji})  & \textnormal{with}~~ b_i \geq 0 ~~\textnormal{and}~~ W_{ii} < 0, \\
\end{array}
\end{equation}
with at least one positive $b_i$ associated to the strong terminal set (i.e.
$i \in {\mathcal{L}}_{\textnormal{p}}$).
\end{definition}
\begin{proposition}\label{lemma:bounded_solution}
Consider system (\ref{eq:Aux_W_diff}) with $\textbf{a} \geq 0$
and nonnegative initial conditions $\mathbf{z}(0) \geq 0$. Then $\mathbf{z}(t) \geq 0$ for every $t >0$. \footnote{This result is actually a special case of Theorem 2 in \cite{Farina2000} (page 14).}
\end{proposition}
\noindent \textbf{Proof:} In order to prove the statement all we
need is to show that the flow associated to the differential system
on the boundary of the positive orthant is either aligned to
the boundary or oriented to the interior of the orthant. Before we
compute the flow, let us define the set $H_{k} = \{\textbf{z} \geq
0  ~ | ~ \bm{\varepsilon}_{k}^{\mathrm{T}} \textbf{z} = 0 \}$ which
characterizes the $k$-th facet of the positive orthant. The inner
product between the flow induced by (\ref{eq:Aux_W_diff_Flux})
(equivalently (\ref{eq:Aux_W_diff})) on any element $\textbf{z} \in
H_{k}$, and the unit vector orthogonal to $H_{k}$ takes the form:
\begin{equation}\label{eq:bound_facet}
\bm{\varepsilon}_{k}^{\mathrm{T}} \dot{\textbf{z}} = {\phi}_{k} (\textbf{z})
 - b_{k} z_{k} + a_k = \sum_{\{j | k \in {\cal{I}}_{j}\}} V_{jk} z_j  + a_k \geq
 0
\end{equation}
where the equivalence at the right hand side holds since $z_k = 0$
in $H_{k}$. Thus, at the boundary of $H_k$, the flow associated to the
differential system will be either aligned to the boundary
($\bm{\varepsilon}_{k}^{\mathrm{T}} \dot{\textbf{z}} = 0$) or
oriented to the interior of the positive orthant (i.e.
$\bm{\varepsilon}_{k}^{\mathrm{T}} \dot{\textbf{z}} > 0$). Repeating
the argument for all values of $k$ completes the proof. \hfill $\Box$

\begin{figure}[ht]
 \centering
\includegraphics[width=0.35\linewidth]{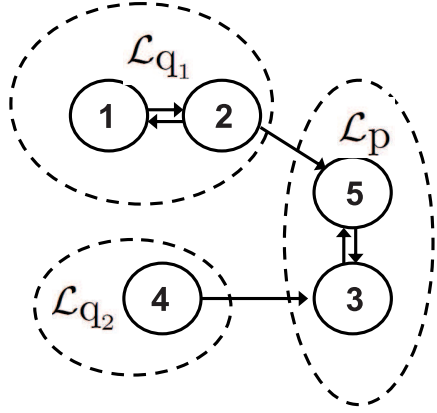}
\caption{ Example of a graph with $\textbf{a}, \textbf{b}=0$,
partitioned into a terminal set ${\mathcal{L}}_{\textnormal{p}}$ and
nonterminal sets ${\mathcal{L}}_{{\textnormal{q}}_{1}}$
${\mathcal{L}}_{{\textnormal{q}}_{2}}$. In this case set ${\mathcal{L}}_{\ell}$ comprises nodes $2$ and $4$.}\label{fig:network_anatomy}
\end{figure}

Next, we will study the equilibrium of system (\ref{eq:Aux_W_diff}) that results from  some constant non-negative vectors  $\textbf{a}$ and  $\textbf{b}$. To that purpose, let ${\mathcal{L}}_{\ell} \subset {\mathcal{L}}_{\textnormal{q}}$ be the set
containing those non-terminal nodes that are directly
linked to any node in the strong terminal set (the different partitions are illustrated
in Figure \ref{fig:network_anatomy}). Let us introduce functions $
\sigma_{\textnormal{p}} (\textbf{z}) = {\bm{\omega}}_{\textnormal{p}}^{T}
\textbf{z}$, $ \sigma_{\textnormal{q}} (\textbf{z})=
{\bm{\omega}}_{\textnormal{q}}^{T} \textbf{z}$ and $\sigma(\textbf{z})  = \sigma_{\textnormal{p}} (\textbf{z}) + \sigma_{\textnormal{q}} (\textbf{z})$, where:
\begin{equation}
{\bm{\omega}}_{\textnormal{p}} = \sum_{i \in
{\mathcal{L}}_{\textnormal{p}}} \bm{\varepsilon}_i,
~~{\bm{\omega}}_{\textnormal{q}} = \sum_{i \in
{\mathcal{L}}_{\textnormal{q}}} \bm{\varepsilon}_i.
\end{equation}
The time derivative of $\sigma_{\textnormal{q}}$ along system
(\ref{eq:Aux_W_diff_Flux}) is of the form:
\begin{eqnarray}
{\dot{\sigma}}_{\textnormal{q}} &=& {\bm{\omega}}_{\textnormal{q}}^{T}
\textbf{a} - \sum_{i \in {\mathcal{L}}_{\textnormal{q}}} b_i z_i -
\sum_{i \in {\mathcal{L}}_{\ell}} ~ \sum_{j \in {\mathcal{I}}_{i}}
 V_{ij} z_i \label{eq:sigma_q} \\
{\dot{\sigma}}_{\textnormal{p}} &=& {\bm{\omega}}_{\textnormal{p}}^{T}
\textbf{a} - \sum_{i \in {\mathcal{L}}_{\textnormal{p}}} b_i z_i +
\sum_{i \in {\mathcal{L}}_{\ell}} ~ \sum_{j \in {\mathcal{I}}_{i}}
 V_{ij} z_i \label{eq:sigma_p}
\end{eqnarray}
\begin{proposition}\label{prop:Zero_Value}
Consider system (\ref{eq:Aux_W_diff}) with $\mathbf{a} = 0$ and $W$ C-Metzler (Definition \ref{def:C-Metzler}). Then, for any $\mathbf{z}(0) \geq 0$, ${\mathbf{z}}^{*} = 0$ is the only equilibrium solution of
(\ref{eq:Aux_W_diff}).
\end{proposition}
\textbf{Proof:} Since $\textbf{z}(0) \geq 0$, by Proposition \ref{lemma:bounded_solution} we have that
$\textbf{z}(t)$ will remain
in the positive orthant,  ${\sigma}_{\textnormal{p}} (\textbf{z})$ and
${\sigma}_{\textnormal{q}} (\textbf{z})$ will be non-negative, and  $\sigma (\textbf{z}) = {\sigma}_{\textnormal{q}} (\textbf{z}) + {\sigma}_{\textnormal{p}} (\textbf{z}) \geq 0$.

For every node $ i \in {\mathcal{L}}_{\ell}$ it is clear that $z_i^* = 0$ is the only possible equilibrium. Otherwise, from (\ref{eq:sigma_q}), it follows that  ${\dot{\sigma}}_{\textnormal{q}} <0$ for all times, what would make  ${\sigma}_{\textnormal{q}} (\textbf{z})$ to become negative. In addition, note that for $z_i^* = 0$ to be an equilibrium point for node $ i \in {\mathcal{L}}_{\ell}$, requires the equilibrium states for all nodes $j$ linked to $i$ (thus $ j \in {\mathcal{L}}_{\textnormal{q}}$) to be zero as well, otherwise ${\dot{z}}_i = {\phi}_{i} (\textbf{z}) + {\varphi}_i (z_i) >0$. Repeating the argument upstream to the nodes linked to $j$,  we conclude that zero is the only possible equilibrium for every node in ${\mathcal{L}}_{\textnormal{q}}$.

Since $W$ is C-Metzler (i.e. there exists at least one index $i\in\mathcal{L}_{\textnormal{p}}$ for which $b_i>0$), zero is also the only possible equilibrium for every node in the terminal set.  Suppose, on the contrary, that there exists a positive equilibrium $z_i^*$ for some
$ i \in \mathcal{L}_{\textnormal{p}}$. In the limit, expression (\ref{eq:sigma_p}) would become:
\[
\lim_{t \rightarrow \infty} {\dot{\sigma}}_{\textnormal{p}} = - \sum_{i \in {\mathcal{L}}_{\textnormal{p}}} b_i z^*_i < 0,
\]
but ${\sigma}_{\textnormal{p}} (\textbf{z})$ cannot become negative, thus $z_i^* = 0$.
Finally, since $z_i^* = 0$ is an equilibrium point for $i\in\mathcal{L}_{\textnormal{p}}$ such that $b_i>0$, the equilibrium states for all nodes $j$ linked to $i$ must be zero as well, otherwise ${\dot{z}}_i = {\phi}_{i} (\textbf{z}) + {\varphi}_i (z_i) >0$.
repeating the argument for every $j\in\mathcal{L}_{\textnormal{p}}$  we have that $\lim_{t \rightarrow \infty} {\textbf{z}} (t) = {\textbf{z}}^{*} = 0$.
\hfill $\Box$

\begin{remark}
Using the terminology of \cite{Fife1972}, $V^T$ defines a compartmental system with one trap, where the notion of trap is equivalent to the definition of strong terminal set used in this paper. Therefore, zero is an eigenvalue of $V^T$ with multiplicity $1$. $W$ is also a compartmental matrix, and due to the choice of $b_i$, the compartmental system corresponding to $W$ contains no traps. Therefore, $W$ is of full rank and thus, assuming $\textbf{a} = 0$, the only equilibrium of (\ref{eq:Aux_W_diff}) is ${\textbf{z}}^{*} = 0$.
\end{remark}
\begin{proposition}\label{prop:equil_WRev}
Consider system (\ref{eq:Aux_W_diff}), with matrix $W$ being C-Metzler (Definition \ref{def:C-Metzler}). Equilibrium ${\mathbf{z}}^{*}$ will be non-negative and globally asymptotically stable. Moreover, let some entry $k$ of vector $\mathbf{a}$ positive (i.e. $a_k >0$). Then, for all nodes $j\in\mathcal{L}$ reached from node $k$ by directed paths, we have that $z^*_j >0$.
\end{proposition}
\textbf{Proof:}
First note that because matrix $W$ in (\ref{eq:Aux_W_diff}) is compartmental, its eigenvalues are either zero or they have negative real parts. Thus, to show that the equilibrium is globally asymptotically stable it only remains to prove that no zero
eigenvalue exists. Actually, this is the case, since from Proposition \ref{prop:Zero_Value}, the only equilibrium solution for which
$W{\textbf{z}}^{*} = 0$ is  ${\textbf{z}}^{*} = 0$. Since all eigenvalues have negative real part, $W$ is invertible and the equilibrium ${\textbf{z}}^{*} = -W^{-1} {\textbf{a}}$ is globally asymptotically stable.

In proving the second part of the statement, we have that since $a_k >0$, the equilibrium in node $k$ must be:
\[
z_k^* = \frac{\sum_{\{j | k \in {\cal{I}}_{j}\}} V_{jk} z^*_j + a_k}{b_k + \sum_{j
\in {\mathcal{I}}_k} V_{kj}} >0
\]
and so is the case for all $j$ reached from $k$, so that $z^*_j >0$, what completes the proof.\hfill $\Box$
\begin{remark}\label{remark:WCM}
The full rank property of $W$ and the uniqueness of the equilibrium  ${\textbf{z}}^{*} = 0$ for $\textbf{a} = 0$ are equivalent, since this latter means that the dimension of the kernel of $W$ is zero. From this, it follows that $W$ cannot have zero eigenvalues, and from the compartmental property of $W$ we obtain that all the real parts of its eigenvalues are negative. This means that $W$ is a stability matrix. Since $\textbf{a}$ can be considered as a bounded input, ${\textbf{z}}^{*} = W^{-1}(-\textbf{a})$ is a unique asymptotically stable equilibrium point of (\ref{eq:Aux_W_diff}). Since $-W$ is a full-rank M-matrix, it is inverse-nonnegative \cite{Plemmons1977}, i.e. all entries of $W^{-1}$ are non-positive. This implies that $\textbf{z}^{*}$ is a nonnegative vector.
\end{remark}
\begin{lemma}\label{prop:E_Metzler}
Any C-Metzler matrix $W$ is non-singular and its inverse
$W^{-1}$ non-positive.  Let its associated graph be numbered so that the first $\textnormal{p}$ nodes are in
${\mathcal{L}}_{\textnormal{p}}$ (thus, the remaining $\textnormal{q} = n - \textnormal{p}$ are in ${\mathcal{L}}_{\textnormal{q}}$). Then $N = W^{-1}$ can be partitioned as:
\begin{equation}\label{eq:structN}
N = \left[
\begin{array}{c|c}
  N_{\textnormal{p}} & N_{\textnormal{pq}} \\
  \hline
  \emptyset & N_{\textnormal{q}}
\end{array}
\right],
\end{equation}
with $\emptyset \in
{\mathbb{R}}^{\textnormal{q} \times \textnormal{p}}$ the zero matrix,  $N_{\textnormal{p}} \in
{\mathbb{R}}^{\textnormal{p} \times \textnormal{p}}$ and  $N_{\textnormal{pq}} \in
{\mathbb{R}}^{\textnormal{p} \times \textnormal{q}}$ strictly negative matrices, and $N_{\textnormal{q}} \in
{\mathbb{R}}^{\textnormal{q} \times \textnormal{q}}$ non-positive. Moreover, for each node $j \in {\mathcal{L}}_{\textnormal{q}}$, every node $i$ not reached from $j$ by directed paths will correspond with an entry $N_{ij} = 0$.
\end{lemma}

\textbf{Proof:} That $W$ (being C-Metzler)  is non-singular and therefore invertible
has been shown in the proof of Proposition \ref{prop:equil_WRev} (see also Remark \ref{remark:WCM}). In addition, note that the equilibrium solution can be written as ${\textbf{z}}^{*} = -N \textbf{a}$, and choose $\textbf{a} = \bm{\varepsilon}_j$. For each $j \in \mathcal{L}$, the corresponding equilibrium will be
${\textbf{z}}^{*} = -(N)_j$, where $(N)_j$ represents the $j$-th column
of matrix $N$.

According to Proposition
\ref{prop:equil_WRev}, the equilibrium state for all nodes $i \in \mathcal{L}$ reached from $j$ by directed paths must be positive, and thus the corresponding entries $N_{ij}$ must be negative.  This holds for all $i, j \in {\mathcal{L}}_{\textnormal{p}}$. Thus, using the order given in the statement of the Lemma, we can conclude that $N_{\textnormal{p}}$ is strictly negative. In addition, every $i \in {\mathcal{L}}_{\textnormal{p}}$ can be reached from $j \in {\mathcal{L}}_{\textnormal{q}}$ what makes $N_{\textnormal{pq}}$ strictly negative as well. The zero matrix $\emptyset$ appears since none of the nodes $i \in {\mathcal{L}}_{\textnormal{q}}$ can be reached from nodes $j \in {\mathcal{L}}_{\textnormal{p}}$. Finally, zero entries in $N_{\textnormal{q}}$ will correspond to those $i$ not reachable from $j$,  with $i,j \in {\mathcal{L}}_{\textnormal{q}}$.
\hfill $\Box$

Lemma \ref{prop:E_Metzler} will be invoked along the sequel on a linkage class basis to study positive equilibrium solutions of chemical reaction networks. The possibility of positive equilibria is at a large extent connected with Proposition 4.1 in Lecture $4$ of Feinberg  \cite{Feinberg:79} (see also proofs in \cite{Horn:72} and \cite{Feinberg-Horn:77}) on the structure of the kernel of $A_k$ defined in (\ref{subeq:canonical_form_AK}). In this regard, it is noted that the arguments behind Proposition \ref{prop:equil_WRev} and Lemma \ref{prop:E_Metzler} can serve as a basis to prove the result in Feinberg Lectures.

{\bf Example:} The graphs in Figure \ref{fig:Figures_graph_ab1214_1}
depict two different scenarios. In Figure
\ref{fig:Figures_graph_ab1214_1}a, input $a_2$ enters a node in the
strong terminal set thus forcing the states $z_i^{*}$ in that set to
be strictly positive, while leaving those that belong to the non-terminal set to reach zero. Figure \ref{fig:Figures_graph_ab1214_1}b, describes the case in
which the input enters a non-terminal node that communicates with
the remaining nodes of the graph (non-terminal and terminal), forcing the system to reach a strictly positive equilibrium state ${\textbf{z}}^{*} > 0$.

Since $b_1 > 0 $ leaves a node from the strong terminal set, the associated matrix $W$ is C-Metzler (Definition \ref{def:C-Metzler}). The sign patterns for the first two columns $(N)_1$ and
$(N)_2$ of its corresponding inverse $N$ are of the form:
$(\begin{array}{rrrr} - & - & 0 & 0
\end{array})^T$. For the $3^{rd}$ and $4^{th}$ columns, the sign patterns
are $(\begin{array}{rrrr} - & - & - & - \end{array})^T$, since every node can be reached from either node $3$ or $4$. The sign pattern of the negative inverse will then be:
\[
(-N) = \left[
\begin{array}{rrrr}
+ & + & + & +\\
+ & + & + & +\\
0 & 0 & + & +\\
0 & 0 & + & +\\
\end{array}
\right].
\]
\hfill $\triangle$

\section{A Canonical Representation of the Equilibrium Set}\label{sec:canonical_representation}

The time
evolution of the concentration vector (\ref{eq:dotC_first}) can be re-written in terms of the $S_{\lambda}$ matrices associated to the stoichiometric subspace $\Xi$, already discussed in Section
\ref{secsec:reaction_simplex}, so that:
\begin{equation}\label{eq:dotC_Sl}
            \dot{\textbf{c}} =  \sum_{\lambda} {S}_{\lambda}
            {\bm{\phi}}_{\lambda} ({\psi}_{j_{\lambda}} (\textbf{c}), {{\bm{\psi}}}_{\lambda} (\textbf{c})),
\end{equation}
where ${\psi}_{j_{\lambda}}: {\mathbb{R}}_{>0}^{m} \rightarrow {\mathbb{R}}_{>0}$,  corresponds to the monomial associated to the reference complex,  and ${{\bm{\psi}}}_{\lambda}: {\mathbb{R}}_{>0}^{m} \rightarrow {\mathbb{R}}_{>0}^{N_{\lambda} -1}$ is the vector function that includes as coordinate functions the monomials associated to the remaining complexes. Finally, vector function ${\bm{\phi}}_{\lambda}: {\mathbb{R}}_{\geq 0}^{N_{\lambda}} \rightarrow {\mathbb{R}}^{N_{\lambda} -1}$ contains the corresponding fluxes. Element-wise, monomials relate to fluxes by expression
(\ref{eq:this_is_flux}), that in matrix form can be written as:

\begin{equation}\label{eq:fundamental_Phi}
\left( \begin{array}{c}
{\phi}_{j_{\lambda}} ({\psi}_{j_{\lambda}}, {{\bm{\psi}}}_{\lambda})\\
{\bm{\phi}}_{\lambda} ({\psi}_{j_{\lambda}}, {{\bm{\psi}}}_{\lambda})
\end{array} \right)  = M_{\lambda}
\left( \begin{array}{c}
{\psi}_{j_{\lambda}} \\
{\bm{\psi}}_{\lambda}
\end{array} \right),
\end{equation}
where the fluxes and monomials are ordered such that for each linkage class, the first elements correspond to the reference complex ${j_{\lambda}}$ (i.e. ${\phi}_{j_{\lambda}}(\textbf{c}), {\psi}_{j_{\lambda}} (\textbf{c})$).
Matrix $M_{\lambda} \in {\mathbb{R}}^{N_{\lambda}\times
N_{\lambda}}$  is a compartmental matrix that has as off-diagonal entries the
corresponding reaction constants. An explicit description of its structure can be given as follows:

Let $\mathcal{L}_{\lambda}(i)$ denote the $i$-th element in the set $\mathcal{L}_{\lambda}$ for $i=1,\dots,N_{\lambda}$.  Without loss of generality, we can assume that the first element of $\mathcal{L}_{\lambda}$ is the index of the reference complex, i.e. $\mathcal{L}_{\lambda}(1)= j_{\lambda}$. In addition, let
$k_{\mathcal{L}_{\lambda}(i),\mathcal{L}_{\lambda}(j)}$ denote the reaction rate coefficient associated to a possible reaction step from complex $\mathcal{L}_{\lambda}(i)$ to $\mathcal{L}_{\lambda}(j)$, being $0$ if such reaction step does not exist. Then:
\begin{align}
(M_{\lambda})_{ij} & = k_{\mathcal{L}_{\lambda}(j),\mathcal{L}_{\lambda}(i)}~~\text{for}~~i,j=1,\dots,N_{\lambda},~ i\ne j\\
(M_{\lambda})_{ii} & = -\sum_{l=1,l\ne i}^{N_{\lambda}} k_{\mathcal{L}_{\lambda}(i),\mathcal{L}_{\lambda}(l)}~~\text{for}~~i=1,\dots,N_{\lambda}.
\end{align}

Note that because of
(\ref{eq:linear_dependent_fluxes}), we have that:
\begin{equation}\label{eq:PhiPsij}
{\phi}_{j_{\lambda}} ({\psi}_{j_{\lambda}}, {{\bm{\psi}}}_{\lambda}) + {\mathbf{1}}_{N_{\lambda} -1}^T  {{\bm{\phi}}}_{\lambda}({\psi}_{j_{\lambda}}, {{\bm{\psi}}}_{\lambda}) = 0,
\end{equation}

so that $M_{\lambda}$ is a Kirchoff (i.e. column
conservation) matrix, which for convenience we re-write as:
\begin{equation}\label{eq:structM_k_l}
M_{\lambda} = \left[
\begin{array}{c|c}
  -({\mathbf{1}}_{N_{\lambda} -1}^T {\textbf{a}}_{\lambda}) & {\textbf{b}}^{T}_{\lambda} \\
  \hline
  {\textbf{a}}_{\lambda} & E_{\lambda}
\end{array}
\right],
\end{equation}
with ${\textbf{a}}_{\lambda},
{\textbf{b}}_{\lambda} \in {\mathbb{R}}^{N_{\lambda} -1}$, $E_{\lambda} \in {\mathbb{R}}^{(N_{\lambda} -1) \times
(N_{\lambda} -1)}$, and ${\textbf{b}}_{\lambda}^{T} = -{\mathbf{1}}_{N_{\lambda} -1}^T E_{\lambda}$.
By construction, the off-diagonal elements of the first column and row in $M_{\lambda}$  correspond to the rate coefficients for reaction steps leaving and entering,
respectively, the reference complex $j_{\lambda}$. Such reference has been chosen to be in the terminal linkage class, what ensures
${\textbf{b}}_{\lambda}$ to be non-zero,
since at least one reaction step is directed to the
reference complex. The off-diagonal elements of matrix $E_{\lambda}$  collect the remaining rate coefficients.

\begin{proposition}\label{eq:ELambdaPropFun}
$E_{\lambda}$ in expression (\ref{eq:structM_k_l})
is C-Metzler, therefore invertible, and its inverse non-positive.
\end{proposition}
\textbf{Proof:}
First, we note that $E_{\lambda}$ complies with Definition \ref{def:C-Metzler} (C-Metzler matrices). This is so because it is associated to a directed graph which coincides with the linkage class (the environment corresponds with the reference complex). By construction, the off-diagonal elements of $E_{\lambda}$ are  either zero or positive. In addition, since ${\textbf{b}}_{\lambda}^{T} = -{\textbf{1}}^{T} E_{\lambda}$, for each diagonal element we have that:
\[
(E_{\lambda})_{ii} = - [ ({\textbf{b}}_{\lambda})_i + {\sum}_{j \neq i} (E_{\lambda})_{ji} ],
\]
where at least one component of ${\textbf{b}}_{\lambda}$ associated to the reference complex (linked to the strong terminal set) is positive. Thus, $E_{\lambda}$ is C-Metzler
and the result then follows by applying Lemma \ref{prop:E_Metzler}.
\hfill $\Box$

Inspection of Eqn (\ref{eq:dotC_Sl}) suggests that apart from complex balanced equilibrium solutions (i.e. those satisfying Definition \ref{def:Complex_Balance_Condition}), non-zero flux combinations can lead to equilibrium, if the corresponding flux vector
${\bm{\phi}} ({{\bm{\psi}}}) = [\begin{array}{ccccc} {\bm{\phi}}_{1}^T ({\psi}_{j_{1}}, {{\bm{\psi}}}_{1})& \cdots & {\bm{\phi}}_{\lambda}^T ({\psi}_{j_{\lambda}}, {{\bm{\psi}}}_{\lambda}) & \cdots & {\bm{\phi}}_{\ell}^T ({\psi}_{j_{\ell}}, {{\bm{\psi}}}_{\ell}) \end{array}]^T$ lies in the kernel of $S$  (\ref{eq:MatrixSNet}). In other words, if the network has non-zero deficiency, vector
${\bm{\phi}} ({{\bm{\psi}}})$ can be written as a linear combination of the set of vectors $\{{\textbf{g}}^{r} ~|~ r=1, \ldots, \delta \}$ that define a basis for the kernel of $S$, so that:
\begin{equation}
{\bm{\phi}} ({{\bm{\psi}}}) = \sum_{r} {\nu}_{r}
{\textbf{g}}^{r},
\end{equation}
for some given scalars ${\nu}_{r}$. Making use of (\ref{eq:BasisKernel}) in the above summation to express each element $r$ of the basis in terms of the sub-vectors ${\textbf{g}}^{r}_{\lambda}$, for $\lambda = 1, \cdots, \ell$, the flux vector for each linkage class can be written as:
\begin{equation}\label{eq:fluxes_def_alpha}
{\bm{\phi}}_{\lambda} ({\psi}_{j_{\lambda}}, {{\bm{\psi}}}_{\lambda}) = \sum_{r} {\nu}_{r}
{\textbf{g}}^{r}_{\lambda}.
\end{equation}
Note that such fluxes lead in fact to an equilibrium solution. This can be verified by substituting (\ref{eq:fluxes_def_alpha}) into (\ref{eq:dotC_Sl}) and re-grouping summations, so that:
\begin{align}
\dot{\textbf{c}} = \sum_{\lambda}  S_{\lambda} \sum_{r} {\nu}_{r} {\textbf{g}}^{r}_{\lambda} = \sum_{r} {\nu}_{r} \sum_{\lambda}  S_{\lambda}  {\textbf{g}}^{r}_{\lambda},
\end{align}
where the right hand side is zero because of (\ref{eq:formalSg}).

\begin{figure}[ht]
\begin{center}
$\begin{array}{@{}cc@{}}
\mathrm{(a)} & \mathrm{(b)} \\
\includegraphics[scale=0.40]{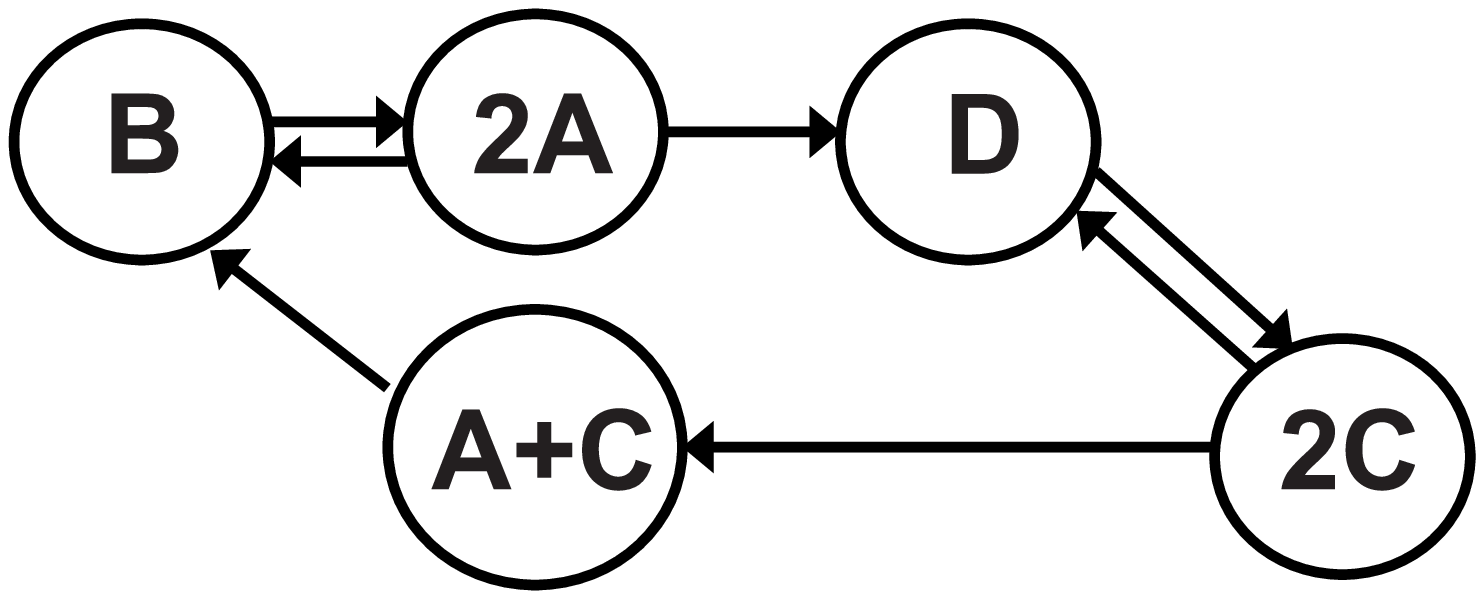} &
\includegraphics[scale=0.40]{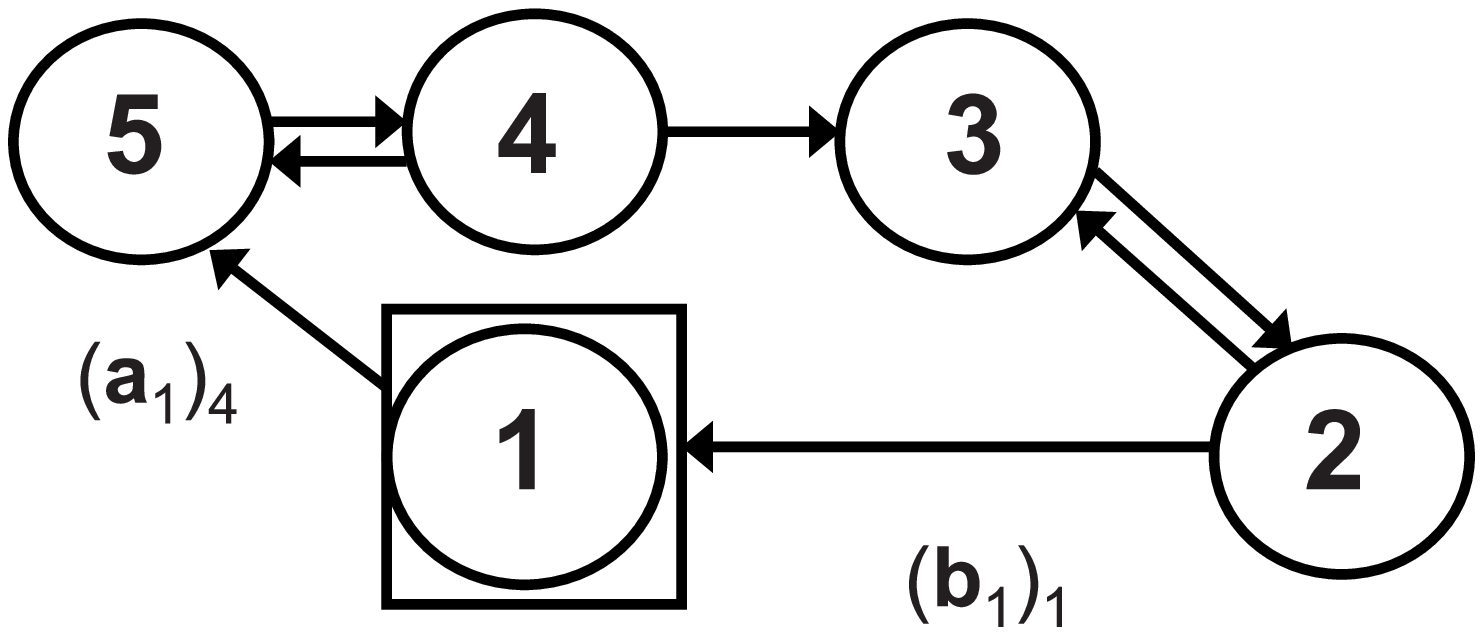}
\end{array}$
\end{center}
\caption{A one linkage class weakly reversible network consisting of
$4$ chemical species $\{ A, B, C, D \}$.  (a) The graph of complexes with
explicit indication of the chemical species. (b) The graph with
numbered complexes, including the reference complex in the
square box and the graph associated to the C-Metzler matrix with
the non-zero coordinates of vectors ${\textbf{a}}_{1}$ and
${\textbf{b}}_{1}$, represented in the diagram as $({\textbf{a}}_{1})_4$ and $({\textbf{b}}_{1})_1$,
respectively.} \label{fig:OneLCN_WWR_ab14}
\end{figure}

{\bf Example:} Let us consider the network depicted in Figure
\ref{fig:OneLCN_WWR_ab14}. Since $\ell = 1$, there is just one matrix $M_{1}$
(\ref{eq:structM_k_l}) that consists of the following sub-matrix
components:
\begin{equation}
E_{1} = \left[ \begin{array}{rrrr}
    -(k_{21} + k_{23}) & k_{32} & 0 & 0\\
    k_{23} & -k_{32} & k_{43} & 0 \\
    0 & 0 & -(k_{43} + k_{45}) & k_{54} \\
    0 & 0 & k_{45} & -k_{54} \\
\end{array} \right],  ~~ {\textbf{a}}_{1} = \left[ \begin{array}{r}
    0 \\
    0 \\
    0 \\
    k_{15} \\
\end{array} \right], ~~ {\textbf{b}}_{1} = \left[ \begin{array}{r}
    k_{21} \\
    0 \\
    0 \\
    0 \\
\end{array} \right].
\end{equation}

The $Y$ and $S$ matrices described in Section \ref{sec:formal_description_RN} are, respectively:
\begin{equation}\label{eq:Molec_Stoich}
Y = \left[
\begin{array}{ccccc}
  1 & 0 & 0 & 2 & 0\\
  0 & 0 & 0 & 0 & 1\\
  1 & 2 & 0 & 0 & 0\\
  0 & 0 & 1 & 0 & 0\\
\end{array}
\right]  ~~ \textnormal{and} ~~ S = \left[ \begin{array}{rrrr}
   -1 & -1 &  1 & -1\\
    0 &  0 &  0 &  1\\
    1 & -1 & -1 & -1\\
    0 &  1 &  0 &  0\\
    \end{array} \right],
\end{equation}
Since $rank(S) = 3$, the network has deficiency $\delta
= 1$. Thus, equilibrium solutions would correspond to fluxes
satisfying relation (\ref{eq:fluxes_def_alpha})  with
${\textbf{g}}^1 \equiv {\textbf{g}}_{1}^1 = \left(
\begin{array}{cccc} 1 & 0 & 1 & 0
\end{array} \right)^T$ defining a one-dimensional subspace.
\hfill $\triangle$

In computing the set of positive equilibrium solutions,
we make use of (\ref{eq:fundamental_Phi}) and (\ref{eq:structM_k_l}) for every linkage class,  to express fluxes ${\bm{\phi}}_{\lambda}$ as:
\begin{align}\label{eq:philambda1}
{\bm{\phi}}_{\lambda} ({\psi}_{j_{\lambda}}, {{\bm{\psi}}}_{\lambda}) = \psi_{j_{\lambda}} \textbf{a}_{\lambda}  + E_{\lambda} {\bm{\psi}}_{\lambda}.
\end{align}
Combining the right hand sides of (\ref{eq:fluxes_def_alpha}) and (\ref{eq:philambda1}) we get:
\begin{equation}\label{eq:family_per_linkage}
E_{\lambda} {\bm{\psi}}_{\lambda} +
 {\psi}_{j_{\lambda}} {{\textbf{a}}}_{\lambda} = \sum_{r} {\nu}_{r}
{\textbf{g}}^{r}_{\lambda}.
\end{equation}
Let $\bm{\nu} = [{\nu}_1~\dots¬{\nu}_{\delta}]^T$, and define a vector $\bm{\chi} \in {\mathbb{R}}^{\delta}$ on the unit sphere as $\bm{\chi}=\frac{1}{\|\bm{\nu} \|} \bm{\nu}$.
Since any positive equilibrium solution requires ${\psi}_{j_{\lambda}} > 0$ and ${\bm{\psi}}_{\lambda} > 0$, we can multiply both sides of Eqn
(\ref{eq:family_per_linkage}) by $1/{\psi}_{j_{\lambda}}$ to obtain,
after some re-arrangement:
\begin{equation}\label{eq:Ef_L}
E_{\lambda} {\textbf{f}}_{\lambda} + {{\textbf{a}}}_{\lambda} =
{x}_{\lambda} {\textnormal{G}}_{\lambda} \bm{\chi}, ~~ \textnormal{with}~~ {\textbf{f}}_{\lambda} \equiv (1/\psi_{j_{\lambda}}){\bm{\psi}}_{\lambda}.
\end{equation}
In the above expression, ${\textnormal{G}}_{\lambda} \in {\mathbb{R}}^{(N_{\lambda} -1) \times \delta}$ is a matrix of the form ${\textnormal{G}}_{\lambda} = [{\textbf{g}}^{1}_{\lambda} \cdots
{\textbf{g}}^{r}_{\lambda} \cdots {\textbf{g}}^{\delta}_{\lambda}]$,
and $x_{\lambda} \equiv \frac{\|\nu \|}{\psi_{j_{\lambda}}}$ a scalar variable. It must be noted that because
$\|\nu \| = x_{\lambda} \psi_{j_{\lambda}}$ for every  $\lambda = 1,\dots \ell$, variables $x_{\lambda}$ are not independent but related to each other through the following equalities:
\begin{equation}\label{eq:TheLinkx_lambda}
x_{1} {\psi}_{j_{1}} = \cdots = x_{\lambda}{\psi}_{j_{\lambda}} = \cdots = x_{\ell} {\psi}_{j_{\ell}}.
\end{equation}
Since $E_{\lambda}$
is invertible (Proposition \ref{eq:ELambdaPropFun}), we can solve (\ref{eq:Ef_L}) to get a vector function ${\textbf{f}}_{\lambda}: \mathbb{R} \times {\mathbb{R}}^{\delta} \rightarrow {\mathbb{R}}^{(N_{\lambda} -1)}$ of the form:
\begin{equation}\label{eq:Fexplicit_first}
{\textbf{f}}_{\lambda} ({x}_{\lambda}; \bm{\chi})  =
{\textbf{f}}_{\lambda}^{*} +  {x}_{\lambda} {\textbf{h}}_{\lambda}
(\bm{\chi}),
\end{equation}
where ${\textbf{f}}_{\lambda}^{*} = -E_{\lambda}^{-1}
{\textbf{a}}_{\lambda}$ and ${\textbf{h}}_{\lambda} (\bm{\chi}) = E_{\lambda}^{-1} {\textbf{g}}_{\lambda} (\bm{\chi})$, with ${\textbf{g}}_{\lambda} (\bm{\chi}) = G_{\lambda} \bm{\chi}$.
Vector fluxes in (\ref{eq:philambda1}) can be expressed as a function of (\ref{eq:Fexplicit_first}) for each linkage class so that:
\begin{equation}\label{eq:FluxZeroProof}
{\bm{\phi}}_{\lambda} (\psi_{j_{\lambda}}, x_{\lambda};\bm{\chi})= \psi_{j_{\lambda}} \left[\textbf{a}_{\lambda}  + E_{\lambda} {\mathbf{f}}_{\lambda}(x_{\lambda};\bm{\chi})\right].
\end{equation}
In this way, for each $\bm{\chi}$ in the unit sphere and $\textbf{x} = ({x}_{1}, \cdots, {x}_{\lambda}, \cdots, {x}_{\ell})^T$, constrained by (\ref{eq:TheLinkx_lambda}) to be either zero or to belong to the interior of the positive orthant ${\mathbb{R}}^{\ell}_{>0}$, the right hand side of (\ref{eq:dotC_Sl}) vanishes.
This follows since for each $\bm{\chi}$ on the unit sphere, the corresponding fluxes (\ref{eq:FluxZeroProof}) become:
\[
{\bm{\phi}}_{\lambda} (\psi_{j_{\lambda}}, x_{\lambda};\bm{\chi}) =  x_{\lambda} \psi_{j_{\lambda}} G_{\lambda} \bm{\chi} .
\]
Substituting the above expressions in (\ref{eq:dotC_Sl}) and making use of (\ref{eq:TheLinkx_lambda}) we get:
\begin{align}
\dot{\textbf{c}} = \left( \sum_{\lambda}  x_{\lambda} \psi_{j_{\lambda}} S_{\lambda} G_{\lambda} \right) \bm{\chi} = x_{1} \psi_{j_{1}} \left( \sum_{\lambda}   S_{\lambda} G_{\lambda} \right) \bm{\chi},
\end{align}
where the right hand side is zero because from  (\ref{eq:formalSg}) we have that:
\[
\sum_{\lambda}   S_{\lambda} G_{\lambda} = 0.
\]
Note that for a vector  ${\bm{\chi}}' = - \bm{\chi}$ we have that $\textbf{x} \in \{0 \} \cup {\mathbb{R}}^{\ell}_{<0}$, so it is enough to study vector functions (\ref{eq:Fexplicit_first}) for  $\textbf{x} \in \{0 \} \cup {\mathbb{R}}^{\ell}_{>0} \cup {\mathbb{R}}^{\ell}_{<0}$ and $\bm{\chi}$ in the set:
\begin{equation}\label{eq:UnitHalfSphere}
{\mathbf{U}} = \{ \bm{\chi} \in {\mathbb{R}}^{\delta} \mid \| \bm{\chi} \| =1,~ \bm{\varepsilon}_1^T {\bm{\chi}}  \geq 0 \}.
\end{equation}
\begin{definition}\label{def:FamilySolutions}
 {\bf (Family of Solutions\footnote{A closely related notion has been proposed in \cite{Otero:12}})} Let ${\mathbb{D}}_{0} \subset {\mathbb{R}}^{\ell} $ be the domain $\{0 \} \cup {\mathbb{R}}^{\ell}_{>0} \cup {\mathbb{R}}^{\ell}_{<0}$, and ${\mathbf{f}}_{\bm{\eta}}: {\mathbb{D}}_{0} \times {\mathbf{U}} \rightarrow {\mathbb{R}}^{(n -\ell)}$ a vector function:
\begin{equation}\label{eq:fNetwork}
{\mathbf{f}}^T_{\bm{\eta}}(\mathbf{x}; \bm{\chi}) = \left[
\begin{array}{ccccc}
{\mathbf{f}}^T_{1} ({x}_{1}; \bm{\chi}) & \cdots & {\mathbf{f}}^T_{\lambda} ({x}_{\lambda}; \bm{\chi}) & \cdots & {\mathbf{f}}_{\ell}^T ({x}_{\ell}; \bm{\chi})
\end{array}
\right],
\end{equation}
with $\mathbf{x} = ({x}_{1}, \cdots, {x}_{\ell})^T$ and ${\mathbf{f}}_{\lambda} ({x}_{\lambda}; \bm{\chi})$ for $\lambda = 1, \cdots, \ell$ as in (\ref{eq:Fexplicit_first}). We will refer to ${\mathbf{f}}_{\bm{\eta}}(\textbf{x}; \bm{\chi})$ as the family of solutions.
\end{definition}
In order to comply with positive equilibrium solutions in the concentration space, each vector function ${\textbf{f}}_{\lambda} ({x}_{\lambda}; \bm{\chi})$ in (\ref{eq:Fexplicit_first}) must be strictly positive, and related to the reaction monomials by the expression:
\begin{equation}\label{eq:vectorfuncfL}
{\textbf{f}}_{\lambda} = \exp \left( \ln
\frac{1}{{\psi}_{j_{\lambda}}} {\bm{\psi}}_{\lambda} \right).
\end{equation}
In what follows, and at the risk of some abuse of notation, when referring to a given linkage class, we will drop subscript $\lambda$, and re-write (\ref{eq:Fexplicit_first}) as:
\begin{equation}\label{eq:General_fN}
\textnormal{\bf{f}} ({x}; \bm{\chi})  = {\textnormal{\bf{f}}}^{*} +
x \textnormal{\bf{h}} (\bm{\chi}).
\end{equation}
If in addition, the discussion concerns a particular vector $\bm{\chi} \in {\mathbf{U}}$, the following simplified expression for Eqn (\ref{eq:General_fN}) will be employed:
\begin{equation}\label{eq:Fexplicit}
{\textbf{f}}(x)   =  {\textbf{f}}^{*} +  x {\textbf{h}}.
\end{equation}

\begin{lemma}\label{eq:Lema_SignfStar}
If the linkage class is weakly reversible, ${\mathbf{f}}^{*}$ in (\ref{eq:Fexplicit}) is strictly positive. If the linkage class is irreversible, ${\mathbf{f}}^{*}$ will be non-negative with zero entries that correspond to the non-terminal complexes.
\end{lemma}

\textbf{Proof:}
If the linkage class is weakly reversible, every complex in the linkage class can be reached from the reference (equivalently, from the complexes associated to positive entries ${\textbf{a}}$ in (\ref{eq:structM_k_l})). Thus  from Lemma \ref{prop:E_Metzler},
${\textbf{f}}^{*} =  -E^{-1}
{\textbf{a}} > 0$.

If the linkage class is irreversible, let  ${\textbf{a}}^T = [{\textbf{a}}^T_{{\textnormal{p}}} ~ {\textbf{a}}^T_{\textnormal{q}}]$ where sub-indexes ${\textnormal{p}}$ and ${\textnormal{q}}$ denote the terminal and non-terminal complexes of the linkage class. Since the reference belongs to the strong terminal linkage class, positive components of vector ${\textbf{a}}$ only enter terminal complexes so that ${\textbf{a}}^T_{\textnormal{q}} = {\textbf{0}}_{\textnormal{q}}$.  Using the inverse (\ref{eq:structN}) from Lemma \ref{prop:E_Metzler},  ${\textbf{f}}^{*} =  -N {\textbf{a}}$ can be written as:
\[
{\textbf{f}}^{*} =
\left[
\begin{array}{c}
  {\textbf{f}}^{*}_{{\textnormal{p}}}\\
{\textbf{f}}^{*}_{{\textnormal{q}}}
\end{array}\right]  =
\left[
\begin{array}{c}
  -N_{\textnormal{p}} {\textbf{a}}_{\textnormal{p}}\\
  {\textbf{0}}_{\textnormal{q}}
\end{array}
\right],
\]
where ${\textbf{f}}^{*}_{{\textnormal{p}}}$ is strictly positive because, by Lemma \ref{prop:E_Metzler}, $N_{\textnormal{p}}$ is strictly negative. Thus ${\textbf{f}}^{*} \geq 0$, with zero entries ${\textbf{f}}^{*}_{{\textnormal{q}}}$ that correspond to the non-terminal complexes.
\hfill $\Box$

Next we present some conditions that ensure positivity of vector functions (\ref{eq:Fexplicit_first}) and the corresponding family of solutions (Definition \ref{def:FamilySolutions}).

\subsection{Positivity conditions for the family of solutions}

At this point, it should be clear that positivity of the family of solutions  in Definition  \ref{def:FamilySolutions} is a necessary condition for positive equilibrium in the concentration space. Next, we discuss how such condition relates to the structure of the network and give some indications on how to construct the domain where (\ref{eq:fNetwork}) remains positive.
\begin{proposition}\label{prop:Family_Solutions}
If a given linkage class is weakly reversible, then for every $\bm{\chi} \in {\mathbf{U}}$ there exists an interval ${\mathbb{X}} (\bm{\chi}) \subset
\mathbb{R}$ (which includes the zero) where the vector function (\ref{eq:General_fN}) remains positive.\\
If the linkage class is irreversible, and there exists a positive vector function (\ref{eq:General_fN}) on ${\mathbb{X}}(\bm{\chi})$, such interval cannot contain the zero.
\end{proposition}
\textbf{Proof:}
%If the linkage class is weakly reversible, every node in the linkage class can be reached from the reference and so the nodes associated to positive entries ${\textbf{a}}$ in (\ref{eq:structM_k_l}). Thus  from Lemma \ref{prop:E_Metzler},
%${\textbf{f}}^{*} \equiv  -E^{-1}
%{\textbf{a}} > 0$.
%
If the linkage class is weakly reversible, by Lemma \ref{eq:Lema_SignfStar} we have that ${\textbf{f}}^{*} > 0$, so the values of the scalar $x$ for
which  ${\textbf{f}}(x)$ in (\ref{eq:Fexplicit}) remains strictly positive will depend on the signs of the entries in ${\textbf{h}}$. For every such entry $i$, define $p_i = h_i/
{\textnormal{f}}^{*}_i$ and introduce two index sets ${\cal{I}}^{+}$
and ${\cal{I}}^{-}$ so that:
\begin{equation}\label{eq:IndexSetS}
\begin{array}{l}
 i \in {\cal{I}}^{+}, ~~ \textnormal{if} ~~ h_i >0, ~~\textnormal{thus} ~~ p_i >0  \\
 i \in {\cal{I}}^{-}, ~~ \textnormal{if} ~~ h_i <0, ~~\textnormal{thus} ~~ p_i <0.  \\
\end{array}
\end{equation}
It is straightforward to see that ${\textbf{f}}(x)$ will be strictly
positive for every $x$ in the open interval:
\begin{equation}\label{eq:domain_LL}
{\mathbb{X}} =
(L^{-},L^{+}), ~ \textnormal{with} ~ L^{-} = {\textnormal{max}}_{i \in
{\cal{I}}^{+}} \{ - 1 / p_{i} \} ~ \textnormal{and,} ~ L^{+} = {\textnormal{min}}_{i \in
{\cal{I}}^{-}} \{ - 1/ p_{i} \}.
\end{equation}
Note that $L^{-} = - \infty$ (respectively, $L^{+} = +
\infty$) provided that  ${\cal{I}}^{+} = \emptyset$ (respectively,
${\cal{I}}^{-} = \emptyset$). In any case, because ${\textbf{f}}^{*} > 0$, the interval ${\mathbb{X}}$ includes the zero.
If the linkage class is irreversible, by using Lemma \ref{eq:Lema_SignfStar}, ${\textbf{f}}(x)$ in (\ref{eq:Fexplicit}) can be written as:
\begin{equation}\label{eq:fS_factor}
\begin{array}{l}
{\textbf{f}}_{\textnormal{p}}(x)   = {\textbf{f}}^{*}_{
{\textnormal{p}}} + x {\textbf{h}}_{\textnormal{p}}\\
{\textbf{f}}_{\textnormal{q}}(x)   =  {\textbf{0}}_{\textnormal{q}} + x
{\textbf{h}}_{\textnormal{q}},
\end{array}
\end{equation}
where sub-indexes ${\textnormal{p}}$ and ${\textnormal{q}}$ denote the terminal and non-terminal nodes of the linkage class. Since ${\textbf{f}}^{*}_{{\textnormal{p}}}$ is strictly positive,
there exists some domain ${\mathbb{X}}_{\textnormal{p}}$ that
includes the zero, for which ${\textbf{f}}_{\textnormal{p}}(x) >
0$. Let ${\mathbb{X}}_{\textnormal{p}} =
{\mathbb{X}}_{\textnormal{p}}^{-} \cup
{\mathbb{X}}_{\textnormal{p}}^{+} \cup \{0 \} $, where
${\mathbb{X}}_{\textnormal{p}}^{-}$ and
${\mathbb{X}}_{\textnormal{p}}^{+}$ are the intervals containing the negative and
positive values, respectively. It is then straightforward to see
from (\ref{eq:fS_factor}) that in order for
${\textbf{f}}_{\textnormal{q}} (x) > 0$,
${\textbf{h}}_{\textnormal{q}}$ must have a definite sign (i.e. all
components either positive or negative). If this is the case, i.e.
if ${\textbf{h}}_{\textnormal{q}} > 0$ (respectively, $< 0$),
we can always find some $x \in {\mathbb{X}}_{\textnormal{p}}^{+}$
 (respectively, $x \in {\mathbb{X}}_{\textnormal{p}}^{-}$), so
 that ${\textbf{f}} (x) > 0$. Otherwise, no positive solution exists. Because ${\mathbb{X}}_{\textnormal{p}}^{+}$
 (respectively, ${\mathbb{X}}_{\textnormal{p}}^{-}$) does not contain the zero, if the linkage class is irreversible, the interval ${\mathbb{X}} (\chi)$ does not contain the zero.\hfill $\Box$
\begin{proposition}\label{prop:PositiveFamily}
Let ${\mathbf{f}}_{\bm{\eta}}: {\mathbb{D}}_{0} \times {\mathbf{U}} \rightarrow {\mathbb{R}}^{(n -\ell)}$ (with ${\mathbb{D}}_{0}  = \{ 0 \} \cup {\mathbb{R}}^{\ell}_{>0} \cup {\mathbb{R}}^{\ell}_{<0}$) be the family of solutions as given in Definition \ref{def:FamilySolutions}. In addition, for each $\lambda = 1, \cdots, \ell$ and $\bm{\chi} \in {\mathbf{U}}$, let ${\mathbb{X}}_{\lambda} (\bm{\chi}) \subset
\mathbb{R}$ be the interval such that
${\mathbf{f}}_{\lambda} ({x}_{\lambda}; \bm{\chi}) > 0$ for every ${x}_{\lambda} \in {\mathbb{X}}_{\lambda} (\bm{\chi})$, and ${\mathbb{X}}_{\bm{\eta}} (\bm{\chi}) = {\mathbb{X}}_1 (\bm{\chi}) \times \cdots {\mathbb{X}}_{\lambda} (\bm{\chi}) \times \cdots {\mathbb{X}}_{\ell} (\bm{\chi})$ an open $\ell$-dimensional domain.
 Then:

If the network is weakly reversible, for every $\bm{\chi} \in {\mathbf{U}}$ there exists a domain ${\mathbb{D}}_{\bm{\eta}} (\bm{\chi}) = {\mathbb{X}}_{\bm{\eta}} (\bm{\chi})\cap {\mathbb{D}}_{0}$ that contains the zero, such that ${\mathbf{f}}_{\bm{\eta}}(\mathbf{x}; \bm{\chi}) > 0$ for every $\mathbf{x} \in {\mathbb{D}}_{\bm{\eta}} (\bm{\chi})$.

If the network is irreversible and the domain
${\mathbb{D}}_{\bm{\eta}} (\bm{\chi}) = {\mathbb{X}}_{\bm{\eta}} (\bm{\chi})\cap {\mathbb{D}}_{0}$ is non-empty, then ${\mathbf{f}}_{\bm{\eta}}(\mathbf{x}; \bm{\chi}) > 0$ for every $\mathbf{x} \in {\mathbb{D}}_{\bm{\eta}} (\bm{\chi})$. Such domain does not contain the zero.
\end{proposition}
\textbf{Proof:} If the network is weakly reversible, Proposition \ref{prop:Family_Solutions} ensures that for every $\lambda = 1, \cdots, \ell$ and $\bm{\chi} \in {\mathbf{U}}$, ${\mathbf{f}}_{\lambda} ({x}_{\lambda}; \bm{\chi}) > 0$ for every  ${x}_{\lambda} \in {\mathbb{X}}_{\lambda} (\bm{\chi})$, where the interval contains the zero. In consequence, a non-empty domain ${\mathbb{D}}_{\bm{\eta}} (\bm{\chi}) = {\mathbb{X}}_{\bm{\eta}} (\bm{\chi})\cap {\mathbb{D}}_{0}$ exists, what proves the first assertion.

If the network is irreversible and ${\mathbb{D}}_{\bm{\eta}} (\bm{\chi})$ is non-empty, the second assertion holds. Because at least one linkage class is irreversible, it follows from Proposition \ref{prop:Family_Solutions} that ${\mathbb{X}}_{\bm{\eta}} (\bm{\chi})$, and therefore ${\mathbb{D}}_{\bm{\eta}} (\bm{\chi})$, does not contain the zero.
Finally, note that for irreversible networks it may well happen that ${\mathbb{D}}_{\bm{\eta}} (\bm{\chi})$, as defined above, is empty so no positive family of solutions exists.
\hfill $\Box$

%\textbf{Remark:}
%\hfill $\triangle$

\subsection{The set of feasible (equilibrium) solutions}

Not all positive elements (vectors) that are part of the family of solutions (\ref{eq:fNetwork}) will necessarily comply with condition
(\ref{eq:vectorfuncfL}) but only a particular subset we will refer
to as the set of \emph{feasible solutions}, that we formally define next.  Using (\ref{eq:lnpsi}) for every complex $i\in\mathcal{L}_{\lambda}$, we have that $\ln
({\psi}_{i} (\textbf{c})/{\psi}_{j_{\lambda}} (\textbf{c})) = (\bm{y}_i -
\bm{y}_{j_{\lambda}})^{T} \ln \textbf{c}$.  Hence, the logarithm at
the right hand side of (\ref{eq:vectorfuncfL}) can be expressed as:
\begin{equation}\label{eq:feasib_by_LC}
\ln \frac{1}{{\psi}_{j_{\lambda}}(\textbf{c})} {\bm{\psi}}_{\lambda}(\textbf{c}) =
S_{\lambda}^{T} \ln \textbf{c},
\end{equation}
so that $\ln {\textbf{f}}_{\lambda} (\bm{\xi}) = S_{\lambda}^{T} \bm{\xi}$ for every $\lambda = 1, ..., \ell$, and $\bm{\xi} \in {\mathbb{R}}^{m}$ ($\bm{\xi} \equiv \ln \textbf{c}$).
\begin{definition}\label{def:SetFeasibleSol}
{\bf (The set of feasible solutions)}
For a given $\bm{\chi} \in {\mathbf{U}}$, let us assume that there exists a non-empty domain ${\mathbb{D}}_{\bm{\eta}} (\bm{\chi}) \subset {\mathbb{D}}_{0}$ such that for every $\mathbf{x} \in {\mathbb{D}}_{\bm{\eta}}$,  ${\mathbf{f}}_{\bm{\eta}}(\mathbf{x}; \bm{\chi}) > 0$. We say that ${\mathbf{f}}_{\bm{\eta}}(\mathbf{x}; \bm{\chi}) > 0$ is a feasible solution if:
\begin{equation}\label{eq:lnfLvsC}
\ln {\mathbf{f}}_{\bm{\eta}}(\mathbf{x}; \bm{\chi}) \in  \textnormal{Im}(S^{T}).
\end{equation}
The set of vectors ${\mathbf{f}}_{\bm{\eta}}(\mathbf{x}; \bm{\chi})$, with $\bm{\chi} \in {\mathbf{U}}$ and $\mathbf{x} \in {\mathbb{D}}_{\bm{\eta}} (\bm{\chi})$ that satisfy (\ref{eq:lnfLvsC}), constitutes the set of feasible solutions.
\end{definition}
Note that (\ref{eq:lnfLvsC}) implies that there exists some $\bm{\xi} \in {\mathbb{R}}^{m}$ such that $\ln {\mathbf{f}}_{\bm{\eta}}(\mathbf{x}; \bm{\chi}) = S^{T} \bm{\xi}$. In this way, each element of the set of feasible solutions relates to a set of equilibrium concentrations of the form $\textbf{c} = \exp(\bm{\xi})$ for system (\ref{eq:dotC_Sl}). The following result gives some conditions to identify the set of feasible equilibrium solutions:
\begin{lemma}\label{def:Feas_sol_gen}
{\bf (Feasibility conditions)}
Every element ${\mathbf{f}}_{\bm{\eta}} (\mathbf{x}; \bm{\chi})$ of the set of feasible solutions satisfies that:
\begin{equation}\label{eq:EfX_formal}
({\mathbf{g}}^{r})^{T} \ln
{\mathbf{f}}_{\bm{\eta}} (\mathbf{x}; \bm{\chi}) = 0, ~~\textnormal{for} ~~ r = 1,
..., \delta.
\end{equation}
\end{lemma}
\textbf{Proof:}
By Definition \ref{def:SetFeasibleSol}, every element of the set of feasible solutions satisfies Eqn (\ref{eq:lnfLvsC}). This implies that $\ln {\mathbf{f}}_{\bm{\eta}} (\mathbf{x}; \bm{\chi})$ is in the range of $S^T$, which in turn is orthogonal to the kernel of $S$. Since $\{{\textbf{g}}^{r} ~|~ r=1, \ldots, \delta \}$ is a basis for the kernel, expressions  in (\ref{eq:EfX_formal}) follow. \hfill $\Box$
\begin{definition}\label{def:Feasibility_Functions}
 {\bf (Feasibility function)} Let ${\textnormal{F}}: \mathbb{X}  \times
{\mathbf{U}} \rightarrow
\mathbb{R}$ (with $\mathbb{X} \subset \mathbb{R}$) be defined as:
\begin{equation}\label{eq:General_FgfN}
{\textnormal{F}}(x; \bm{\chi}) =
{\textnormal{\bf{g}}}^{T}({\bm{\chi}}) \ln \mathbf{f} (x;
\bm{\chi}),
\end{equation}
where  $\bm{\chi} \in {\mathbf{U}}$ and $\mathbb{X} (\bm{\chi})$ is the interval in which
$\mathbf{f} (x;
\bm{\chi})$, of the form (\ref{eq:General_fN}), remains positive
\end{definition}

We make use of the above definition to present an immediate consequence of Lemma \ref{def:Feas_sol_gen}.
\begin{proposition}\label{Corlry:Feas_sol_gen}
For every element ${\mathbf{f}}_{\bm{\eta}} (\mathbf{x}; \bm{\chi})$ of the set of feasible solutions, the following relation holds:
\begin{equation}\label{eq:Gen_ort_STSlnf_SUM}
\sum_{\lambda} {\textnormal{F}}_{\lambda} (x_{\lambda};
{\bm{\chi}}) = 0, ~~\textnormal{where} ~~
{\textnormal{F}}_{\lambda} (x_{\lambda}; {\bm{\chi}}) =
{\textbf{g}}_{\lambda}^{T} (\bm{\chi}) \ln {\mathbf{f}}_{\lambda}
(x_{\lambda}; {\bm{\chi}}).
\end{equation}
\end{proposition}
\textbf{Proof:}
Pre-multiplying each equality in (\ref{eq:EfX_formal}) (Lemma \ref{def:Feas_sol_gen}) by the corresponding coordinate ${\chi}_r$, taking the summation to $\delta$ and expanding over linkage classes, we get:
\[
\sum_{r} {\chi}_r ({\textbf{g}}^{r})^{T} \ln
{\textbf{f}}_{\eta} (\textbf{x}; \bm{\chi}) = 0, ~~\textnormal{and} ~~
\sum_{\lambda} {\textnormal{\bf{g}}}_{\lambda}^{T}({\bm{\chi}}) \ln {\textnormal{\bf{f}}}_{\lambda} (x_{\lambda};
\bm{\chi}) = 0.
\]
The result then follows by using Definition \ref{def:Feasibility_Functions} to re-write the above expression as in (\ref{eq:Gen_ort_STSlnf_SUM}).
\hfill $\Box$

As we will see in the next sections, feasibility functions
${\textnormal{F}}_{\lambda} (x; \bm{\chi})$ will prove to be
fundamental to characterize the structure of equilibrium, allowing in some instances to conclude uniqueness of equilibrium in each positive stoichiometric compatibility class.

\subsection{Example: feasibility for a one linkage class weakly reversible network}
\label{sec:The4RevNetwork}
\begin{figure}[ht]
\begin{center}
$\begin{array}{@{}ccc@{}}
\mathrm{(a)} & \qquad & \mathrm{(b)} \\
\includegraphics[scale=0.50]{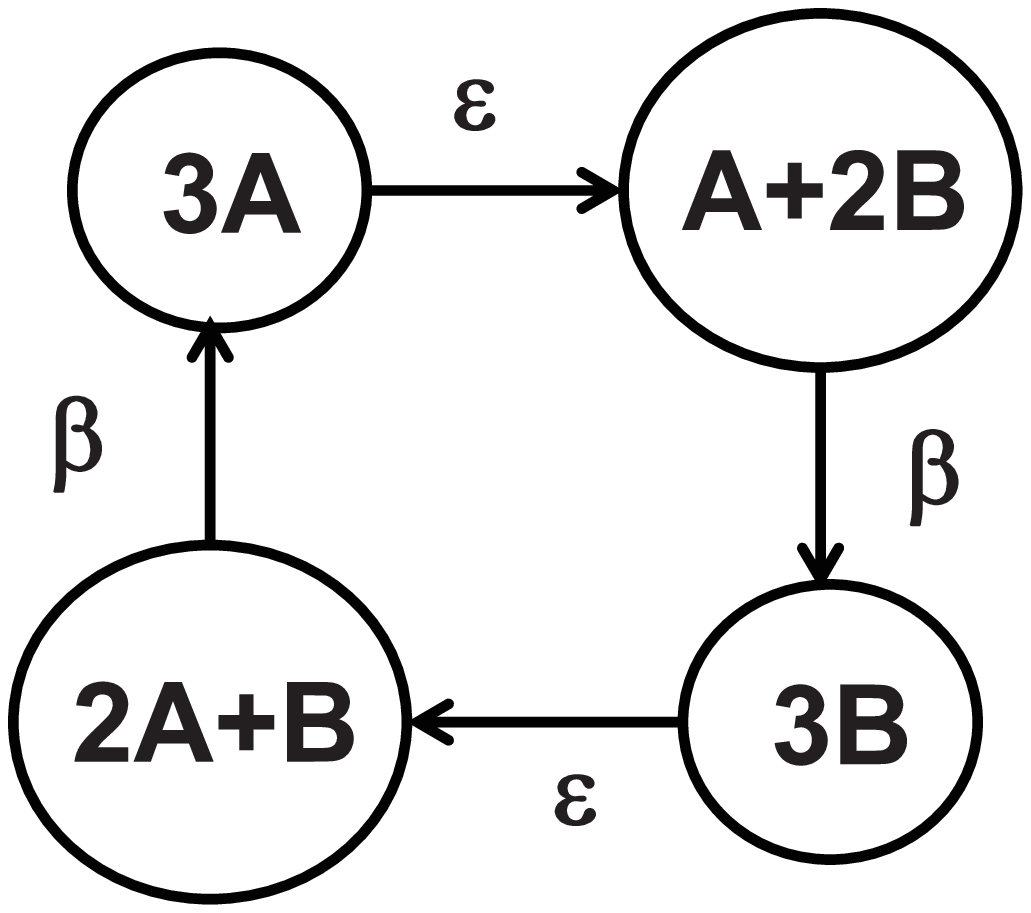} &
\qquad &
\includegraphics[scale=0.50]{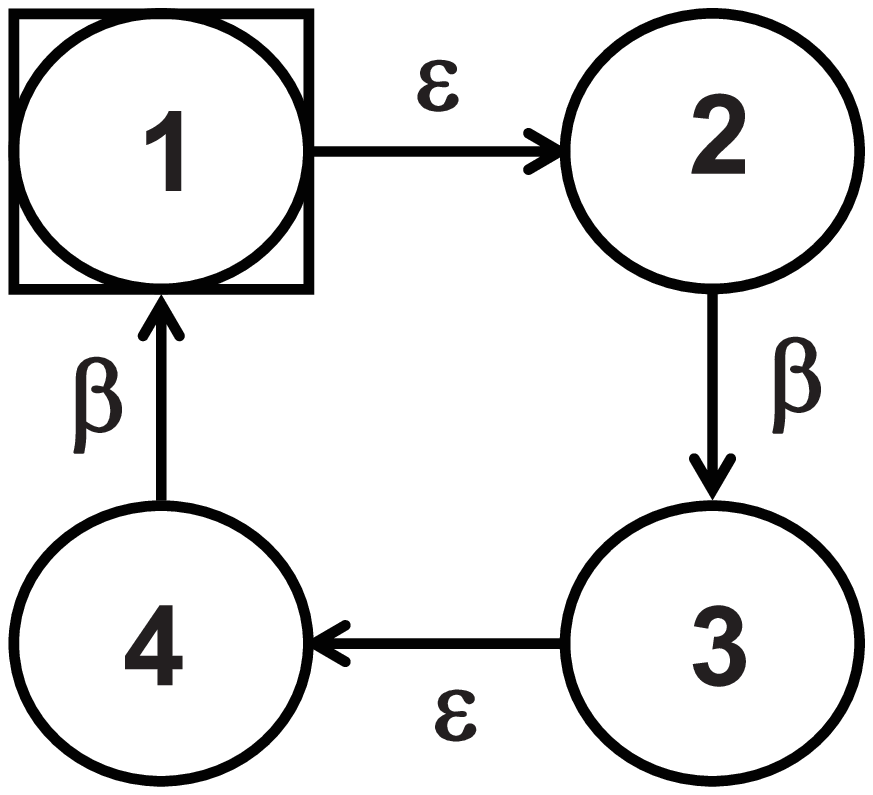}
\end{array}$
\end{center}
\caption{A weakly reversible one-linkage class network
with reaction constants $k_{12} = k_{34} = \epsilon$ and $k_{23} =
k_{41} = \beta$. (a) The network with explicit indication of the
chemical species. (b) The corresponding graph of complexes with the
reference complex in the squared box.} \label{fig:UNSTOneLCN_WWR}
\end{figure}
\begin{figure}[ht]
\begin{center}
$\begin{array}{@{}cc@{}}
\mathrm{(a)} & \mathrm{(b)} \\
\includegraphics[scale=0.33]{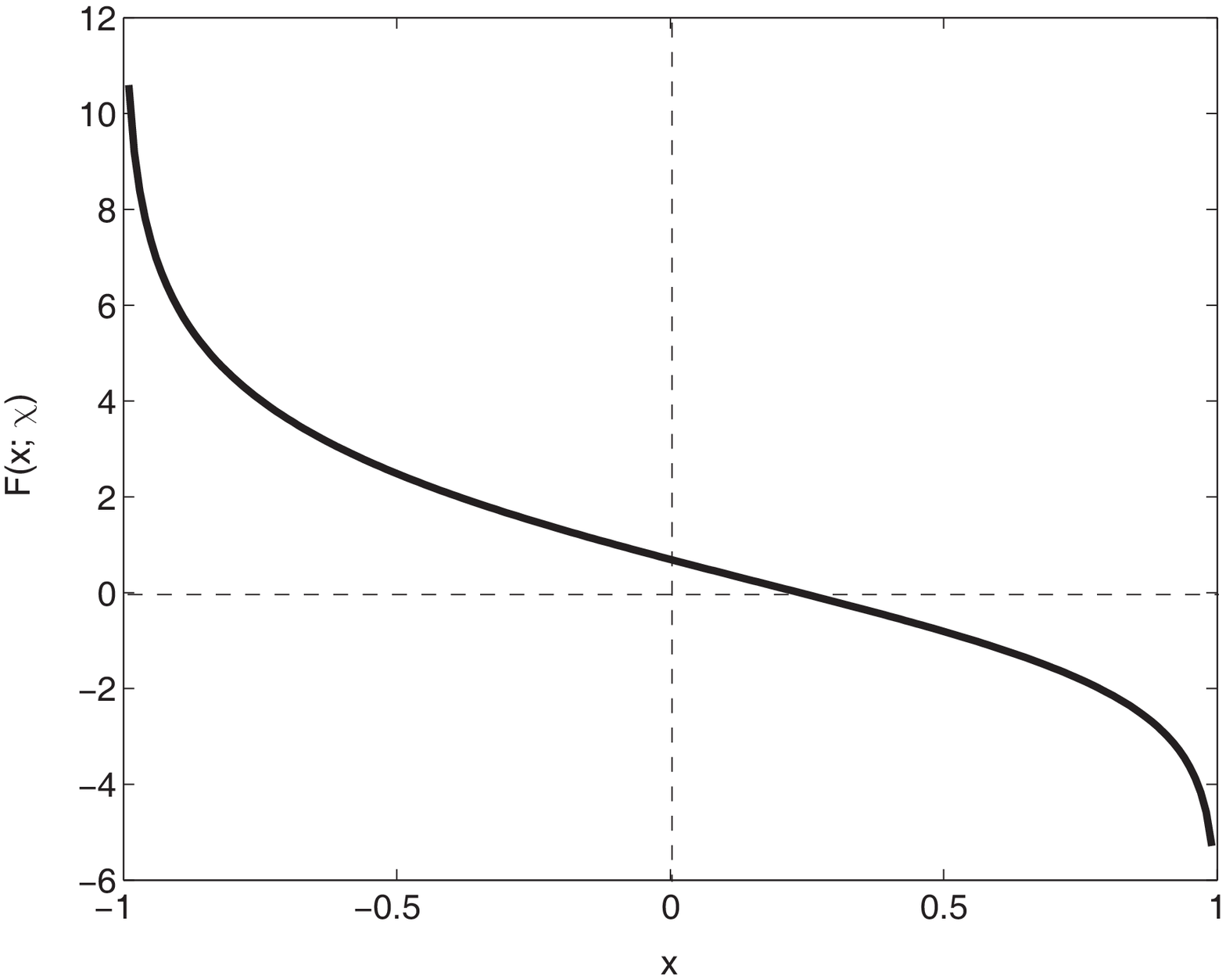} &
\includegraphics[scale=0.33]{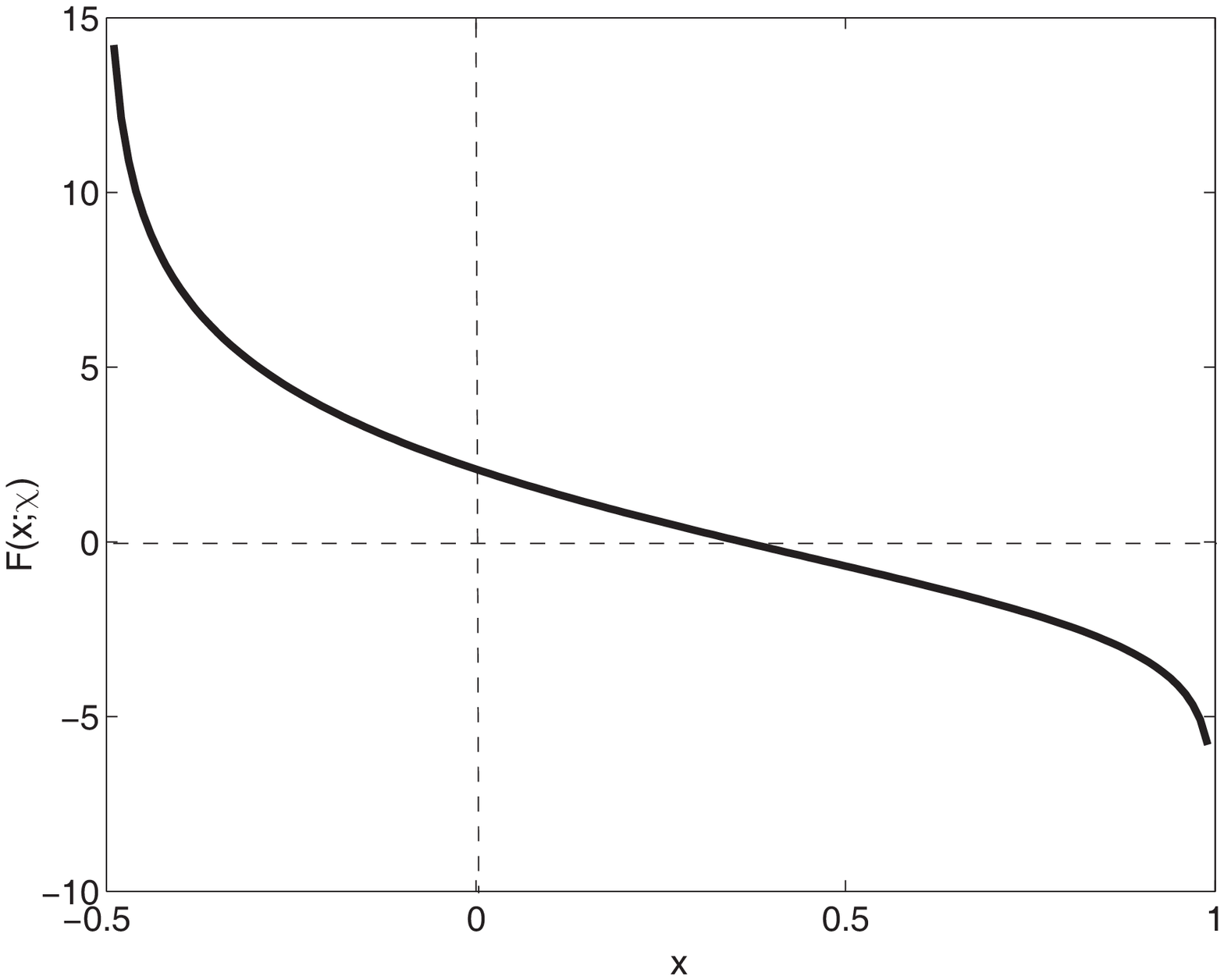} \\
\mathrm{(c)} & \mathrm{(d)} \\
\includegraphics[scale=0.33]{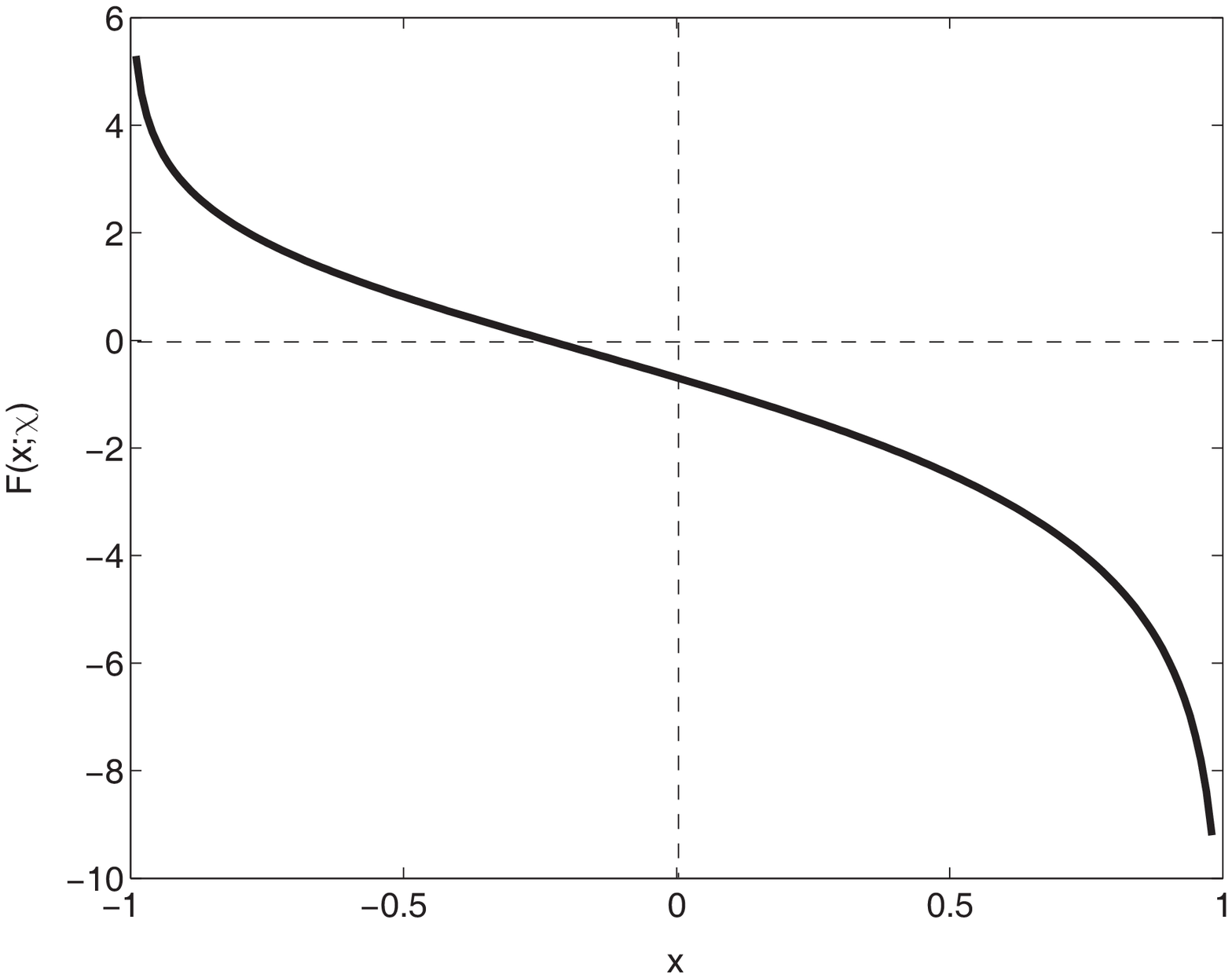} &
\includegraphics[scale=0.33]{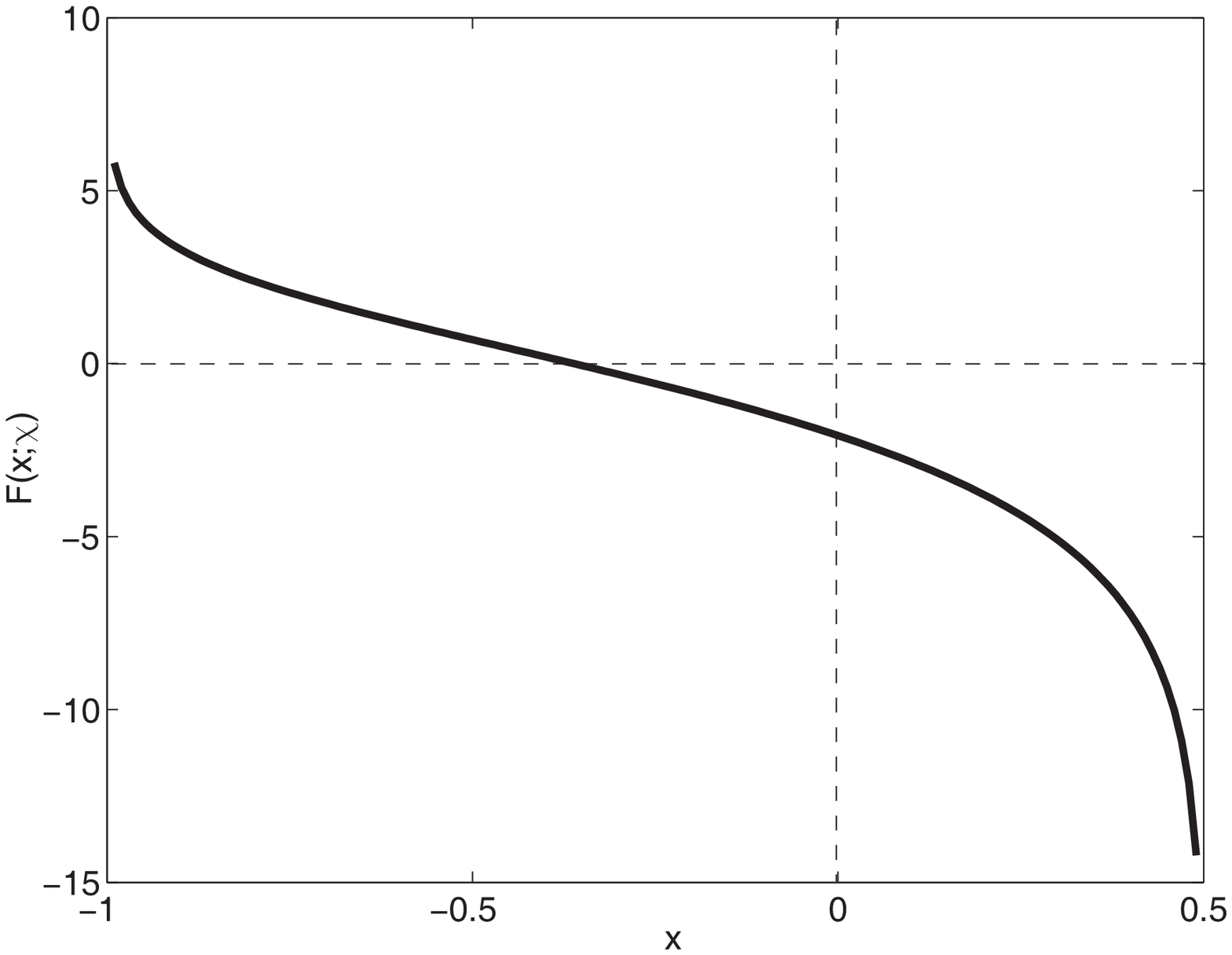} \\
\end{array}$
\end{center}
\caption{Feasibility functions $F(x;{\chi})$ associated to the
network presented in Figure \ref{fig:UNSTOneLCN_WWR} for different
vectors $\bm{\chi}$ in the unit sphere. (a) ${\bm{\chi}} = (1 ~  0)^T$, (b) ${\bm{\chi}} = (0 ~
1)^T$, (c) ${\bm{\chi}} = (-1 ~  0)^T$, (d) ${\bm{\chi}} = (0 ~  -1)^T$. Note that since vectors in plots (a)-(c), and (b)-(d) relate as ${\bm{\chi}}' = -{\bm{\chi}}$, the corresponding functions relate as $\textnormal{F} (x;{\bm{\chi}}) = -\textnormal{F} (-x;{\bm{\chi}}')$.}
\label{fig:Feasib_x_Chi}
\end{figure}

Let us consider the weakly reversible reaction network taken from \cite{Feinberg:87} and presented in Figure
\ref{fig:UNSTOneLCN_WWR}.  Matrix (\ref{eq:structM_k_l}) for this network takes the
form:
\begin{equation}
M = \left[ \begin{array}{rrrr}
    -k_{12} & 0 & 0 & k_{41}\\
    k_{12} & -k_{23} & 0 & 0 \\
    0 & k_{23} & -k_{34} & 0 \\
    0 & 0 & k_{34} & -k_{41} \\
\end{array} \right].
\end{equation}
Substituting the reaction constants given in Figure
\ref{fig:UNSTOneLCN_WWR}, leads to the following sub-matrices
in $M$:
\begin{equation}
E = \left[ \begin{array}{rrr}
    -\beta & 0 & 0\\
     \beta & -\epsilon & 0 \\
     0 & \epsilon & -\beta \\
\end{array} \right],  ~~ {\textbf{a}} = \left[ \begin{array}{r}
    \epsilon \\
    0 \\
    0 \\
\end{array} \right], ~~ {\textbf{b}}  = \left[ \begin{array}{r}
    0\\
    0 \\
    \beta \\
\end{array} \right].
\end{equation}
For this example, matrix $S$ and a basis for its kernel,
expressed as columns of matrix $G$ read:
\[
S =\left[\begin{array}{rrr}
        -2 &  -3 & -1 \\
         2 &   3 &  1 \\
     \end{array} \right] ,  ~~ G = \left[ \begin{array}{rrr}
    1 & 0 & -2 \\
    0 & 1 & -3 \\
\end{array} \right]^T.
\]
In order to compute the feasibility function (\ref{eq:General_FgfN}), we have that:
\begin{equation}
{\textbf{f}}^{*} = -E^{-1} \textbf{a} = \left(
\begin{array}{rrr} \epsilon / \beta & 1 & \epsilon / \beta
\end{array} \right)^T
,  ~~ E^{-1} = \left[ \begin{array}{rrr}
    -\frac{1}{\beta} & 0 & 0\\
   -\frac{1}{\epsilon} & -\frac{1}{\epsilon} & 0 \\
   -\frac{1}{\beta} & -\frac{1}{\beta} & -\frac{1}{\beta} \\
\end{array} \right],
\end{equation}
and
\[
\textbf{h} (\bm{\chi}) = E^{-1} \left[ {\chi}_1 \left(
\begin{array}{r} 1 \\ 0 \\ -2 \end{array} \right) + {\chi}_2 \left(
\begin{array}{r} 0 \\ 1 \\ -3 \end{array} \right) \right] =
{\chi}_1 \left( \begin{array}{r} - 1/ \beta \\ -1/ \epsilon \\ 1/
\beta \end{array} \right) + {\chi}_2 \left( \begin{array}{r} 0 \\ -1
/ \epsilon \\ 2/\beta \end{array} \right).
\]
For parameters $\beta =
2$ and $\epsilon = 1$ we have that:
\[
{\textbf{f}}^{*} = \left( \begin{array}{r} 1/ 2
\\ 1  \\ 1/ 2 \end{array} \right)
~~ \textnormal{and}  ~~ \textbf{h} (\bm{\chi}) = {\chi}_1 \left(
\begin{array}{r} - 1/ 2
\\ -1  \\ 1/ 2 \end{array} \right) + {\chi}_2 \left(
\begin{array}{r} 0 \\ -1  \\  1 \end{array} \right)
\]
For ${\bm{\chi}}_1 = (1 ~ 0)^T$, the entries of the vector function
${\textbf{f}} (x; {\bm{\chi}}_1)$ will remain all positive as long
as the values taken by $x$ will lie in the open interval $(-1, +1)$.
Hence, its domain ${\mathbb{X}} ({\bm{\chi}}_1) = (-1, +1)$. In
the same way, for ${\bm{\chi}}_2 = (0 ~ 1)^T$, ${\mathbb{X}}
({\bm{\chi}}_2) = (-1/2, +1)$. Because the logarithm is defined
for positive values, both domains ${\mathbb{X}} ({\bm{\chi}}_1)$ and
${\mathbb{X}} ({\bm{\chi}}_2)$ coincide also with the domains for
$\textnormal{F} (x;{\bm{\chi}}_1)$ and $\textnormal{F}
(x;{\bm{\chi}}_2)$. The explicit expressions become:
\[
\textnormal{F} (x;{\bm{\chi}}_1) = \ln \frac{2 (1-x)}{(1+x)^2}, ~~~~
\textnormal{and}  ~~~~ \textnormal{F} (x;{\bm{\chi}}_2) = \ln
\frac{8 (1-x)}{(1+2x)^3}.
\]
Feasibility functions for some vectors $\bm{\chi}$ in the unit sphere are presented in Figure \ref{fig:Feasib_x_Chi}. It must be
observed that each vector $\bm{\chi}$ leads to a different domain ${\mathbb{X}}
(\bm{\chi})$. Finally note that since ${\textbf{f}}^{*}$ is strictly
positive, all possible domains will include the zero. \hfill
$\triangle$

It must be noted that the functions depicted in the above example
are monotonous decreasing in their respective domains. Remarkably,
this will be the case for any feasibility function $\textnormal{F}
(x;{\bm{\chi}})$, despite network structure or stoichiometry. Next section provides a formal proof of this
fact which will turn out to be central in exploring the nature of
equilibrium solutions.

\section{Monotonicity of feasibility functions}
\label{sec:Domain_properties_Feasible}

Here, we study the properties of function (\ref{eq:General_FgfN}), presented in Definition \ref{def:Feasibility_Functions}. In particular, it will be shown that it is monotonous decreasing in its domain $\mathbb{X} (\bm{\chi})$ and crosses the $x$-axis.

Let us consider a weakly reversible linkage class and a given vector $\bm{\chi} \in \mathbf{U}$. Without loss of generality, assume that the  $N-1$ components of the vector ${\textbf{h}}$ in (\ref{eq:Fexplicit}), being $m$ of them positive, $r$ zero and the remaining negative, are ordered so
that:
\begin{equation}\label{eq:Orderh}
\begin{array}{l}
  h_1 \geq \cdots \geq h_k \geq \cdots \geq h_m > 0 > h_{m+r+1} \geq
\cdots \geq h_{\ell} \geq \cdots  \geq h_{N-1},
\\
h_{m+1} = \cdots = h_{m + r} = 0.
\end{array}
\end{equation}
Note that such order can always be induced by a suitable row and
column permutation in equations ${\textbf{g}} = E{\textbf{h}}$ and
$E {\textbf{f}}^{*} = -{\textbf{a}}$.
%Making use of the index sets
%(\ref{eq:IndexSetS}) to define the interval ${\mathbb{X}} = (L^{-},L^{+})$ in (\ref{eq:domain_LL}), we have that ${\cal{I}}^{+} = \{1, \cdots, m
%\}$, ${\cal{I}}^{-} = \{m+r+1, \cdots, N -1 \}$.
%
In order to simplify notation, let us re-write function (\ref{eq:General_FgfN}) for a fixed ${\bm{\chi}}$ as:
\begin{equation}\label{eq:Fx_decrecente}
 {\textnormal{F}}(x)   = {\textbf{g}}^T \ln {\textbf{f}}(x).
\end{equation}
The main result on monotonicity is presented in the
theorem below.
\begin{theorem}\label{lem:Lemma_on_F}
Let ${\mathbb{X}} \subset
\mathbb{R}$ and consider the function ${\textnormal{F}}(x): {\mathbb{X}} \mapsto \mathbb{R}$ defined in (\ref{eq:Fx_decrecente}).
${\textnormal{F}}(x) $ is monotonous decreasing in the interval
${\mathbb{X}}$, defined as in (\ref{eq:domain_LL}). Moreover,
\begin{equation}\label{eq:RightLeftLimFx}
\lim_{x^+ \rightarrow L^{-}} {\textnormal{F}}(x) = + \infty ~~\textnormal{and} ~~
\lim_{x^- \rightarrow L^{+}} {\textnormal{F}}(x) = - \infty.
\end{equation}
\end{theorem}
\textbf{Proof:} Function (\ref{eq:Fx_decrecente}) is continuous differentiable in
the the interval ${\mathbb{X}}$ since ${\textbf{f}}(x)$ is strictly
positive (see Proposition \ref{prop:Family_Solutions}). Thus, the first part of the proof reduces to computing the
first derivative with respect to $x$ and studying its
sign in the interval $\mathbb{X}$. For every entry $i$ of
vectors ${\textbf{h}}$ and ${\textbf{f}}^{*}$,  let us define $p_i = h_i
/{\textnormal{f}}^{*}_i$ as in the proof of Proposition \ref{prop:Family_Solutions}, and re-write ${\textbf{h}}$ as:
\begin{equation}\label{eq:hDp}
{\textbf{h}} = {\cal{D}} ({\textbf{f}}^{*}) {\textbf{p}},
\end{equation}
where vector $\textbf{p} \in {\mathbb{R}}^{N-1}$ includes the elements $p_i$  and ${\cal{D}} ({\textbf{f}}^{*})$ represents a diagonal matrix with the components of
${\textbf{f}}^{*}$ in the diagonal. Let us also re-order the $p_i$
elements so that:
\begin{equation}\label{eq:orderpj}
\begin{array}{l}
  p_1 \geq \cdots \geq p_k \geq \cdots \geq p_m > 0 > p_{m+r+1} \geq
\cdots \geq p_{\ell} \geq \cdots  \geq p_{N-1},
\\
p_{m+1} = \cdots = p_{m + r} = 0.
\end{array}
\end{equation}
Note that the number of positive, negative and zero elements must
coincide with those in (\ref{eq:Orderh}), although not necessarily  in
the same order. Define functions $Q_i (x):
{\mathbb{X}} \mapsto {\mathbb{R}}$ as:
\begin{equation}\label{eq:the function QjX}
Q_i (x) = \frac{p_i}{1 + x p_i}.
\end{equation}
For every $k$ such that $p_k >0$ and $x \in {\mathbb{X}}$, we have that $x > -(1/p_k)$, since from (\ref{eq:domain_LL}):
\[
x > {\textnormal{max}}_{i \in
{\cal{I}}^{+}} \{ - 1 / p_{i} \} (= -1/p_{1}).
\]
In turns, this implies that $x + (1/p_k)
> 0$, and:
\[
Q_k (x) = \frac{1}{x + (1/p_k)}
>0.
\]
Using the same argument for the negative elements, we have that $x < -1/p_{\ell}$ so that $Q_{\ell}(x) <0$ for any $\ell=m+r+1,\dots, N-1$. In
addition, for any $p_i \geq p_j$ (both,  either positive or negative), we have
that $Q_i(x) \geq Q_j(x)$. Consequently, from (\ref{eq:orderpj}),
for every $x \in {\mathbb{X}}$ we have that:
\begin{equation}\label{eq:orderQxSS}
\begin{array}{l}
  Q_1(x) \geq \cdots \geq Q_k(x) \geq \cdots \geq Q_m (x) > 0 > Q_{m+r+1}(x) \geq
\cdots \geq Q_{\ell} (x) \geq \cdots  \geq Q_{N-1} (x),
\\
Q_{m+1} (x) = \cdots = Q_{m + r} (x) = 0.
\end{array}
\end{equation}
Keeping the order established in (\ref{eq:orderpj}), the first derivative can be written as:
\begin{equation}\label{eq:DerF_1Ori}
{\textnormal{F}'}(x)   = \sum_{i=1}^{N-1} \frac{g_i
h_i}{{\textnormal{f}}_i (x)},
\end{equation}
where  $g_i$ is the $i$ coordinate of vector ${\textbf{g}} =
E{\textbf{h}}$, and ${{\textnormal{f}}_i (x)} $ represents the $i$ component of vector function (\ref{eq:Fexplicit}). The derivative is well defined and continuous on ${\mathbb{X}}$,
since ${\textnormal{f}}_i (x) >0$ for every $i$ and $x \in
{\mathbb{X}}$. By dividing every element of the summation by
${\textnormal{f}}^{*}_i$, and using $p_i = h_i
/{\textnormal{f}}^{*}_i$, we can re-write (\ref{eq:DerF_1Ori}) in term of functions $Q_i(x)$ (\ref{eq:the function QjX}) as:
\begin{equation}\label{eq:DerF}
{\textnormal{F}'}(x)   = \sum_{i=1}^{N-1} g_i Q_i (x).
\end{equation}
Let us define a matrix $H \in {\mathbb{R}}^{(N-1) \times (N-1)}$ as:
\begin{equation}\label{eq:HED}
H = E {\cal{D}} ({\textbf{f}}^{*}),
\end{equation}
which by construction is C-Metzler (Definition \ref{def:C-Metzler}), since $E$ is C-Metzler, and the columns of $E$ are scaled by a positive diagonal matrix. By means of $H$ and Eqn (\ref{eq:hDp}), we re-write $E
{\textbf{f}}^{*} = -{\textbf{a}}$ and ${\textbf{g}} = E
{\textbf{h}}$, respectively, as:
\begin{equation}\label{eq:A1aandgHp}
H {\mathbf{1}}_{N -1} = -{\textbf{a}}, ~~ \textnormal{and} ~~ {\textbf{g}} = H
{\textbf{p}}.
\end{equation}
Because $H$ is C-Metzler,
and the relations (\ref{eq:orderQxSS}) and (\ref{eq:A1aandgHp})
hold, we are under the conditions of Lemmas \ref{lem:HQ_result} and
\ref{lem:G_result} in Appendix \ref{sec:appendixAAA}. In particular, the right hand side of (\ref{eq:DerF}) has the same structure as $G(x)$ in Lemma \ref{lem:G_result}.
Consequently, the first derivative is strictly negative on the interval
$\mathbb{X}$ and monotonicity follows.

In order to prove (\ref{eq:RightLeftLimFx}), we note that
each entry of ${\textbf{f}} (x)$ can be expressed as ${\textnormal{f}}_i (x) = {\textnormal{f}}^{*}_i (1 + x p_i)$,
and re-write (\ref{eq:Fx_decrecente}) as:
\begin{equation}\label{eq:OrigFx}
{\textnormal{F}}(x)   = \sum_{i=1}^{N-1} g_i \ln
{\textnormal{f}}_i (x) = \sum_{i=1}^{N-1} g_i \ln
{\textnormal{f}}^{*}_i + \sum_{i=1}^{N-1} g_i {\Pi}_i (x),
\end{equation}
where ${\Pi}_i (x) = \ln (1 + x p_i)$. In addition, let us re-write the sequence of positive parameters in (\ref{eq:orderpj}), in the equivalent form:
\[
p_1 = \cdots = p_s > p_{s+1} \geq \cdots p_k \geq \cdots \geq p_m >0,
\]
with $s$ being an integer that denotes the first strict inequality in the sequence (counted starting from the largest element), and can take any value between $1$ and $m$. Similarly, let us re-write the sequence of negative parameters in (\ref{eq:orderpj}) as:
\[
0 > p_{m+r+1} \geq \cdots \geq p_{\ell} \geq \cdots \geq  p_{t-1} > p_{t} = \cdots = p_{N-1},
\]
with $t$ being an integer that denotes the first strict inequality in the sequence (counted from the smallest element), and can take any value between $m+r+1$ and $N-1$.

The first term at the right
hand side of (\ref{eq:OrigFx}) is constant, while the second term can be expanded as in
(\ref{eq:Gx_equiv}) (proof of Lemma \ref{lem:G_result}) with
${\Pi}_i (x)$ instead of $Q_i (x)$. Taking into account the above sequences, the expansion can be written as:
\begin{equation}\label{eq:Fx_equiv}
{\textnormal{F}}(x) = \sum_{i=1}^{N-1} g_i \ln {\textnormal{f}}^{*}_i + {\textnormal{F}}^{+}(x) + {\textnormal{F}}^{-}(x),
\end{equation}
with:
\begin{eqnarray}
{\textnormal{F}}^{+}(x) &=& ({\Pi}_1 (x)
-{\Pi}_{s+1} (x)) \sum_{i=1}^{s} g_i + \cdots + ({\Pi}_k (x)
- {\Pi}_{k+1} (x))\sum_{i=1}^{k} g_i + \cdots \nonumber \\
&+& {\Pi}_m (x) \sum_{i=1}^{m}
g_i, \label{eq:Fx_equiv_Fplus} \\
{\textnormal{F}}^{-}(x) &=& {\Pi}_{m + r + 1} (x) \sum_{i=m + r +1 }^{N-1} g_i + \cdots +
({\Pi}_{\ell} (x) - {\Pi}_{\ell -1} (x)) \sum_{i=\ell }^{N-1} g_i + \cdots \nonumber \\
 &+& ({\Pi}_{N-1} (x) - {\Pi}_{t-1} (x)) \sum_{i= t}^{N-1} g_i, \label{eq:Fx_equiv_Fminus}
\end{eqnarray}
where terms of the form ${\Pi}_{i} (x) - {\Pi}_{j} (x)$ such that $p_i = p_j$ have been dropped from the expansion, as they are zero.
In computing the left and right limits of ${\textnormal{F}}(x)$, these are the possible scenarios:

1. If $s = m$ or $t = m +r +1$, expansions (\ref{eq:Fx_equiv_Fplus}) or  (\ref{eq:Fx_equiv_Fminus}) reduce to:
\[
{\textnormal{F}}^{+}(x) = {\Pi}_m (x) \sum_{i=1}^{m} g_i,  ~~ \textnormal{or} ~~
{\textnormal{F}}^{-}(x) =  {\Pi}_{m + r + 1} (x) \sum_{i=m + r +1 }^{N-1} g_i.
\]
By Lemma \ref{lem:HQ_result} (Expression in (\ref{eq:defsign_Final})), we have that $\sum_{i=1}^{m} g_i <0$, and $\sum_{i=m + r +1 }^{N-1} g_i > 0$. We also have that $p_m = p_1$ and $p_{m+r+1} = p_{N-1}$. Thus, using the limits (\ref{eq:Limits1}) in Proposition \ref{lem:Limits_Sequences} we obtain (\ref{eq:RightLeftLimFx}).

2. If $s < m$ and $t > m +r +1$, there are both, positive, zero and negative entries so the interval becomes
${\mathbb{X}} = \left( - 1/p_1, -1/p_{N-1} \right)$. In the limit as $x^+ \rightarrow - 1/p_1$ all terms in (\ref{eq:Fx_equiv}) are constant (see limits (\ref{eq:Limits4}) in Proposition \ref{lem:Limits_Sequences}) except the first term associated to ${\textnormal{F}}^{+}(x)$. Concerning this term, we have that $\sum_{i=1}^{s} g_i < 0$ (i.e. strictly negative according to Proposition \ref{prop:DefSigns_s_t}) so by using (\ref{eq:Limits2}) of Proposition \ref{lem:Limits_Sequences},  we get:
\[
\lim_{x^+ \rightarrow L^{-}} {\textnormal{F}}(x) = \lim_{x^+ \rightarrow -(\frac{1}{p_1})} ({\Pi}_1 (x) - {\Pi}_{s+1} (x)) \sum_{i=1}^{s} g_i = + \infty.
\]
In the limit as $x^- \rightarrow - 1/p_{N-1}$, all terms in (\ref{eq:Fx_equiv}) are constant, but the last one associated to ${\textnormal{F}}^{-}(x)$. Again, using Propositions \ref{prop:DefSigns_s_t} and \ref{lem:Limits_Sequences}, we have that:
\[
\lim_{x^- \rightarrow L^{+}} {\textnormal{F}}(x) = \lim_{x^- \rightarrow -(\frac{1}{p_{N-1}})} ({\Pi}_{N-1} (x) - {\Pi}_{t-1} (x)) \sum_{i= t}^{N-1} g_i = - \infty.
\]

3. All entries are positive or zero, so that $m + r = N-1$. In this case,
the interval ${\mathbb{X}} = \left( -1/p_1, + \infty  \right)$, and (\ref{eq:Fx_equiv})
reduces to:
\[
{\textnormal{F}}(x) = \sum_{i=1}^{N-1} g_i \ln {\textnormal{f}}^{*}_i + {\textnormal{F}}^{+}(x).
\]
As in the previous case, $\lim_{x^+ \rightarrow L^{-}} {\textnormal{F}}(x) = + \infty$. On the other hand, in the limit as $x \rightarrow + \infty$, by Proposition \ref{lem:Limits_Sequences}, all terms but the last one in ${\textnormal{F}}^{+}(x)$ are constant. Because $\sum_{i=1}^{m} g_i < 0$ (Lemma \ref{lem:HQ_result}), and $\lim_{x \rightarrow + \infty} {\Pi}_m (x) = + \infty$, we have that:
\[
\lim_{x \rightarrow + \infty} {\textnormal{F}}(x) = \lim_{x \rightarrow + \infty} {\Pi}_m (x) \sum_{i=1}^{m}
g_i = - \infty.
\]

4. All entries are negative or zero so that $m = 0$, the interval ${\mathbb{X}} = \left( - \infty, -1/p_{N-1} \right)$ and (\ref{eq:Fx_equiv})
reduces to:
\[
{\textnormal{F}}(x) = \sum_{i=1}^{N-1} g_i \ln {\textnormal{f}}^{*}_i
+{\textnormal{F}}^{-}(x).
\]
As in case $2$, $\lim_{x^- \rightarrow L^{+}} {\textnormal{F}}(x) = - \infty$.  On the other hand, in the limit as $x \rightarrow - \infty$, by Proposition \ref{lem:Limits_Sequences}, all terms but the first one in ${\textnormal{F}}^{-}(x)$ are constant, and because $\sum_{i= m+r+1}^{N-1} g_i > 0 $ (Lemma \ref{lem:HQ_result}) and $\lim_{x \rightarrow - \infty} {\Pi}_{m + r + 1} = + \infty$, we have that:
\[
\lim_{x \rightarrow -\infty} {\textnormal{F}}(x) = \lim_{x \rightarrow - \infty} {\Pi}_{m + r + 1} (x) \sum_{i=m + r +1 }^{N-1} g_i = + \infty.
\]
\hfill $\Box$

\section{Some network classes with unique equilibrium solutions}
\label{sec:SNetClUnique}

Chemical reaction network structure, with its associated C-Metzler
matrices, influences the set of feasible equilibrium solutions that we compute
by applying conditions in Lemma \ref{def:Feas_sol_gen} and are related to the feasibility functions given in Definition
\ref{def:Feasibility_Functions}. In this section, we exploit
monotonicity of these functions to explore existence and uniqueness of equilibria
within positive stoichiometric compatibility classes for weakly reversible reaction networks.

Monotonicity will be used to identify sub-sets within the space of possible reaction rate coefficients leading to complex balancing and in line with the classical works in \cite{Horn-Jackson:72,Horn:72,Feinberg:79}, to conclude existence and uniqueness of equilibria within positive stoichiometric compatibility classes.

The monotonicity argument will also be employed to show existence and uniqueness of equilibrium solutions for a class of positive deficiency networks. This might support the construction of an alternative proof of the well-known deficiency one theorem \cite{Feinberg:95,Boros:12} for weakly reversible reaction networks.

\subsection{Zero flux conditions and complex balanced equilibrium}

Here we examine the equilibrium solutions ${\textbf{c}}^{*}$  for the dynamic system  (\ref{eq:dotC_Sl}),  that result from all fluxes in the network to be zero, namely ${\bm{\phi}}_{\lambda} ({\psi}_{j_{\lambda}}, {{\bm{\psi}}}_{\lambda}) = 0$ for every $\lambda = 1, \ldots, \ell$. In exploring such a case (we will refer to as the zero flux condition), we first note that ${\mathbf{f}}_{\bm{\eta}}(\textbf{0}; \bm{\chi})$ (i.e. the family of solutions (Definition \ref{def:FamilySolutions}) evaluated at $\mathbf{x} = 0$) corresponds to a zero flux condition.
This can be shown by substituting $x_{\lambda} = 0$ for every $\lambda = 1, \ldots, \ell$ in (\ref{eq:FluxZeroProof}), and
%
%This can be shown by employing the relationship in (\ref{eq:Ef_L})
%%between ${\mathbf{f}}_{\lambda}$ and ${{\bm{\psi}}}_{\lambda}$
%to express the flux vector (\ref{eq:philambda1}) for each linkage class as:
%%
%\begin{equation}\label{eq:FluxZeroProof}
%{\bm{\phi}}_{\lambda} (\psi_{j_{\lambda}}, x_{\lambda};\bm{\chi})= \psi_{j_{\lambda}} \left[\textbf{a}_{\lambda}  + E_{\lambda} {\mathbf{f}}_{\lambda}(x_{\lambda};\bm{\chi})\right].
%\end{equation}
%%
%Substituting $x_{\lambda} = 0$ for every $\lambda = 1, \ldots, \ell$ in the above expression,
and using the fact that ${\mathbf{f}}_{\bm{\eta}}(\textbf{0}; \bm{\chi})$ implies that ${\mathbf{f}}_{\lambda} (0; \bm{\chi}) = {\mathbf{f}}_{\lambda}^{*}$, so that Eqn (\ref{eq:FluxZeroProof}) becomes:
\[
{\bm{\phi}}_{\lambda} (\psi_{j_{\lambda}}, 0 ;\bm{\chi})= \psi_{j_{\lambda}}(\textbf{a}_{\lambda}  + E_{\lambda} {\mathbf{f}}_{\lambda}^{*}) = 0,
\]
where the zero flux condition follows since ${\mathbf{f}}_{\lambda}^{*} = - E_{\lambda}^{-1} \textbf{a}_{\lambda}$ for every $\lambda=1, \ldots, \ell$. Let us denote  ${\mathbf{f}}_{\bm{\eta}}(\textbf{0}; \bm{\chi})$ by ${\mathbf{f}}_{\bm{\eta}}^{*}$, which by construction is of the form:
\begin{equation}\label{eq:E_eta_Ref}
({\mathbf{f}}_{\bm{\eta}}^{*})^T = \left[
\begin{array}{ccccc}
({\mathbf{f}}^{*}_{1})^T  & \cdots & ({\mathbf{f}}^{*}_{\lambda})^T & \cdots & ({\mathbf{f}}^{*}_{\ell})^T
\end{array}
\right].
\end{equation}
If the network is irreversible, it follows from Proposition \ref{prop:PositiveFamily} that the domain ${\mathbb{D}}_{\bm{\eta}} (\bm{\chi})$, is either empty for some $\bm{\chi} \in \mathbb{U}$, or if not, it does not contain the zero.  Hence, ${\mathbf{f}}_{\bm{\eta}}^{*}$ cannot be a strictly positive vector, what in turns results in some species concentrations (associated to the zero entries of ${\mathbf{f}}_{\bm{\eta}}^{*}$) to be zero. Since there is no strictly positive equilibrium vector ${\textbf{c}}^{*}$ complying with a zero flux condition, irreversible networks do not accept complex balanced equilibrium, according to Definition \ref{def:Complex_Balance_Condition}.

On the other hand, if the network is weakly reversible, Proposition \ref{prop:PositiveFamily} asserts that for every $\bm{\chi} \in {\mathbf{U}}$, there exists a domain ${\mathbb{D}}_{\eta} (\bm{\chi})$, which contains the zero, such that ${\mathbf{f}}_{\bm{\eta}}(\mathbf{x}; \bm{\chi})$ is strictly positive, and consequently ${\mathbf{f}}_{\bm{\eta}}^{*} > 0$.
If in addition, ${\mathbf{f}}_{\bm{\eta}}^{*}$ belongs to the set of feasible solutions (Definition \ref{def:SetFeasibleSol}), there exist strictly positive vectors ${\textbf{c}} = \exp ({\bm{\xi}})$, such that $\ln {\mathbf{f}}_{\bm{\eta}}^{*} = S^{T} {\bm{\xi}}$, which are equilibrium solutions of system (\ref{eq:dotC_Sl}).
%\footnote{if $s = m$ there will be exactly one positive vector ${\textbf{c}}^{*}$. If $s \leq m$ the vectors ${\bm{\xi}}^{*}$ that correspond with (positive) equilibrium concentrations will lie on a $m-s$-dimensional manifold}.
According to Definition \ref{def:Complex_Balance_Condition}, those equilibria are complex balanced.

We recall from Eqns (\ref{eq:structM_k_l}), (\ref{eq:Ef_L}) and (\ref{eq:Fexplicit_first}) that ${\mathbf{f}}_{\bm{\eta}}^{*}$ depends on the reaction rate coefficients through $E_{\lambda}$ and $a_{\lambda}$ for $\lambda=1,\dots,\ell$, which in the last instance determine feasibility, in the sense of Definition \ref{def:SetFeasibleSol}.
For convenience, we collect the set of reaction rate coefficients of the network into a vector ${\bm{k}} \in \mathbb{R}_{>0}^{\rho}$, where  $\rho$ denotes the number of irreversible reaction steps in the network, and introduce the so-called \emph{Horn set} \cite{alonso_szederkenyi:12}, that is formally defined as:
\begin{equation}\label{eq:eqpolyhedron_Horn}
{\cal{H}} = \{{\bm{k}} \in \mathbb{R}_{>0}^{\rho}  ~~ | ~~ \ln
{\mathbf{f}}_{\bm{\eta}}^{*} \in \textnormal{Im}(S^{T}) \}.
\end{equation}
Note that the set is only meaningful for ${\mathbf{f}}_{\bm{\eta}}^{*} > 0$, which as discussed above requires the network to be weakly reversible. The result we present next shows that the set ${\cal{H}}$ contains all possible reaction rate coefficients leading to complex balanced equilibrium solutions.
\begin{proposition}\label{the:ComplexBalanceUniqueStab}
Any chemical reaction network with
${\bm{k}} \in {\cal{H}}$ will only accept complex balanced equilibrium solutions.
\end{proposition}
\textbf{Proof:}
For any ${\bm{k}} \in {\cal{H}}$, ${\mathbf{f}}_{\bm{\eta}}^{*}$ is an element of the set of feasible solutions (Definition \ref{def:SetFeasibleSol}),
since there exist vectors ${\bm{\xi}} \in {\mathbb{R}}^{m}$ such that:
\begin{equation}
\ln {\mathbf{f}}_{\bm{\eta}}^{*}  = S^{T} {\bm{\xi}},
\end{equation}
and as discussed above, the corresponding strictly positive vectors ${\textbf{c}} = \exp ({\bm{\xi}})$ must be complex balanced equilibrium solutions satisfying:
\begin{equation}\label{eq:FeasRel_fAndXi}
\ln {\mathbf{f}}_{\bm{\eta}}^{*}  = S^{T} \ln {\textbf{c}}.
\end{equation}
In fact, as we will prove next, these are the only possible equilibrium solutions.
First, we note that by Proposition \ref{Corlry:Feas_sol_gen}, for any
${\bm{\chi}} \in {\mathbf{U}}$ we have that:
\begin{equation}\label{eq:HereWeSumAgain_Balance}
\sum_{\lambda} {\textnormal{F}}_{\lambda} (0;
{\bm{\chi}}) = 0.
\end{equation}
Suppose that for a ${\bm{k}} \in {\cal{H}}$ there exists a non-zero flux condition that lead to an equilibrium solution. This implies that there exists at least one ${\bm{\chi}}^{*} \in {\mathbf{U}}$ and
${\mathbf{x}}^{*} \in {\mathbb{D}}_{\bm{\eta}} ({\bm{\chi}}^{*})$ with
${\mathbf{x}}^{*} \neq \mathbf{0}$ such that ${\mathbf{f}}_{\bm{\eta}}({\mathbf{x}}^{*}; {\bm{\chi}}^{*})$ is feasible, and therefore relation (\ref{eq:Gen_ort_STSlnf_SUM})  applies, so that:
\begin{equation}\label{eq:HereWeSumAgain}
\sum_{\lambda} {\textnormal{F}}_{\lambda} (x^{*}_{\lambda};
{\bm{\chi}}^{*}) = 0,
\end{equation}
Since the network is weakly reversible, by Proposition \ref{prop:PositiveFamily} we have that  the domain ${\mathbb{D}}_{\bm{\eta}} ({\bm{\chi}}^{*}) = {\mathbb{X}}_{\bm{\eta}} ({\bm{\chi}}^{*})\cap {\mathbb{D}}_{0}$ is non-empty, contains the zero, and can be partitioned as:
\[
{\mathbb{D}}_{\bm{\eta}} ({\bm{\chi}}^{*}) = {\mathbb{D}}^{-}_{\bm{\eta}} ({\bm{\chi}}^{*}) \cup \{ 0\} \cup {\mathbb{D}}^{+}_{\bm{\eta}} ({\bm{\chi}}^{*}),
\]
with
${\mathbb{D}}^{-}_{\bm{\eta}} ({\bm{\chi}}^{*}) = {\mathbb{X}}_{\bm{\eta}} ({\bm{\chi}}^{*})\cap {\mathbb{R}}^{\ell}_{<0}$ and
${\mathbb{D}}^{+}_{\bm{\eta}} ({\bm{\chi}}^{*}) = {\mathbb{X}}_{\bm{\eta}} ({\bm{\chi}}^{*})\cap {\mathbb{R}}^{\ell}_{>0}$. Since ${\mathbf{x}}^{*} \in {\mathbb{D}}_{\bm{\eta}} ({\bm{\chi}}^{*})$ and ${\mathbf{x}}^{*} \neq \mathbf{0}$, then it either belongs to ${\mathbb{D}}^{-}_{\bm{\eta}} ({\bm{\chi}}^{*})$ or to ${\mathbb{D}}^{+}_{\bm{\eta}} ({\bm{\chi}}^{*})$.

Suppose that ${\mathbf{x}}^{*} \in {\mathbb{D}}^{-}_{\bm{\eta}} ({\bm{\chi}}^{*})$, then we have that $x^{*}_{\lambda} < 0$ for every $\lambda = 1, \cdots, \ell$. By Theorem \ref{lem:Lemma_on_F}, every function ${\textnormal{F}}_{\lambda} (x_{\lambda};
{\bm{\chi}}^{*})$  in (\ref{eq:Gen_ort_STSlnf_SUM}) is monotonous decreasing. Thus, for every  $\lambda$, the following strict inequalities hold:
\[
 {\textnormal{F}}_{\lambda} (0;
{\bm{\chi}}^{*}) < {\textnormal{F}}_{\lambda} (x^{*}_{\lambda};
{\bm{\chi}}^{*}).
\]
The summation over  $\lambda$ results in:
\[
\sum_{\lambda} {\textnormal{F}}_{\lambda} (0;
{\bm{\chi}}^{*}) < \sum_{\lambda} {\textnormal{F}}_{\lambda} (x^{*}_{\lambda};
{\bm{\chi}}^{*}) = 0,
\]
what is in contradiction with expression (\ref{eq:HereWeSumAgain_Balance}). A similar argument for ${\mathbf{x}}^{*} \in {\mathbb{D}}^{+}_{\bm{\eta}} ({\bm{\chi}}^{*})$ leads to:
\[
\sum_{\lambda} {\textnormal{F}}_{\lambda} (0;
{\bm{\chi}}^{*}) > \sum_{\lambda} {\textnormal{F}}_{\lambda} (x^{*}_{\lambda};
{\bm{\chi}}^{*}) = 0,
\]
which again is in contradiction with (\ref{eq:HereWeSumAgain_Balance}). This proves that for any ${\bm{k}} \in {\cal{H}}$, the set of feasible solutions contains just one element ${\mathbf{f}}_{\bm{\eta}}^{*}$, which corresponds to complex balanced equilibrium solutions.
\hfill $\Box$
\begin{proposition}\label{prop:the:ComplexBalanceUniqueStabII}
For each ${\bm{k}} \in {\cal{H}}$, there exists exactly one complex balanced equilibrium in each positive stoichiometric compatibility class, and this equilibrium is locally asymptotically stable within the corresponding stoichiometric compatibility class. Furthermore, if the reaction network consists of one linkage class, then the asymptotic stability of the equilibrium point corresponding to any $k\in \mathcal{H}$ within its stoichiometric compatibility class is global.
\end{proposition}
\textbf{Proof:}
As discussed above, for a given ${\bm{k}} \in {\cal{H}}$, the set of feasible solutions contains just one element ${\mathbf{f}}_{\bm{\eta}}^{*}$. Let ${\textbf{c}}_{0} \in {\mathbb{R}}^{m}_{>0}$ be one (complex balanced) equilibrium point so that according to (\ref{eq:FeasRel_fAndXi}) we have that $\ln {\mathbf{f}}_{\bm{\eta}}^{*}  = S^{T} \ln {\textbf{c}}_{0}$. Then, any (complex balanced) equilibrium satisfies:
\begin{equation}\label{eq:FeasRel_fAndXi_inC}
S^{T} (\ln {\textbf{c}} - \ln {\textbf{c}}_{0}) = 0,
\end{equation}
which coincides with the set ${\cal{U}} (\mathbf{c}_0)$ defined as (\ref{eq:TheFamousSetUE}) in Proposition \ref{prop:Unique_in_CC}. From this proposition, it follows that the set contains exactly one element in each positive stoichiometric  compatibility class.

That each equilibrium is locally asymptotically stable follows from a standard result presented in \cite{Feinberg:79} (Lecture 5, Proposition 5.3) and summarized in Appendix \ref{sec:appendixBBB} (Proposition \ref{prop:B_Lyapunov}).
The global stability of the complex balanced equilibria in the single linkage class case is proved in \cite{anderson:2011}.
\hfill $\Box$

\begin{remark}
For weakly reversible deficiency zero reaction networks, $\textnormal{Im}(S^{T})$ spans $\mathbb{R}^{n - \ell}$, what in turn implies that any vector of reaction rate coefficients $\bm{k} \in \mathbb{R}_{>0}^{\rho}$ will
be an element of the Horn set (\ref{eq:eqpolyhedron_Horn}). Thus, from
Propositions \ref{the:ComplexBalanceUniqueStab} and \ref{prop:the:ComplexBalanceUniqueStabII}, it follows that any equilibrium solution will be unique (one in each positive stoichiometric compatibility class) and locally asymptotically stable. If the network is irreversible, positive equilibrium does not exist. Such conclusions
have been formally stated in the so-called Deficiency Zero Theorem.  \cite{Feinberg:87}.
Additionally, we remark here that according to recent results, the stability of any complex balanced equilibrium point is most probably global \cite{craciun:2015}.
\end{remark}

\begin{remark}
If the reaction network is elementary and compatible with thermodynamics (this implying detailed balancing), any allowed reaction constant for the network must be in ${\cal{H}}$. In this way, the definition of the Horn set could be considered as an alternative statement of the Wegscheider conditions (see \cite{Gorban_etal:13} for a classical statement of the conditions).
\end{remark}
{\bf Example: The Horn set for a weakly reversible reaction}

Let us consider the example discussed in Subsection \ref{sec:The4RevNetwork} of a one linkage class network of deficiency $\delta = 2$. The corresponding graph structure, stoichiometry, and the set of possible parameters is depicted in Figure \ref{fig:UNSTOneLCN_WWR}. For this $2$-parameter network, the Horn set (\ref{eq:eqpolyhedron_Horn}) is obtained by finding those $(\epsilon, \beta)$ that make $\ln {\textbf{f}}^{*}$ orthogonal to ${\textbf{g}}^1$ and ${\textbf{g}}^2$. Note that this implies that $\ln {\textbf{f}}^{*}$ lies in $\textnormal{Range}(S^{T})$. The conditions can be written as:
\begin{equation}
\left[ \begin{array}{rrr}
    1 & 0 & -2 \\
    0 & 1 & -3 \\
\end{array} \right]
\ln \left(
\begin{array}{r} \epsilon / \beta \\
1  \\
\epsilon / \beta
\end{array} \right) = 0.
\end{equation}
This results in a set with parameters satisfying $\alpha = \beta$, that lead to complex balanced solutions. By construction, $\ln {\textbf{f}} (0; \bm{\chi}) = \bm{0}$ for every $\bm{\chi}$, and therefore $\textnormal{F} (0; \bm{\chi}) = 0$. Since $\textnormal{F} (x; \bm{\chi})$  is monotonous decreasing, only complex balanced solutions exist for parameters in the Horn set.
%%\hfill $\triangle$
%
%
\subsection{The Deficiency One Theorem revisited}
Next, we present a result that might be a basis for an alternative
proof of the deficiency one theorem for weakly reversible reaction networks. The theorem was  originally proposed by
\cite{Feinberg:95} and recently discussed by \cite{Boros:12}, employing in both cases
a graph theoretical formalism. The argument we propose builds on the following observations: On the one hand, the existence of a basis in the kernel of $S$ that has at most one vector ${\textbf{g}}^{r}$ per linkage class. On the other hand, a particular orthogonal structure for such basis.

\begin{figure}[ht]
\begin{center}
$\begin{array}{@{}c@{}}
\includegraphics[scale=0.45]{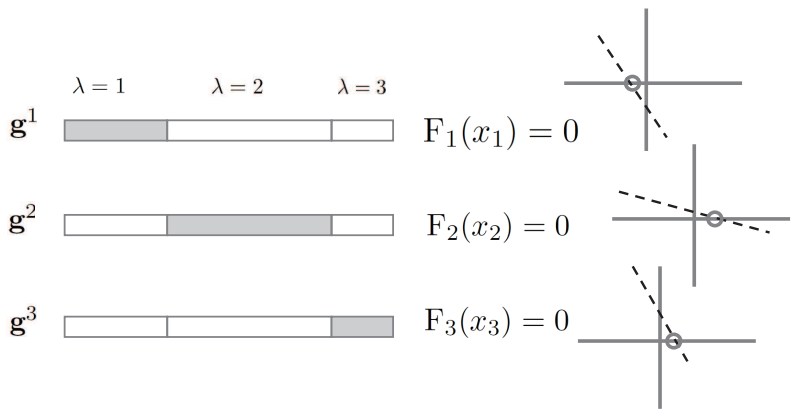}
\end{array}$
\end{center}
\caption{Vectors ${\textnormal{\bf{g}}}^{r}$ form a basis of the kernel of $S$ where the grey areas indicate the only possible non-zero coordinates. This structure decouples feasibility conditions (at most one per linkage class). Monotonicity of  ${\textnormal{F}}_{\lambda} (x_{\lambda})$, schematically represented as discontinuous lines at the right of the figure, then leads to just one solution per linkage class.} \label{fig:Cond_Def1}
\end{figure}

The structure is such that for each element of the basis, the only possible nonzero coordinates must be at the location of the complexes that correspond to the linkage class the vector is associated to. Orthogonality on a normalized basis is formally expressed as:
%Formally, this last condition can be expressed as:
%%
\begin{equation}\label{eq:OrthgBasis}
({\mathbf{g}}^{i})^T {\mathbf{g}}^{j} = {\delta}_{ij},
\end{equation}
where ${\delta}_{ij}$ denotes the Kronecker delta. The structure of the vectors is illustrated in Figure \ref{fig:Cond_Def1}, for a network consisting of $3$ linkage classes, with the grey areas representing the non-zero vector coordinates. As sketched in the same figure, that particular structure of the ${\textbf{g}}^{r}$ vectors decouples the corresponding feasibility conditions (Lemma \ref{def:Feas_sol_gen}) along linkage classes, so to have one feasibility function per linkage class. As we will see, monotonicity of such functions (Theorem \ref{lem:Lemma_on_F}) will ensure uniqueness.

%***!!!

\begin{theorem}\label{prop:Def1_theorem_Alt}
Let us consider a weakly reversible
reaction network with $\ell$ linkage classes such that $\delta =
\sum_{\lambda} {\delta}_{\lambda}$, where ${\delta}_{\lambda}$ is
either $0$ or $1$. Then, there will be a unique equilibrium in each
positive stoichiometric compatibility class.
\end{theorem}
\textbf{Proof:}
Let $t$ be the number of linkage classes having deficiency 1. Since for each linkage class, ${\delta}_{\lambda}$ can be either zero or $1$, then $t \leq \ell$. Assume, without loss of generality, that the deficiencies of linkage classes $\lambda = 1, \ldots, t$ are $1$, whereas for the remaining $\lambda = t+1, \ldots, \ell$, linkage deficiencies are $\delta_{\lambda}= 0$.

For the first $t$ linkage classes, let the non-zero vectors ${\mathbf{p}}^{\lambda} \in {\mathbb{R}}^{N_{\lambda} -1}$ (for $\lambda = 1, \ldots, t$)  be the basis associated to the kernel of $S_{\lambda}$ so that $S_{\lambda} {\mathbf{p}}^{\lambda} = 0$. From the assumption, we have that:
\[
\delta =
\sum_{\lambda} {\delta}_{\lambda} = t,
\]
and a basis $\{{\mathbf{g}}^{\lambda} ~|~ \lambda=1, \ldots, t \}$ for the kernel of $S$ satisfying (\ref{eq:OrthgBasis}) under proper normalization can be constructed as:
\begin{equation}\label{eq:OrthBasisA}
\begin{array}{cccccccc}
({\mathbf{g}}^{1})^T  =  [\;
           ({\mathbf{p}}^{1})^T & \cdots & ({\textbf{0}}_{\lambda})^T & \cdots &
           & ({\textbf{0}}_{t})^T & \cdots & ({\textbf{0}}_{\ell})^T \; ]\\
\,\vdots\qquad\qquad\, \vdots\quad\; & & \vdots\quad\; & &  & \vdots\quad\; & & \vdots\quad\;\,\\
({\mathbf{g}}^{\lambda})^T  =  [\;
           ({\textbf{0}}_{1})^T & \cdots & ({\mathbf{p}}^{\lambda})^T & \cdots &
           & ({\textbf{0}}_{t})^T & \cdots & ({\textbf{0}}_{\ell})^T \; ]\\
\,\vdots\qquad\qquad\, \vdots\quad\; & & \vdots\quad\; & &  & \vdots\quad\; & & \vdots\quad\;\,\\
({\mathbf{g}}^{t})^T  =  [\;
           ({\textbf{0}}_{1})^T & \cdots & ({\textbf{0}}_{1})^T & \cdots &
           & ({\mathbf{p}}^{t})^T & \cdots & ({\textbf{0}}_{\ell})^T \; ]\\
\end{array}
\end{equation}
where ${\textbf{0}}_{\lambda} \in {\mathbb{R}}^{N_{\lambda} -1}$ for $\lambda = 1, \ldots,\ell$ are zero vectors. Note that under the assumptions of the theorem, for any equivalent basis $\{{\mathbf{\widehat{g}}}^{\lambda} ~|~ \lambda=1, \ldots, t \}$ of the kernel of $S$, the sub-vectors ${\mathbf{\widehat{g}}}_{\lambda}^{\lambda}$ in (\ref{eq:BasisKernel}) must satisfy that $S_{\lambda} {\mathbf{\widehat{g}}}_{\lambda}^{\lambda} = 0$ for $\lambda = 1, \ldots, t$. This is so since each element ${\mathbf{\widehat{g}}}^{\lambda}$ of the basis can be expressed as a linear combination of (\ref{eq:OrthBasisA}), what in addition implies that the vectors ${\mathbf{\widehat{g}}}_{\lambda}^{\lambda}$ and ${\mathbf{p}}^{\lambda}$ are parallel. Equivalently, there exists a non-zero scalar ${\mu}_{\lambda}$ such that
${\mathbf{\widehat{g}}}_{\lambda}^{\lambda} = {\mu}_{\lambda} {\mathbf{p}}^{\lambda}$.

Using the basis $\{{\mathbf{g}}^{\lambda} ~|~ \lambda=1, \ldots, t \}$, conditions (\ref{eq:EfX_formal}) in Lemma \ref{def:Feas_sol_gen} can be written as:
\begin{equation}\label{eq:theaboveexpr}
({\mathbf{g}}^{\lambda})^{T} \ln
{\mathbf{f}}_{\bm{\eta}} (\mathbf{x}; \bm{\chi}) = 0 , ~~\textnormal{for} ~~ \lambda = 1, \cdots, t.
\end{equation}
Taking into account the structure of the basis components (\ref{eq:OrthBasisA}), conditions in (\ref{eq:theaboveexpr}) reduce to:
\begin{equation}\label{eq:FeasCondDef1Final}
\begin{array}{cccc}
({\mathbf{p}}^{1})^{T} & \ln &
( {\textbf{f}}_{1}^{*} +  {z}_{1} E_{1}^{-1} {\textbf{p}}^{1}) = & 0\\
 & \vdots & & \vdots\\
({\mathbf{p}}^{\lambda})^{T} & \ln &
( {\textbf{f}}_{\lambda}^{*} +  {z}_{\lambda} E_{\lambda}^{-1} {\textbf{p}}^{\lambda}) = & 0\\
 & \vdots & & \vdots\\
({\mathbf{p}}^{t})^{T} & \ln &
( {\textbf{f}}_{t}^{*} +  {z}_{t} E_{t}^{-1} {\textbf{p}}^{t}) = & 0\\
\end{array}
\end{equation}
where $z_{\lambda} = {x}_{\lambda} {\chi}_{\lambda}$ for $\lambda = 1, \ldots, t$. The left hand side of each of the expressions in (\ref{eq:FeasCondDef1Final}), that we denote as ${\textnormal{F}}_{\lambda} (z_{\lambda})$ for ($\lambda = 1, \ldots, t$), is of the form (\ref{eq:Fx_decrecente}) (Section \ref{sec:Domain_properties_Feasible}). According to Theorem \ref{lem:Lemma_on_F}, each of those functions is monotonous decreasing and because of (\ref{eq:RightLeftLimFx}), it becomes zero at a point
$z^{*}_{\lambda} \in {\mathbb{X}}_{\lambda}$.
Therefore, the set of feasible solutions contains just one element:
\begin{equation}
{\mathbf{f}}_{\bm{\eta}}^T (z_{1}^{*}, \ldots, z_{t}^{*}) = \left[
\begin{array}{cccccccc}
{\textbf{f}}_{1}^T (z_{1}^{*})  & \cdots & {\textbf{f}}^T_{\lambda}(z_{\lambda}^{*}) & \cdots & {\textbf{f}}^T_{t}(z_{t}^{*}) & ({\textbf{f}}_{t+1}^{*})^T
& \cdots & ({\textbf{f}}_{\ell}^{*})^T
\end{array}
\right],
\end{equation}
with ${\textbf{f}}_{\lambda}(z_{\lambda}^{*}) = {\textbf{f}}_{\lambda}^{*} +  {z}^{*}_{\lambda}  {\textbf{h}}_{\lambda}$ for $\lambda =1, \cdots, t$.

Similarly to the proof of Proposition \ref{prop:the:ComplexBalanceUniqueStabII}, let ${\textbf{c}}_{0} \in {\mathbb{R}}^{m}_{>0}$ be one equilibrium point so that according to (\ref{eq:FeasRel_fAndXi}), ${\mathbf{f}}_{\bm{\eta}}^T (z_{1}^{*}, \ldots, z_{t}^{*}) = S^{T} \ln {\textbf{c}}_{0}$. Then, any equilibrium satisfies:
\[
S^{T} (\ln {\textbf{c}} - \ln {\textbf{c}}_{0}) = 0,
\]
which coincides with the set (\ref{eq:TheFamousSetUE}). Finally, from Proposition \ref{prop:Unique_in_CC}, it follows that the set contains exactly one element in each positive stoichiometric  compatibility class.
\hfill $\Box$

As a final remark, we note that this result allows us to conclude uniqueness of equilibria in each positive stoichiometric compatibility class (although not stability), for networks with deficiency other than zero, provided that feasibility conditions in Lemma \ref{def:Feas_sol_gen} can be decoupled along linkage classes, as discussed above.

\subsection{A complex network satisfying the deficiency one theorem}

\begin{figure}[ht]
\begin{center}
$\begin{array}{@{}cc@{}}
\mathrm{(a)} & \mathrm{(b)} \\
\includegraphics[scale=0.33]{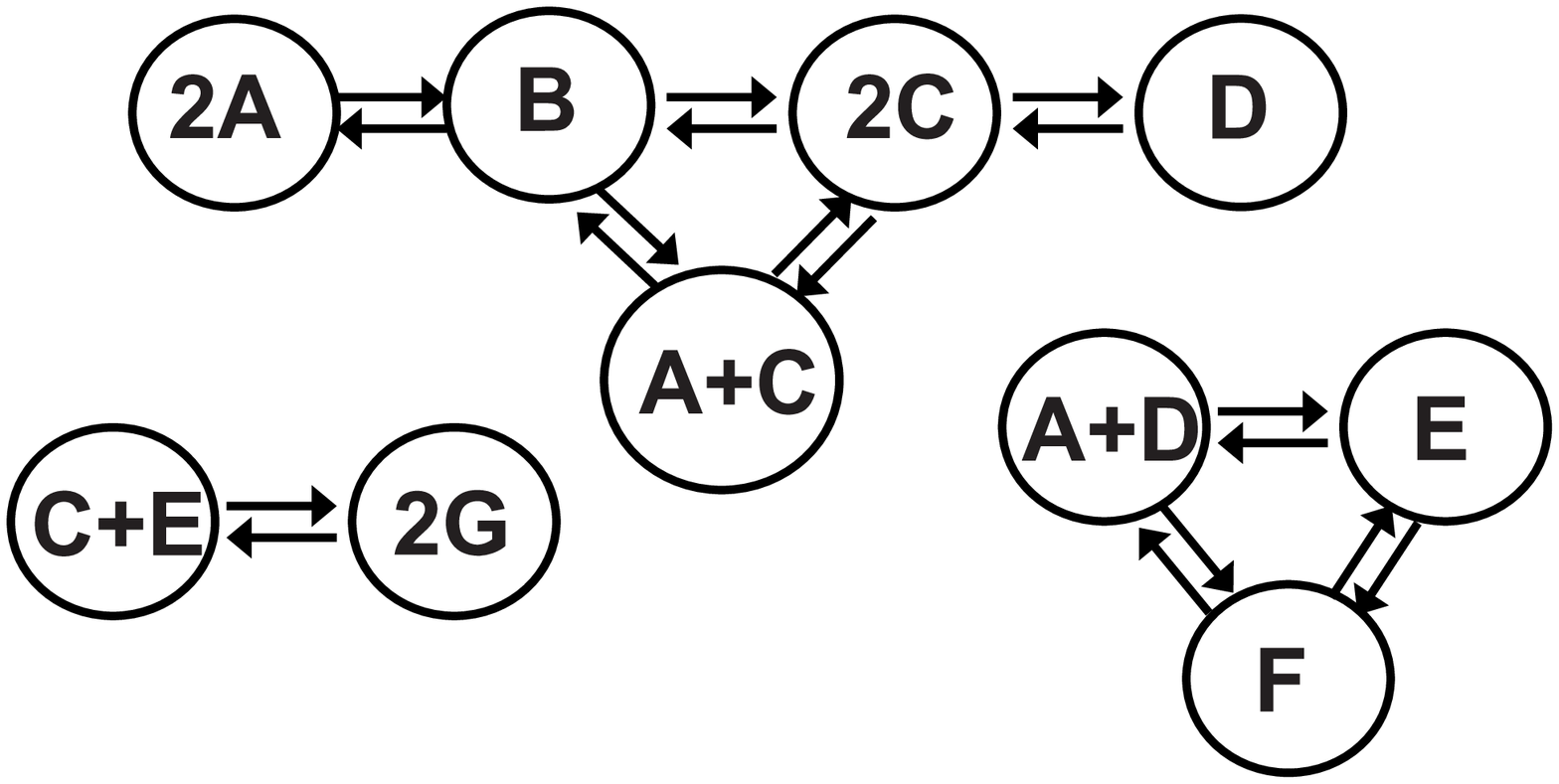} &
\includegraphics[scale=0.33]{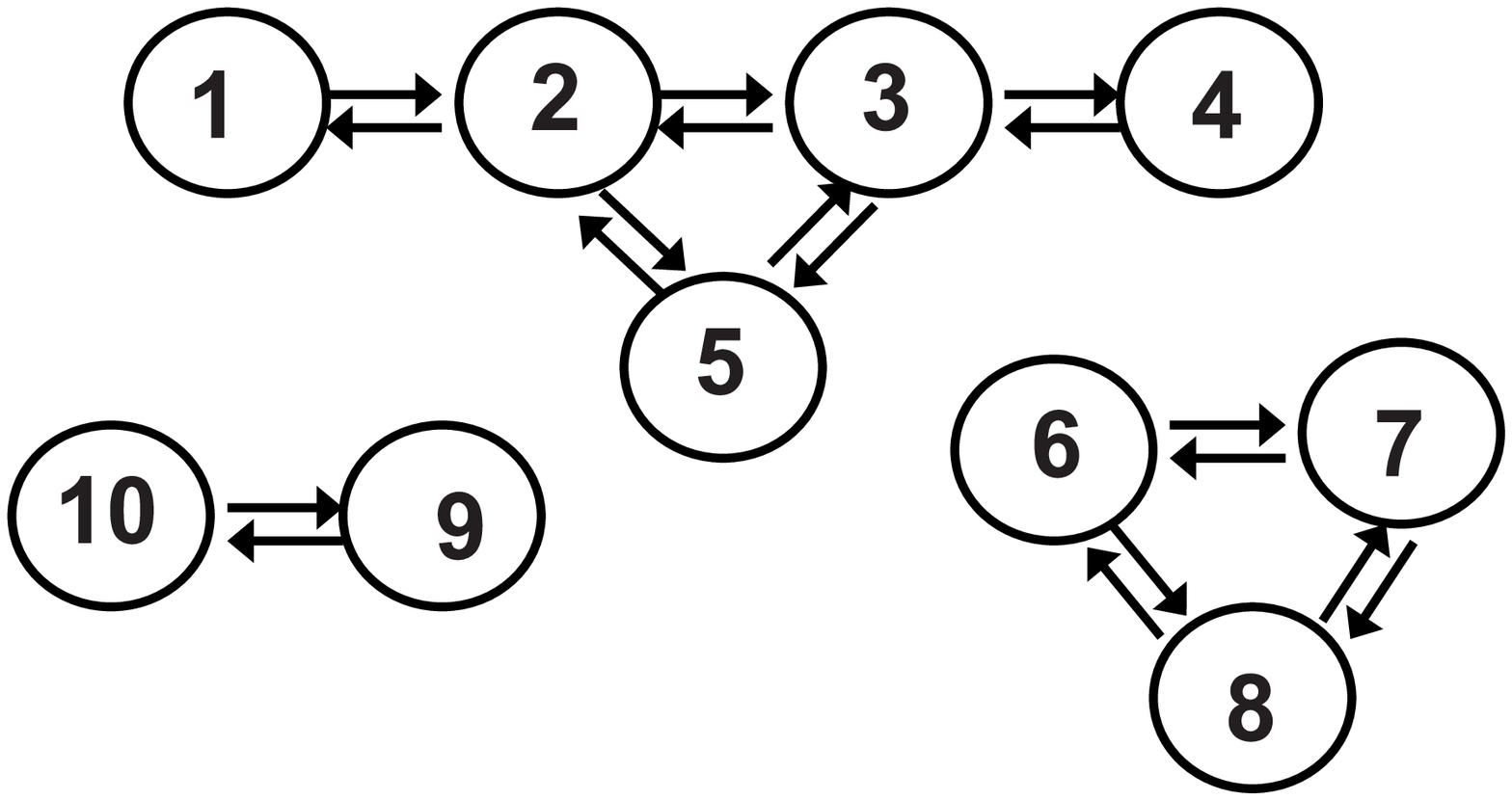}
\end{array}$
\end{center}
\caption{Graph representation for the reaction network. (a) The
species that are part of each complex are explicitly indicated. (b)
The same graph described in terms of numbered complexes.} \label{fig:Example_1}
\end{figure}

Let us consider a reaction network involving $m=7$ chemical species,
we label with capital letters from $A$ to $G$. The (reversible)
reaction steps that take place are:
\begin{equation}
\begin{array}{lclcl}
  2 A \leftrightarrows B &\quad & B \leftrightarrows 2C & \quad & 2 C \leftrightarrows D \\
  B \leftrightarrows A + C &\quad & 2 C \leftrightarrows A + C & \quad & \quad \\
  C + E \leftrightarrows 2 G  &\quad & \quad & \quad & \quad \\
  A + D \leftrightarrows E &\quad & E \leftrightarrows F & \quad & A + D \leftrightarrows F\\
\end{array}
\end{equation}
This particular reaction network comprises $n=10$ complexes and $\ell = 3$ linkage classes, we represent in graph form in Figure \ref{fig:Example_1}, with explicit indication of the species (Figure \ref{fig:Example_1}a) as well as in terms of numbered complexes (Figure
\ref{fig:Example_1}b). %
%\begin{equation}
%\begin{array}{ll}
%  {\cal{I}}_1 = \{2\}& {\cal{I}}_2 = \{1,3,5\} \\
%  {\cal{I}}_3 = \{2,4,5 \} & {\cal{I}}_4 = \{3 \} \\
%  {\cal{I}}_5 = \{2,3\} & {\cal{I}}_6 = \{7,8\} \\
%  {\cal{I}}_7 = \{6,8\} & {\cal{I}}_8 = \{6,7\}\\
%  {\cal{I}}_9 = \{10 \} & {\cal{I}}_{10} = \{9 \}
%\end{array}
%\end{equation}

Complexes are grouped by linkage class in the sets ${\cal{L}}_{1} = \{ 1, 2, 3, 4, 5\}$, ${\cal{L}}_{2} = \{ 6, 7, 8\}$ and ${\cal{L}}_{3} = \{ 9, 10 \}$. The stoichiometry associated to the complexes is given (column-wise) in  the following molecularity matrix (Section \ref{secsec:dynamics_RN}):
\begin{equation}
Y = \left[
\begin{array}{cccccccccc}
  2 & 0 & 0 & 0 & 1 & 1 & 0 & 0 & 0 & 0\\
  0 & 1 & 0 & 0 & 0 & 0 & 0 & 0 & 0 & 0\\
  0 & 0 & 2 & 0 & 1 & 0 & 0 & 0 & 0 & 1\\
  0 & 0 & 0 & 1 & 0 & 1 & 0 & 0 & 0 & 0\\
  0 & 0 & 0 & 0 & 0 & 0 & 1 & 0 & 0 & 1\\
  0 & 0 & 0 & 0 & 0 & 0 & 0 & 1 & 0 & 0\\
  0 & 0 & 0 & 0 & 0 & 0 & 0 & 0 & 2 & 0
\end{array}
\right].
\end{equation}
Choosing $j_1 = 1$, $j_2 = 6$ and $j_3 = 9$, as the reference complexes, matrices $S_{\lambda}$, at the right of expression (\ref{eq:dotC_Sl}), become:
\[
S_{1} = \left[\begin{array}{rrrr}
  -2 & -2 & -2 & -1 \\
   1 &  0 &  0 &  0 \\
   0 &  2 &  0 &  1 \\
   0 &  0 &  1 &  0 \\
   0 &  0 &  0 &  0 \\
   0 &  0 &  0 &  0 \\
   0 &  0 &  0 &  0
\end{array} \right], ~~ S_2 =\left[\begin{array}{rr}
       -1 & -1 \\
        0 &  0 \\
        0 &  0 \\
       -1 & -1 \\
        1 &  0 \\
        0 &  1 \\
        0 &  0
     \end{array} \right], ~~ S_3 = \left[\begin{array}{r}
                0 \\
                0 \\
                1 \\
                0 \\
                1 \\
                0 \\
               -2
\end{array}\right].
\]
Net reaction fluxes take the form:
\begin{equation}\label{eq:example_fluxes}
\begin{array}{l}
  {\phi}_2 (\bm{\psi})= k_{1,2} {\psi}_1 + k_{3,2} {\psi}_3 + k_{5,2} {\psi}_5 - (k_{2,1} + k_{2,3} + k_{2,5}){\psi}_2\\
  {\phi}_3 (\bm{\psi})= k_{2,3} {\psi}_2 + k_{4,3} {\psi}_4 + k_{5,3} {\psi}_5 - (k_{3,2} + k_{3,4} + k_{3,5}){\psi}_3 \\
  {\phi}_4 (\bm{\psi}) = k_{3,4} {\psi}_3  - k_{4,3}{\psi}_4 \\
  {\phi}_5 (\bm{\psi}) = k_{2,5} {\psi}_2 + k_{3,5} {\psi}_3 - (k_{5,2} + k_{5,3}){\psi}_5\\
  {\phi}_7 (\bm{\psi}) = k_{6,7} {\psi}_6 + k_{8,7} {\psi}_8 - (k_{7,6} + k_{7,8}){\psi}_7\\
  {\phi}_8 (\bm{\psi}) = k_{6,8} {\psi}_6 + k_{7,8} {\psi}_7 - (k_{8,6} + k_{8,7}){\psi}_8\\
  {\phi}_{10} (\bm{\psi}) = k_{9,10} {\psi}_9 -k_{10,9}{\psi}_{10}
\end{array}
\end{equation}
The remaining fluxes ${\phi}_1 (\bm{\psi})$, ${\phi}_6 (\bm{\psi})$ and
${\phi}_9 (\bm{\psi})$, associated to the reference complexes, are obtained by means of relation (\ref{eq:linear_dependent_fluxes}).

\begin{figure}[ht]
\begin{center}
$\begin{array}{@{}cc@{}}
\mathrm{(a)} & \mathrm{(b)} \\
\includegraphics[scale=0.33]{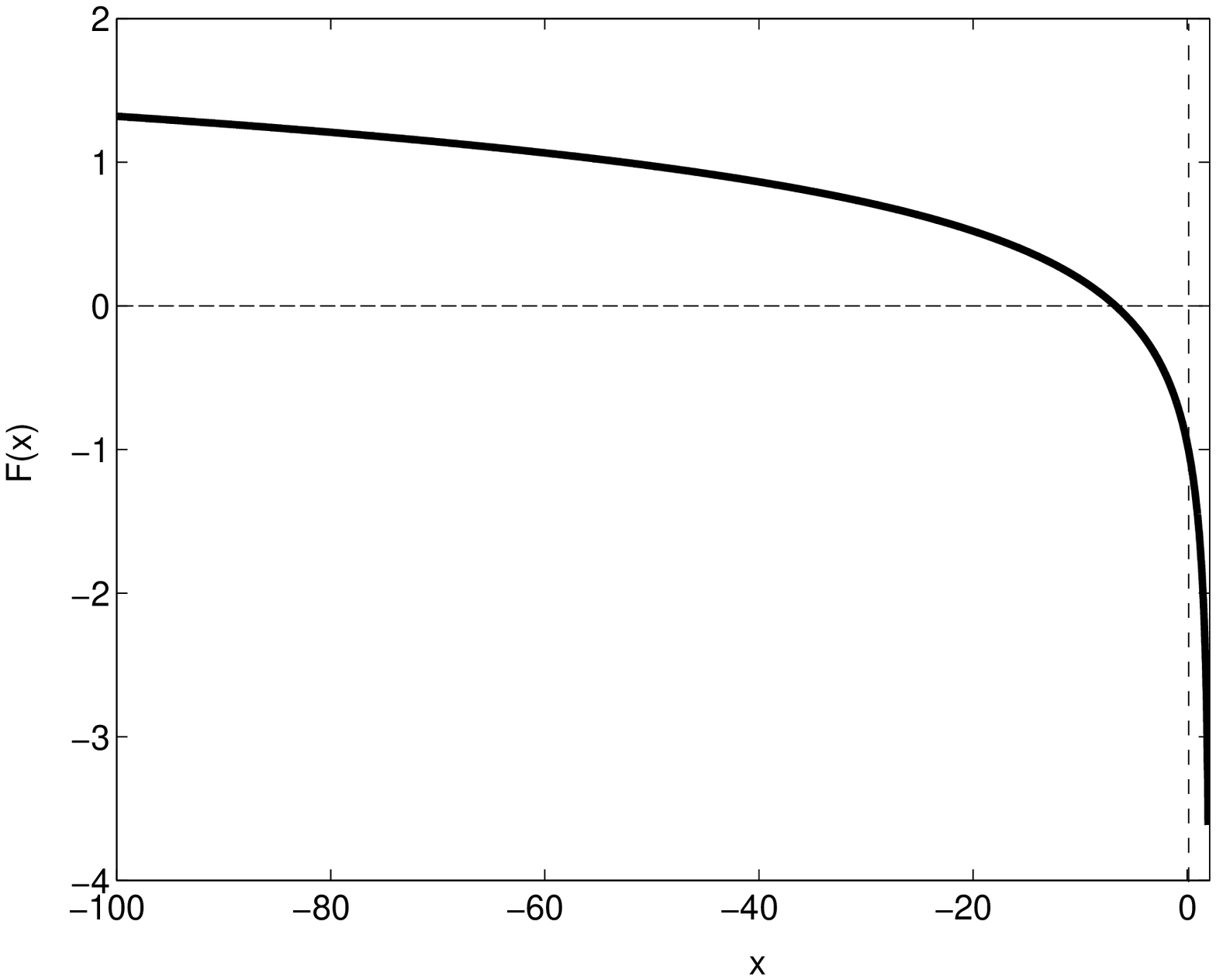} &
\includegraphics[scale=0.33]{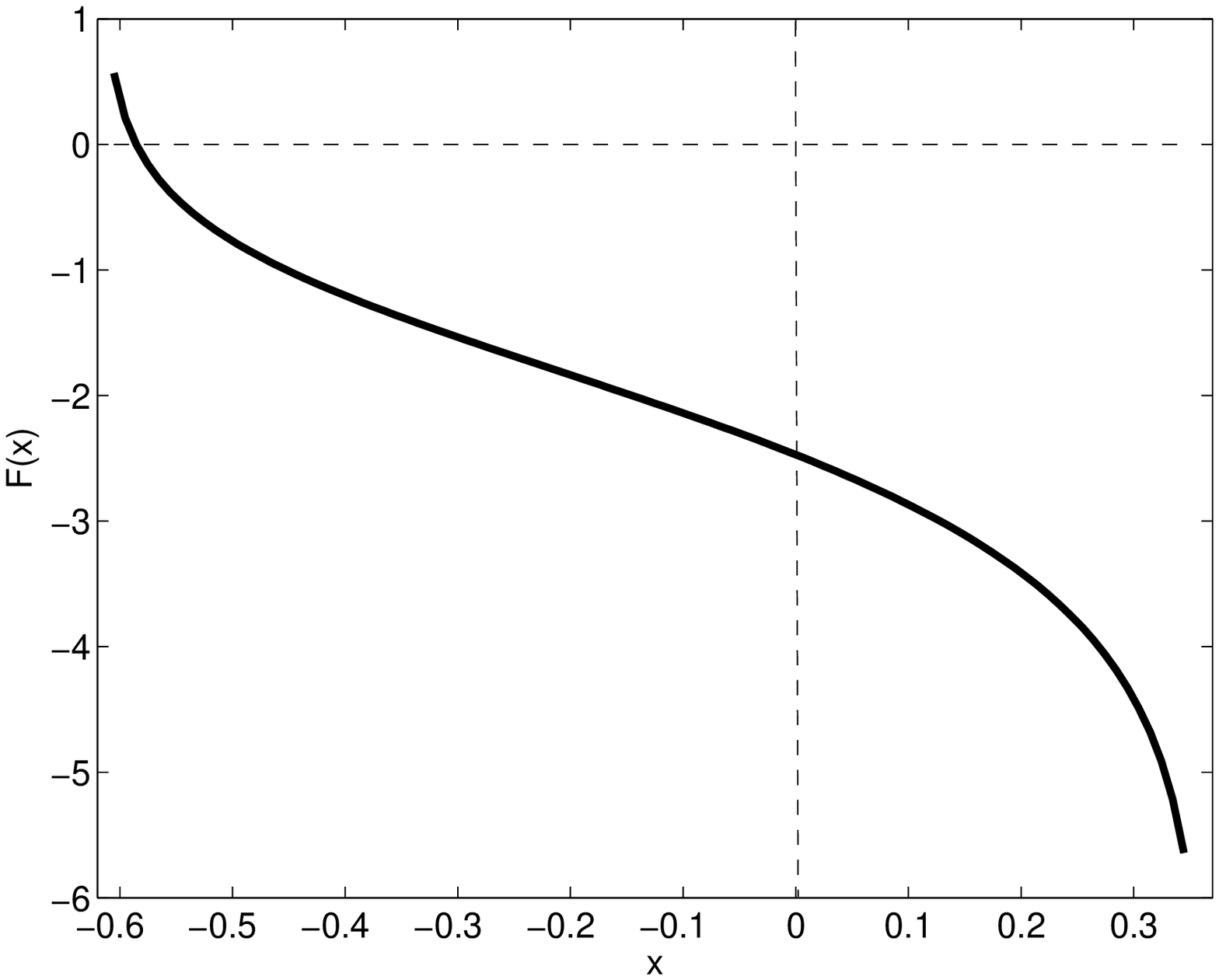}
\end{array}$
\end{center}
\caption{Function ${\textnormal{F}}_{1} (x) = (g_{1}^{1})^{T} \ln
{\textnormal{f}}_{1} (x)$ corresponding to linkage class 1. Note
that for the other linkage classes the functions are zero since
vectors $g_2^1$ and $g_3^1$ are identically zero. In (a) the
function is represented for the set of rate coefficients in Table
\ref{tab:table_1}. In (b) the function is represented for two modified rate
constants, $k_{2,1} = k_{5,2} = 10.0$.} \label{fig:Fx_plot}
\end{figure}

The dimension of the stoichiometric subspace, which coincides with the rank of
matrix $S = \left[\begin{array}{ccc}
              S_1 & S_2 & S_3
            \end{array} \right]$,
is $s = 6$, and renders a network deficiency $\delta = 10 - 3 - 6 = 1$. Hence, the kernel of $S$ is one dimensional with a basis
${\textbf{g}}^1 = \left(\begin{array}{rrrrrrr}
                0 & -1/2 & 0 & 1 & 0 & 0 & 0
              \end{array} \right)^T $.
As shown in Section \ref{secsec:reaction_simplex}, we identify the following three sub-vectors in ${\textbf{g}}^1$ that solve (\ref{eq:formalSg}):
\[\begin{array}{l}
g_{1}^{1} = \left(\begin{array}{rrrr}
                    0 & -1/2 & 0 & 1
                  \end{array} \right)^{T}, \\
g_{2}^{1} = \left(\begin{array}{rr}
                    0 & 0
                  \end{array} \right)^{T}, \\
g_{3}^{1} = \left(
                    0 \right)^{T}.
\end{array}
\]
The canonical representation of the equilibrium set will be
expressed in terms of matrices $M_{\lambda}$ that appear in Eqn (\ref{eq:fundamental_Phi}). From the expressions (\ref{eq:example_fluxes}) for the fluxes, we
have that:
\begin{equation}
M_{1} = \left[ \begin{array}{rrrrr}
    -k_{1,2} & k_{2,1} & 0 & 0 & 0\\
    k_{1,2} & -(k_{2,1} + k_{2,3} + k_{2,5}) & k_{3,2} & 0 & k_{5,2}\\
    0 & k_{2,3} & -(k_{3,2} + k_{3,4} + k_{3,5}) & k_{4,3} & k_{5,3}\\
    0 & 0 & k_{3,4} & -k_{4,3} & 0\\
    0 & k_{2,5} & k_{3,5} & 0 & -(k_{5,2} + k_{5,3})
\end{array} \right]
\end{equation}
\begin{equation}
M_{2} = \left[ \begin{array}{rrr}
    -(k_{6,7}  + k_{6,8}) & k_{7,6} & k_{8,6}\\
    k_{6,7} & -(k_{7,6} + k_{7,8}) & k_{8,7} \\
    k_{6,8} & k_{7,8} & -(k_{8,6} + k_{8,7}) \\
\end{array} \right], ~~ M_{3} = \left[ \begin{array}{rr}
    - k_{9,10} & k_{10,9} \\
    k_{9,10} & - k_{10,9}
\end{array} \right].
\end{equation}
Comparing each matrix with the structure given in (\ref{eq:structM_k_l}), we get
for each linkage class:
\begin{equation}
E_{1} = \left[ \begin{array}{rrrr}
    -(k_{2,1} + k_{2,3} + k_{2,5}) & k_{3,2} & 0 & k_{5,2}\\
    k_{2,3} & -(k_{3,2} + k_{3,4} + k_{3,5}) & k_{4,3} & k_{5,3}\\
    0 & k_{3,4} & -k_{4,3} & 0\\
    k_{2,5} & k_{3,5} & 0 & -(k_{5,2} + k_{5,3})
\end{array} \right]
\end{equation}
\begin{equation}
E_{2} = \left[ \begin{array}{rr}
    -(k_{7,6} + k_{7,8}) & k_{8,7} \\
    k_{7,8} & -(k_{8,6} + k_{8,7}) \\
\end{array} \right], ~~ E_{3} = \left[ \begin{array}{r}
    - k_{10,9}
\end{array} \right]
\end{equation}
\begin{equation}\begin{array}{lll}
a_{1} = \left[ \begin{array}{rrrr}
    k_{1,2} & 0 & 0 & 0
\end{array} \right]^T  & a_{2} = \left[ \begin{array}{rr}
    k_{6,7} & k_{6,8}
\end{array} \right]^T  & a_{3} = \left[ \begin{array}{r}
    k_{9,10}
\end{array} \right]^T \\
b_{1} = \left[ \begin{array}{rrrr}
    k_{2,1} & 0 & 0 & 0
\end{array} \right]^T & b_{2} = \left[ \begin{array}{rr}
    k_{7,6} & k_{8,6}
\end{array} \right]^T & b_{3} = \left[ \begin{array}{r}
    k_{10,9}
\end{array} \right]^T
\end{array}
\end{equation}
Expressions of the form (\ref{eq:Fexplicit_first}), which describe the family of
solutions, become as follows:
\begin{equation}\begin{array}{l}
{\textbf{f}}_{1} (x_1) = {\textbf{f}}_{1}^{*} + x_1 {\textbf{h}}_1\\
{\textbf{f}}_{2} (x_2) = {\textbf{f}}_{2}^{*} \\
{\textbf{f}}_{3} (x_3) = {\textbf{f}}_{3}^{*}
\end{array}
\end{equation}
where vectors at the right hand side, for the parameters given in
Table \ref{tab:table_1}, become:
\begin{equation}\begin{array}{ll}
 {\textbf{f}}_{1}^{*} = (1.0000 ~~ 30.1818 ~~ 75.4545 ~~ 2.1091 )^T, & {\textbf{f}}_{2}^{*} = ( 0.2000 ~~ 0.8571)^T \\
 {\textbf{h}}_1 = (  -0.2500 ~~  -7.5455 ~~ -18.8636 ~~  -1.0273)^T & {\textbf{f}}_{3}^{*}= ( 0.2500)
\end{array}
\end{equation}
\begin{table}[h!]
\caption{Reaction rate coefficients for the network}
\label{tab:table_1}
\begin{tabular}{l l l l l l}
\hline
 $k_{1,2} = 2.0$  & $k_{2,1} = 2.0$ & $k_{2,3} = 16.0 $ & $k_{3,2} = 0.5 $  & $k_{3,4}=2.5 $ & $k_{4,3} =1.0 $ \\
 $k_{2,5} =1.2 $  & $k_{5,2} = 1.0 $ & $k_{3,5} = 0.1 $ & $k_{5,3} = 1.0 $  & $k_{6,7}=1.0 $ & $k_{7,6} =1.0 $ \\
 $k_{7,8} =10.0  $  & $k_{8,7} = 1.4 $ & $k_{8,6} = 2.1 $ & $k_{6,8} = 1.0 $  & $k_{9,10}=1.0 $ & $k_{10,9} =4.0$\\
\hline
\end{tabular}
\end{table}
As it can be seen in Figure \ref{fig:Fx_plot}a, function
${\textnormal{F}}_{1} (x_1) = (g_{1}^{1})^{T} \ln
{\textnormal{f}}_{1} (x_1) $ is monotonous decreasing, with one
solution (intersection with the $x$-axis) at $x_1^{*} = -6.69$ as asserted by Theorem \ref{prop:Def1_theorem_Alt}. Another example is represented in Figure
\ref{fig:Fx_plot}b, for a network with reaction rates as in Table
\ref{tab:table_1}, except for constants  $k_{2,1}$ and  $k_{5,2}$
which now take the value $10$. The domain of the function is now
constrained to the interval ${\mathbb{X}}_1 = (L_1^{-},L_1^{+})$,
with $L_1^{-} = -0.6154$ and $L_1^{+} = + 0.3636$. Vector ${\textbf{h}}_1$ for
this parameter set becomes ${\textbf{h}}_1 = (-0.0500 ~~ 0.1857 ~~ 0.1857 ~~
-0.0786)^T$. For this case the function crosses the $x$-axis at
$x_1^{*} = -0.5854$.

In order to compute all possible equilibrium solutions in the
concentration space, we make use of ${\mathbf{f}}_{\bm{\eta}}^T ({\textbf{x}}^{*}) =
\left[\begin{array}{ccc} {\textbf{f}}_{1}^T (x_1^{*}) &
{\textbf{f}}_{2}^{*T} & {\textbf{f}}_{3}^{*T}\end{array}\right]$,
where ${\textbf{f}}_{1}^T (x_1^{*})$ for $x_1^{*} = -6.69$ and $x_1^{*} = -0.5854$ takes, respectively, the
values:
\begin{eqnarray*}
 {\textbf{f}}_{1} (x_1^{*}) &=& (\begin{array}{cccc}    2.6725 &
80.6609 & 201.6523  &  8.9815 \end{array})^T,  \\
  {\textbf{f}}_{1} (x_1^{*}) &=& (\begin{array}{cccc} 0.2293 & 0.0056 &
0.0056 & 0.0746 \end{array})^T.
\end{eqnarray*}
The set of equilibrium concentrations is then computed by solving
$\ln {\mathbf{f}}_{\bm{\eta}} = S^T \ln \textbf{c}$. In particular, for this
network we can express the first $6$ chemical species in terms of
species $G$, what leads to a straight line in the $\ln \textbf{c}$-space, which intersects the interior of any positive stoichiometric compatibility class defined in
(\ref{eq:eqpolyhedron}) (with $B = \left( \begin{array}{rrrrrrr}
             1/2 & 1 & 1/2 & 1 & 3/2 & 3/2 & 1
           \end{array}\right)^T$) in just one point.
\section{Conclusions}
This contribution concentrates on the study of feasibility conditions to identify admissible equilibria for weakly reversible mass action law (MAL) systems. To that purpose, a flux-based form of the model equations describing the time evolution of the species concentration has been exploited, in combination
with results from the theory of linear compartmental systems to develop a canonical representation of the equilibrium set. Ingredients of such representation include the so-called family of solutions, with the corresponding positivity conditions, and the feasibility functions employed to characterize the set of feasible (equilibrium) solutions.

One main result of this contribution is that the introduced feasibility functions are monotonously decreasing on their domain. This allows us to establish connections with classical results in CRNT related to the existence and uniqueness of equilibria within positive stoichiometric compatibility classes.
In particular, we employ monotonicity to identify regions in the set of possible reaction rate coefficients leading to complex balancing, and to conclude uniqueness of equilibria for a class of positive deficiency networks. It is our hope that the proposed results might support the understanding of the deficiency one theorem from a different point of view, with the possibility of an alternative proof.

A number of examples of different complexity are employed to illustrate the notions presented and their relations. As the examples show, all components used for the characterization of equilibria, in particular the family of solutions and the feasibility functions, can be computed efficiently in an algorithmic way, even for large kinetic models. Future work will be focused on the constructive application of these functions for the computational search or design of networks with unique equilibria.

{\bf Acknowledgements}

AAA acknowledges partial financial support by grants PIE201230E042 and grant Salvador de Madariaga (PR2011-0363). GS acknowledges the support of the grants NF104706 from the Hungarian Scientific Research Fund, and KAP-1.1-14/029 from Pazmany Peter Catholic University.

\bibliographystyle{plain}
%%\bibliographystyle{unsrt}
%%\bibliography{CRNT_refs}       % expects file "myrefs.bib"

\newpage
\appendix
\numberwithin{equation}{section}

\section{Some results required to prove Theorem \ref{lem:Lemma_on_F}}
\label{sec:appendixAAA}

%%The following results directly relate to the properties of C-Metzler matrices.
\begin{lemma}\label{lem:HQ_result}
Let $H \in {\mathbb{R}^{n \times n}}$ be C-Metzler and such that:
\begin{equation}\label{eq:H1a}
H {\mathbf{1}}_{n} = -\mathbf{a},
\end{equation}
with $\mathbf{a} \geq 0$. Let $\mathbf{p}\in {\mathbb{R}}^n$ (with $\mathbf{p}\ne 0$)  be a vector with
$m$ positive, $r$ zero and $n-m-r$ negative components satisfying:
\begin{equation}
\begin{array}{l}
  p_1 \geq \cdots \geq p_k \geq \cdots \geq p_m > 0 > p_{m+r+1} \geq
\cdots \geq p_{\ell} \geq \cdots  \geq p_n,
\\
p_{m+1} = \cdots = p_{m + r} = 0,
\end{array}
\end{equation}
and
\begin{equation}\label{eq:gHp}
\mathbf{g} = H \mathbf{p}.
\end{equation}
Then:
\begin{equation} \label{eq:FirstR}
\sum_{i=1}^{k} g_i \leq 0 ~~ \textnormal{for every }~~ 1 \leq k \leq
m,
\end{equation}
and
\begin{equation}\label{eq:SecondR}
\sum_{i=\ell }^{n} g_i \geq 0 ~~ \textnormal{for every }~~ m+r+1
\leq \ell \leq n.
\end{equation}
Moreover:
\begin{equation}\label{eq:defsign_Final}
\sum_{i=1}^{m} g_i < 0 ~~ \textnormal{and}~~ \sum_{i=m+r+1 }^{n} g_i
> 0.
\end{equation}
\end{lemma}

\textbf{Proof:} Multiplying both sides of (\ref{eq:H1a}) by the
scalar $p_k >0$ and subtracting the result from (\ref{eq:gHp}), we
get:
\begin{equation}
H (\textbf{p} - p_k {\mathbf{1}}_{n} ) = \textbf{g} + p_k \textbf{a}.
\label{eq:Hpk}
\end{equation}
Summing the first $k$ elements and reordering terms results in:
\begin{equation}\label{eq:The_Equivalence}
\sum_{i=1}^{k} g_i = \sum_{j=1}^{k-1}  \left(\sum_{i=1}^k
H_{ij} \right) (p_j-p_k)
 + \sum_{j=k +1}^{n} \left( \sum_{i=1}^{k} H_{ij} \right) (p_j -
p_k) + (- p_k) \sum_{i=1}^{k} a_i.
\end{equation}
Since $H$ is C-Metzler, according to Definition \ref{def:C-Metzler}, for every $j = 1, \ldots, k$, with  $k = 1, \ldots, n$, we have that:
\[
\sum_{i=1}^k H_{ij} = - b_{j} - \sum_{i=k+1}^n H_{ij} \le 0.
\]
The first term at the right hand side of (\ref{eq:The_Equivalence}) is non-positive since,  by construction,  $p_j
-p_k \geq 0$ for $j = 1, \ldots, k-1$, and the above summations are non-positive.
The second term is non-positive since for every $j = k+1, \ldots, n$ and $i \neq j$,
$H_{ij} \geq 0$ and $p_j -p_k \leq 0$.  Thus,
relation (\ref{eq:FirstR}) follows, since $\textbf{a}$ is a nonnegative vector and $p_k$ is positive for $k = 1, \ldots, m$, so the third term at the right hand side of (\ref{eq:The_Equivalence}) is also non-positive.

In a similar way we prove (\ref{eq:SecondR}). Substituting
$p_{\ell}<0$ for $p_k$ in \eqref{eq:Hpk}, we get:
\begin{equation}
H (\textbf{p} - p_{\ell} {\mathbf{1}}_{n} ) = \textbf{g} + p_{\ell}
\textbf{a}. \label{eq:Hpl}
\end{equation}
Summing the elements of $\textbf{g}$ from $\ell = m + r +1, \ldots, n $ gives:
\begin{equation}\label{eq:sumHpl}
\sum_{i=\ell}^{n} g_i = \sum_{j= \ell +1}^{n}  \left(\sum_{i= \ell}^n
H_{ij} \right) (p_j-p_{\ell})
 + \sum_{j=1}^{\ell -1} \left( \sum_{i=\ell}^{n} H_{ij} \right) (p_j -
p_{\ell}) + (- p_{\ell}) \sum_{i= \ell}^{n} a_i.
\end{equation}
Because $H$ is C-Metzler, we have that:
\[
\sum_{i=\ell}^n H_{ij} \le 0,~\text{for any}~ j = \ell, \ldots, n.
\]
Thus, the first term at the right hand side of of \eqref{eq:sumHpl} is non-negative,
since $(p_j-p_{\ell})\le 0$ for $j = \ell +1, \ldots, n$. The second term in the expression is also non-negative, since the
off-diagonal elements of $H$ are non-negative and $(p_j-p_{\ell})\ge
0$ for $j =1, \ldots,  \ell -1$. Finally, the last term in
\eqref{eq:sumHpl} is non-negative due to the negativity of
$p_{\ell}$ and the non-negativity of $\textbf{a}$.

Strict inequalities (\ref{eq:defsign_Final}) can be proven in a straightforward manner from expressions (\ref{eq:The_Equivalence}) and (\ref{eq:sumHpl}),
if the non-zero components of vector $\textbf{a} \geq 0$ are within the first $m$ and last $n-m-r$ entries. This would be the case, since the last terms at the right hand side in both equations would be strictly negative (with $k=m$), and positive (with $\ell = m +r +1$), respectively.

If the non-zero components are not within the first $m$, nor within the last $n-m-r$ entries, the strict inequalities still hold. In order to prove this point, we express $H$ as:
\begin{equation}\label{eq:H_Factor}
H = \left[
\begin{array}{c|c}
  H_{11} & H_{12} \\
  \hline
  H_{21} & H_{22}
\end{array}
\right],
\end{equation}
where $H_{11} \in {\mathbb{R}}^{m \times m}$.  Let the first $m$ components of vector $\textbf{a}$ to be zero. Then,  $H_{12} \in {\mathbb{R}}^{m \times (n-m)}$ in (\ref{eq:H_Factor}) must necessarily have at least one positive element (any non-zero element must be positive because $H$ is C-Metzler). Suppose, on the contrary, that $H_{12}$ is a zero matrix. Then, by using (\ref{eq:H1a}) we have that:
\[
H_{11}{\bm{1}}_m  = 0,
\]
which means that $H_{11}$, and consequently $H$, are not invertible, contradicting the fact  that $H$ is C-Metzler and therefore, non-singular. Since at least one entry of $H_{12}$ is positive, the second term at the right hand side of (\ref{eq:The_Equivalence}) for $k = m$ must be strictly negative.

A similar line of arguments can be employed if the last $n-m-r$ components of $\textbf{a}$ are zero, with matrix $H_{22} \in {\mathbb{R}}^{(n-m-r) \times (n-m-r)}$ and $H_{21} \in {\mathbb{R}}^{(n-m-r) \times (m+r)}$, instead of $H_{11}$ and $H_{12}$.
Now, we suppose that $H_{21}$ is a zero matrix, what combined with (\ref{eq:H1a}) leads to:
\[
H_{22}{\bm{1}}_{n-m-r}  = 0.
\]
Thus $H_{22}$, and consequently $H$, are not invertible, what is in contradiction with the fact that $H$ is C-Metzler and therefore, non-singular. Since at least one entry of $H_{21}$ must be positive, the second term at the right hand side of (\ref{eq:sumHpl}), for $\ell = n-m-r$, must be strictly positive, completing the proof. \hfill $\Box$

\begin{lemma}\label{lem:G_result}
Let ${\mathbb{X}} \subset \mathbb{R}$ and consider the function $G(x):{\mathbb{X}} \mapsto \mathbb{R}$ defined as:
\begin{equation}\label{eq:Gx}
G(x) = \sum_{i=1}^{n} g_i Q_i (x),
\end{equation}
where $g_i$ are the coordinates of the vector $\textbf{g} = H \textbf{p}$,
with $H$ and $\textbf{p}$ as in Lemma \ref{lem:HQ_result}. For every
$i=1, \ldots, n$ and $x \in {\mathbb{X}}$, let also have that:
\begin{equation}\label{eq:PutoOrdenMCGED}
\begin{array}{l}
  Q_1 (x) \geq \cdots \geq Q_k (x) \geq \cdots \geq Q_m (x) > 0 > Q_{m+r+1} (x) \geq
\cdots \geq Q_{\ell} (x) \geq \cdots  \geq Q_n (x),
\\
Q_{m+1} (x) = \cdots = Q_{m + r} (x) = 0.
\end{array}
\end{equation}
Then, $G(x) <0$ for every $x \in {\mathbb{X}}$.
\end{lemma}

\textbf{Proof:} First, we note that (\ref{eq:Gx}) can be re-written
as:
\[
G(x) = (Q_1 (x) -Q_2 (x)) g_1 + (Q_2 (x) - Q_3 (x) ) (g_1 + g_2) + \cdots + (Q_k (x) -
Q_{k+1} (x))\sum_{i=1}^{k} g_i + \cdots + Q_m (x) \sum_{i=1}^{m} g_i +
\]
\begin{equation}\label{eq:Gx_equiv}
 Q_{m + r + 1} (x) \sum_{i=m + r +1 }^{n} g_i + \cdots +
(Q_{\ell} (x) - Q_{\ell -1} (x) ) \sum_{i=\ell }^{n} g_i + \cdots + (Q_n (x) -
Q_{n-1} (x) ) g_n,
\end{equation}
where implicitly, each $Q_i$ is assumed to be a function of $x$. From (\ref{eq:PutoOrdenMCGED}), we have that $Q_{i} (x) - Q_{j} (x)
\geq 0$ for every $Q_{i} (x) \geq Q_{j} (x)$ and $x \in \mathbb{X}$,
what implies that $(Q_k (x) - Q_{k+1} (x)) \geq 0 $ for every $k = 1,
\ldots, m-1$, and $(Q_{\ell} (x) - Q_{\ell -1} (x)) \leq 0$ for every $\ell =
m + r +1, \ldots, n$. Thus, from Lemma \ref{lem:HQ_result}, we have
that:
\[
G(x) \leq  Q_m (x) \sum_{j=1}^{m} g_j + Q_{m + r + 1} (x) \sum_{j=m + r
+1 }^{n} g_j.
\]
The signs $Q_m (x) > 0$, $Q_{m + r + 1} (x) <0$ as well as inequalities
(\ref{eq:defsign_Final}), from Lemma \ref{lem:HQ_result}, make the right hand side
of the above expression strictly negative, what completes the proof.
\hfill $\Box$

\begin{proposition}\label{prop:DefSigns_s_t}
Under the conditions of Lemma \ref{lem:HQ_result}, let the $m$ positive and the $n-m-r$ negative components of $\textbf{p}\in {\mathbb{R}}^n$ satisfy:
\begin{eqnarray}
p_1 &\geq &  \cdots \geq p_{s-1} \geq p_s > p_{s+1} \geq \cdots \geq p_m > 0,
\label{eq:Seq_s_positive}\\
0 & > & p_{m+r+1} \geq \cdots \geq p_{t-1} > p_t \geq p_{t+1} \geq  \cdots  \geq p_n.
\label{eq:Seq_t_negative}
\end{eqnarray}
Then:
\begin{equation}\label{eq:DefiniteSign_s_t}
\sum_{i=1}^{s} g_i < 0 ~~ \textnormal{and}~~ \sum_{i=t}^{n} g_i
> 0,
\end{equation}
for some $s=1, \ldots, m-1$ and $t= m+r+2, \ldots, n$.
\end{proposition}

\textbf{Proof:}
The line of arguments is similar to that employed in Lemma \ref{lem:HQ_result} to prove (\ref{eq:defsign_Final}). If the non-zero components of vector $\textbf{a} \geq 0$ are within the first $s$ and the last $n-t +1$ entries, it is straightforward to prove strict inequalities  from expressions (\ref{eq:The_Equivalence}) and (\ref{eq:sumHpl}),
for the last terms at the right hand side in both equations (with $k=s$ and $\ell = t$) are strictly negative, and positive, respectively.

If, on the other hand, the first $s$ and the last $n-t +1$ entries of $\textbf{a} \geq 0$ are zero, then matrix $H$, which can be expressed as in (\ref{eq:H_Factor}) with $H_{11} \in {\mathbb{R}}^{s \times s}$, must have for $H_{12} \in {\mathbb{R}}^{s \times (n-s)}$ at least one positive element. Otherwise, from (\ref{eq:H1a}), we would have that:
\[
H_{11}{\bm{1}}_s  = 0,
\]
what contradicts the hypothesis that $H$ is C-Metzler and therefore, non-singular. Since at least one entry of $H_{12}$ must be positive, and because of  (\ref{eq:Seq_s_positive}), $p_j -p_s > 0$ (for $j = s +1 , \ldots, n$) the second term at the right hand side of (\ref{eq:The_Equivalence}), for $k = s$, is strictly negative, what proves the first inequality in (\ref{eq:DefiniteSign_s_t}).

In order to prove the second inequality, we make use of a similar argument with $H_{22} \in {\mathbb{R}}^{(n-t+1) \times (n-t+1)}$ in (\ref{eq:H_Factor}), to show that $H_{21} \in {\mathbb{R}}^{(n-t+1) \times (m+r)}$ must have at least one positive entry. From (\ref{eq:Seq_t_negative}), we also have that $p_j -p_t >0$ (for $j = 1, \ldots, t-1$) so the second term at the right hand side of (\ref{eq:sumHpl}), for $\ell = t$, must be strictly positive. This proves the  second inequality in (\ref{eq:DefiniteSign_s_t}). \hfill $\Box$

\begin{proposition}\label{lem:Limits_Sequences}
Let us consider the following set of ordered parameters $p_1 > p_k \geq p_{k+1} >0 > p_{\ell -1 } \geq p_{\ell} > p_n$,
%and the intervals ${\mathbb{X}}_k = (-1/p_k, + \infty)$ for $p_k >0$, and  ${\mathbb{X}}_{\ell} = (-\infty, -1/p_{\ell})$ for $p_{\ell} <0$.
%
and functions  ${\Pi}_{j}:\mathbb{R}  \rightarrow \mathbb{R}$ of the form  ${\Pi}_j (x) = \ln (1 +x p_j)$, with $p_j$ being a given parameter within the ordered set. Then, we have that:
\begin{equation}\label{eq:Limits1}
\lim_{x^+ \rightarrow -(\frac{1}{p_1})} {\Pi}_1 (x) = \lim_{x^- \rightarrow -(\frac{1}{p_n})} {\Pi}_n (x) = - \infty,
\end{equation}
%%\emptyset
\begin{equation}\label{eq:Limits2}
\lim_{x^+ \rightarrow -(\frac{1}{p_1})} ({\Pi}_1 (x) - {\Pi}_{k} (x) ) = \lim_{x^- \rightarrow -(\frac{1}{p_n})} ({\Pi}_{n} (x) - {\Pi}_{\ell} (x)) = - \infty,
\end{equation}
\begin{equation}\label{eq:Limits3}
\lim_{x^+ \rightarrow -(\frac{1}{p_1})} ({\Pi}_k (x) - {\Pi}_{k+1} (x) ) = C_1, ~~
\lim_{x^- \rightarrow -(\frac{1}{p_n})} ({\Pi}_{\ell -1} (x) - {\Pi}_{\ell} (x) ) = C_2,
\end{equation}
\begin{equation}\label{eq:Limits4}
\lim_{x \rightarrow + \infty} ({\Pi}_k (x) - {\Pi}_{k+1} (x) ) = C_3
, ~~
\lim_{x \rightarrow - \infty} ({\Pi}_{\ell -1} (x) - {\Pi}_{\ell} (x) ) = C_4,
\end{equation}
where $C_1$, $C_2$, $C_3$ and $C_4$ are constants.
\end{proposition}

\textbf{Proof:}
The limits in  (\ref{eq:Limits1}) follow since  ${\Pi}_1$ increases and ${\Pi}_n$ decreases monotonically in their respective domains $(-1/p_1, +\infty)$, $ (- \infty , - 1/p_n)$. In order to prove (\ref{eq:Limits2}), we have that:
\[
\lim_{x^+ \rightarrow -(\frac{1}{p_1})} {\Pi}_k (x)  = {\Pi}_k(-1/p_1),
~\text{and}~ \lim_{x^- \rightarrow -(\frac{1}{p_n})} {\Pi}_{\ell} (x) = {\Pi}_{\ell}(-1/p_n),
\]
which are (negative) constants, because $p_1 > p_k$ and $p_{\ell} > p_n$ . Using (\ref{eq:Limits1}), we then get (\ref{eq:Limits2}).

In order to compute the limits in (\ref{eq:Limits3}),  we have that:
\[
0 < 1 - \frac{p_k}{p_1} \leq 1 - \frac{p_{k+1}}{p_1}. ~\text{Thus,}~
\lim_{x^+ \rightarrow -(\frac{1}{p_1})} ({\Pi}_{k} (x) - {\Pi}_{k+1} (x)) = \ln \frac{1 - (p_k/p_1)}{1 - (p_{k+1}/p_1)} \leq 0.
\]
Similarly:
\[
0 < 1 - \frac{p_{\ell}}{p_n} \leq 1 - \frac{p_{\ell -1}}{p_n}. ~\text{Thus,}~
\lim_{x^- \rightarrow -(\frac{1}{p_n})} ({\Pi}_{\ell -1} (x) - {\Pi}_{\ell} (x) )
 = \ln \frac{1 -(p_{\ell -1}/p_n)}{1 - (p_{\ell}/p_n)} \geq 0.
\]
In proving (\ref{eq:Limits4}), we have that:
\[
\lim_{x \rightarrow + \infty} ({\Pi}_k (x) - {\Pi}_{k+1} (x) ) = \ln \lim_{x \rightarrow + \infty} \frac{1 + x p_{k}}{1 + x p_{k+1}},
~\text{and}~
\lim_{x \rightarrow - \infty} ({\Pi}_{\ell -1} (x) - {\Pi}_{\ell} (x) ) = \ln \lim_{x \rightarrow - \infty} \frac{1 + x p_{\ell -1}}{1 + x p_{\ell}}.
\]
Thus, by the theorem of l'Hopital, we have that:
\[
\lim_{x \rightarrow + \infty} ({\Pi}_k (x) - {\Pi}_{k+1} (x) ) = \ln \frac{p_{k}}{p_{k+1}} \geq 0,
~\text{and}~
\lim_{x \rightarrow - \infty} ({\Pi}_{\ell -1} (x) - {\Pi}_{\ell} (x) ) = \ln \frac{p_{\ell -1}}{p_{\ell}} \leq 0.
\]
\hfill $\Box$

\section{Some convenient results on uniqueness and stability} \label{sec:appendixBBB}

For the sake of completeness, here we summarize in the form of propositions, two  fundamental results from CRNT on uniqueness and stability. The complete set of arguments can be found in \cite{Feinberg:79}.

\begin{lemma}[see also \cite{alonso_ydstie:01}]\label{lemma:convexity_SH}
Let $V(x): \mathbb{X} \rightarrow \mathbb{R}$, with
$\mathbb{X}\subseteq {\mathbb{R}}^n$ its domain, a convex function
with continuous derivatives in $\mathbb{X}$, and $\nu(x): \mathbb{X}
\rightarrow {\mathbb{R}}^n$ be the gradient of $V(x)$. Then, the
following inequalities hold for every $x \in \mathbb{X}$:
\begin{description}
  \item[(i)] ${\nu}^{T} (x_1)(x-x_1) \leq V(x) - V(x_1)$, for any
  $x_1 \in \mathbb{X}$.
  \item[(ii)] $\left[ \nu(x_2) - \nu(x_1) \right]^{T} (x_2 - x_1) \geq
  0$, for any $x_1, x_2 \in \mathbb{X}$.
\end{description}
Inequalities are strict whenever $x \neq x_1$ or $x_1 \neq x_2$ in
(i) and (ii), respectively.
\end{lemma}

\textbf{Proof:} In order to prove the first part, choose any $x_1 \in
\mathbb{X}$ and construct a function $B_{1}(x; x_1)$ as the
difference between $V(x)$ and its supporting hyperplane at $x_1$.
The supporting hyperplane is of the form $H(x; x_1) = V(x_1) + {\nu}^{T}(x_1) (x-x_1) $, and $B_{1}(x; x_1) =
V(x) - H(x;x_1)$.
By construction, the function is strictly positive, i.e. it is
positive for all $x \in \mathbb{X}$ other than $x_1$, and result
(i) follows in a straightforward manner, since $B_{1}(x; x_1) = V(x) - V(x_1) - {\nu}^{T}(x_1) (x-x_1) \geq
0$, what implies that $V(x) - V(x_1) \geq {\nu}^{T}(x_1) (x-x_1)$.

To prove the second part, we note that $B_{1}(x;x_1)$ is itself a
convex function since $ {\nabla}_x B_{1} = \nu(x) -\nu(x_1)$, so its
Hessian coincides with that of the convex function $V(x)$. By using
the same supporting hyperplane argument, we construct the following
strictly positive definite function around some $x_2 \in
\mathbb{X}$:
\[
B_{2}(x; x_1, x_2) = B_{1}(x; x_1) - B_{1}(x_2; x_1) -
\left[\nu(x_2) - \nu(x_1) \right]^{T} (x - x_2) \geq 0,
\]
where the inequality holds for any $x \in \mathbb{X}$. In particular,
it holds for $x = x_1$, and therefore:
\[
B_{1}(x_2; x_1) + \left[\nu(x_2) - \nu(x_1) \right]^{T} (x_1 - x_2)
\leq 0,
\]
which implies that $B_{1}(x_2; x_1) \leq \left[\nu(x_2) - \nu(x_1)
\right]^{T} (x_2 - x_1)$, and proves (ii).
\hfill $\Box$

\begin{proposition}[Corollary 4.14 \cite{Feinberg:79}]\label{prop:Unique_in_CC}
Let $\mathbf{c}_0 \in {\mathbb{R}}^{m}_{>0}$ be a fixed reference. The set
\begin{equation}\label{eq:TheFamousSetUE}
{\cal{U}} (\mathbf{c}_0) = \{ \mathbf{c} \in \mathbb{R}_{>0}^{m}
~ | ~  S^\mathrm{T} (\ln
\mathbf{c}- \ln \mathbf{c}_0) = 0 \} ,
\end{equation}
contains exactly one element in
each positive stoichiometric compatibility class.
\end{proposition}
\textbf{Proof:} In proving uniqueness, suppose that there are two elements: ${\textbf{c}}^{*}, {\textbf{c}}^{**} \in {\cal{U}} (\textbf{c}_0)$, that belong to the same stoichiometric compatibility class.
Then, we have that $S^{T} (\ln {\textbf{c}}^{*} - \ln
{\textbf{c}}^{**}) = 0$, what implies that $(\ln {\textbf{c}}^{*} -
\ln {\textbf{c}}^{**})$ is orthogonal to the stoichiometric subspace
$\Xi$. Because ${\textbf{c}}^{*}$ and
${\textbf{c}}^{**}$ are assumed to be in the same compatibility
class, the vector ${\textbf{c}}^{*} -{\textbf{c}}^{**}$ must belong to the
stoichiometric subspace $\Xi$, and the following relation hold:
\begin{equation}\label{eq:lncSORTcS}
(\ln {\textbf{c}}^{*} - \ln {\textbf{c}}^{**})^{T}({\textbf{c}}^{*}
-{\textbf{c}}^{**}) = 0.
\end{equation}
Using the convex function $V(\textbf{c}) = \textbf{c}^{T} (\ln
\textbf{c} - \textbf{1})$, with gradient $\nu(\textbf{c}) = \ln \textbf{c}$, and applying Lemma \ref{lemma:convexity_SH} (condition (ii)), it follows that equality (\ref{eq:lncSORTcS}) holds if and only if ${\textbf{c}}^{*} = {\textbf{c}}^{**}$, what proves that the set ${\cal{U}} (\mathbf{c}_0)$
can have at most one element in each positive stoichiometric compatibility class.

As pointed out in \cite{Feinberg:79}, the question of existence (i.e. that each (positive) stoichiometric compatibility class in fact meets ${\cal{U}} (\mathbf{c}_0)$) is somewhat more difficult to answer than uniqueness. The complete argument can be found in \cite{Feinberg:79} (Proposition 4.13). \hfill $\Box$

\begin{proposition}[see also \cite{Feinberg:79}]\label{prop:B_Lyapunov}
Complex balanced equilibria are locally asymptotically stable in all positive stoichiometric compatibility classes.
\end{proposition}
\textbf{Proof:} First of all, let us make use of
Eqn (\ref{eq:Ak_by_fluxes_A_Feinb}) to write the right hand side of
system (\ref{subeq:canonical_form_AK}) as a summation over $\lambda$, of
functions:
\begin{equation}
           {\cal{R}}^{\lambda} (\textbf{c}) \equiv
           \sum_{i \in {\cal{L}}_{\lambda}} \psi_i (\textbf{c}) \sum_{j \in \mathcal{I}_i} k_{ij} \cdot
(\bm{y}_j - \bm{y}_i).
\end{equation}
Select some positive reference ${\textbf{c}}^{*} > 0$ (its associated
vector ${\bm{\psi}}^{*}$ is strictly positive) and re-write the
previous expression in the equivalent form:
\begin{equation} \label{eq:sff_per_linkage}
 {\cal{R}}^{\lambda} (\overline{\nu})   =\sum_{i \in {\cal{L}}_{\lambda}} {\textnormal{e}}^{\bm{y}^{T}_i
\overline{\nu}} \sum_{j \in \mathcal{I}_i} \psi_{i}^{*} k_{ij} \cdot
(\bm{y}_j - \bm{y}_i),
\end{equation}
where $\overline{\nu} = \ln \textbf{c} -\ln {\textbf{c}}^{*}$. The
inner product between $\overline{\nu}$ and ${\cal{R}}^{\lambda}
(\textbf{c})$ (\ref{eq:sff_per_linkage}) results into the following
scalar function:
\begin{equation}\label{eq:function_g}
{\overline{\nu}}^{T}  {\cal{R}}^{\lambda}  (\overline{\nu}) = \sum_{i
\in {\cal{L}}_{\lambda}} {\textnormal{e}}^{z_i (\overline{\nu})} \sum_{j \in
\mathcal{I}_i} \psi_{i}^{*} k_{ij} \cdot (z_j (\overline{\nu}) - z_i (\overline{\nu})),
\end{equation}
where $z_i (\overline{\nu}) = \bm{y}^{T}_i \overline{\nu}$. In order to get an upper bound for
(\ref{eq:function_g}), we make use of Lemma
\ref{lemma:convexity_SH} (condition (i)), with the convex function $V(z)={\textnormal{e}}^{z}$, to obtain:
\begin{equation}\label{eq:ineq_exp}
{\textnormal{e}}^{z_i} (z_j - z_i) \leq {\textnormal{e}}^{z_j} -
{\textnormal{e}}^{z_i}.
\end{equation}
For any scalars $z_i$ and $z_j$. Strict convexity of $V(z)$ ensures
that the equality holds only if $z_i = z_j$. We also have that:
\begin{equation}\label{eq:factor_exp}
{\textnormal{e}}^{z_j} - {\textnormal{e}}^{z_i} =
(\bm{\varepsilon}_j - \bm{\varepsilon}_i)^{T} \sum_{k =1}^{n}
\bm{\varepsilon}_k {\textnormal{e}}^{z_k}.
\end{equation}
Combining (\ref{eq:factor_exp}) with (\ref{eq:ineq_exp}), and
substituting the resulting expression in (\ref{eq:function_g}), we
get:
\begin{small}
\begin{align}\label{eq:upper_bound_g}
& {\overline{\nu}}^{T}  {\cal{R}}^{\lambda}  (\overline{\nu}) \leq
\left( \sum_{i =1}^{n} {\textnormal{e}}^{z_i (\overline{\nu})}
\bm{\varepsilon}_{i}^{T} \right) \left[ \sum_{i \in
{\cal{L}}_{\lambda}} \psi_{i}^{*} \sum_{j \in \mathcal{I}_i} k_{ij}
\cdot (\bm{\varepsilon}_j - \bm{\varepsilon}_i) \right] = \\ &
\left(\sum_{i =1}^{n} {\textnormal{e}}^{z_i (\overline{\nu}) }
\bm{\varepsilon}_{i}^{T} \right) A^{\lambda}_{k}
({\bm{\psi}}^{*}). \nonumber
\end{align}
\end{small}
If the reference corresponds with a complex balanced equilibrium, then for every $\lambda = 1, \ldots, \ell$, $A^{\lambda}_{k}
({\bm{\psi}}^{*}) = 0$ and so is the right hand side of (\ref{eq:upper_bound_g}). Note that inequality is strict, in the sense that it holds whenever
$z_i \neq z_j$, for every $i,j \in {\cal{L}}_{\lambda}$.

Local asymptotic stability is proved by the standard Lyapunov stability method (see for instance \cite{Khalil1996}) with the following Lyapunov function candidate, constructed as in the proof of Lemma
\ref{lemma:convexity_SH}:
\[
B (\textbf{c}; {\textbf{c}}^{*}) = V(\textbf{c}) -
V({\textbf{c}}^{*}) - {\nu}^{T}({\textbf{c}}^{*})
({\textbf{c}}-{\textbf{c}}^{*}) \geq 0,
\]
with $V(\textbf{c})$, being a convex function of the form:
\[
V(\textbf{c}) =  \sum_{i=1}^{m} {\textnormal{c}}_i (\ln
{\textnormal{c}}_i -1).
\]
Computing the derivative of $B$ along
(\ref{subeq:canonical_form_AK}), and using (\ref{eq:upper_bound_g}), we get:
\begin{equation}
\dot{B} =  \sum_{\lambda = 1}^{\ell} {\overline{\nu}}^{T}
{\cal{R}}^{\lambda} (\overline{\nu}) \leq \left(\sum_{i =1}^{n}
{\textnormal{e}}^{z_i (\overline{\nu})} \bm{\varepsilon}_{i}^{T} \right)\sum_{\lambda
= 1}^{\ell} A^{\lambda}_{k} ({\bm{\psi}}^{*}) = 0.
\end{equation}
The result then follows, since $B(\textbf{c}; {\textbf{c}}^{*})
\geq 0$ and $\dot{B}(\textbf{c}; {\textbf{c}}^{*}) \leq 0$, with equality only if $\textbf{c} = {\textbf{c}}^{*}$.  \hfill $\Box$
\end{document}